\documentclass[aps,pra,reprint,superscriptaddress,10pt,footinbib,longbibliography]{revtex4-2}

\usepackage[english]{babel}
\usepackage{graphicx}
\usepackage{color}
\usepackage[usenames,dvipsnames]{xcolor}
\usepackage{amsmath,bbm,amssymb, amsthm}
\usepackage[normalem]{ulem}
\usepackage{multirow}
\usepackage{hyperref}
\usepackage{tikz}
\usetikzlibrary{quantikz}
\usepackage{thmtools}

\newcommand{\ketbra}[2]{{|#1 \rangle \langle #2|}}
\newcommand{\Xgen}{\mathcal{X}_\mathrm{gen}}
\newcommand{\Zgen}{\mathcal{Z}_\mathrm{gen}}
\newcommand{\upi}{\mathrm{i}}

\newcommand{\cnot}{\textsc{cnot}}
\newcommand{\tauP}{\tau_\mathrm{phys}}
\newcommand{\tauL}{\tau_\mathrm{L}}
\newcommand{\tauG}{\tau_\mathrm{G}}
\newcommand{\tauAnc}{\tau_\mathrm{anc}}
\newcommand{\tauMeas}{\tau_\mathrm{M}}

\newcommand{\cC}{\mathcal{C}}
\newcommand{\cG}{\mathcal{G}}
\newcommand{\Gphys}{\cG_\mathrm{phys}}
\newcommand{\GL}{\cG_\mathrm{L}}
\newcommand{\cS}{\mathcal{S}}
\newcommand{\Sphys}{\cS_\mathrm{phys}}
\newcommand{\SL}{\cS_\mathrm{L}}
\newcommand{\cM}{\mathcal{M}}
\newcommand{\Mphys}{\cM_\mathrm{phys}}
\newcommand{\ML}{\cM_\mathrm{L}}
\newcommand{\psiL}{\ket{\psi}_{\!\mathrm{L}}}
\newcommand{\Qphys}{Q^\mathrm{phys}}
\newcommand{\Qalgo}{Q^\mathrm{algo}}
\newcommand{\muCl}{\mu_{0/\!+}}
\newcommand{\muClFlag}{\mu^{\text{flag}}_{0/\!+}}
\newcommand{\Lmax}{\mathcal{L}_{\rm max}}

\newcommand{\total}{{\operatorname{total}}} 

\newcommand{\opnormal}{{\operatorname{normal}}}
\newcommand{\opmagic}{{\operatorname{magic}}}

\newcommand{\id}{\mathbbm{1}}

\begin{document}

\title{Magic states are rarely the best resource to optimize: \\
An analytical tool for qubit resource estimation in concatenated codes
}




\author{Marco \surname{Fellous-Asiani}}
\email{fellous.asiani.marco@gmail.com}
\affiliation{Université de Lorraine, CNRS, INRIA, LORIA, 54000 Nancy, France}
\affiliation{Centre for Quantum Optical Technologies, Centre of New Technologies, University of Warsaw, Banacha 2c, 02-097 Warsaw, Poland}

\author{Hui Khoon \surname{Ng}}
\email{huikhoon.ng@nus.edu.sg}
\affiliation{Department of Physics, National University of Singapore, Singapore}
\affiliation{Centre for Quantum Technologies, National University of Singapore, Singapore}
\affiliation{MajuLab, CNRS-UCA-SU-NUS-NTU International Joint Research Laboratory, Singapore}

\author{Robert S.~\surname{Whitney}}
\email{robert.whitney@grenoble.cnrs.fr}
\affiliation{Universit\'e Grenoble Alpes, CNRS, LPMMC, 38000 Grenoble, France.}

\date{May 5, 2026}
 
\begin{abstract}
Concatenated error-correction schemes are well-understood routes to fault-tolerant quantum computing, and research on such schemes continues, including recent claims that they may be competitive with surface codes, and show potential when combined with high-rate Quantum Low Density Parity Check 
codes.
However, there are few tools to evaluate the qubit resources required by concatenated schemes. We propose such a tool here. Its equations are closed-form and remain simple for an arbitrary number of levels of concatenation, making it ideal for comparing and minimizing the resource costs of such schemes.
We use this tool to evaluate the resources for gate operations that require the injection of so-called ``magic states'', 
needed to complete the set of logical operations.
It was expected that the complexity of such ``magic operations" would make them dominate the resource costs of a calculation, with numerous works proposing optimizations of these cost. Our work reveals that this expectation is often inaccurate: Magic operations are rarely the dominant cost of concatenated schemes, mirroring similar conclusions from past work for surface codes.
Optimizations affecting all operations 
naturally have more impact
than those on magic operations alone, yet we unexpected find that the former can reduce qubit resources by a few orders of magnitude while the latter give only marginal reductions.  We show this in detail for a 7-qubit concatenated scheme with Steane error-correction gadgets
or flag-qubits gadgets, and argue that our findings are representative of most concatenated schemes.
\end{abstract}

\maketitle


\section{Introduction}
\label{sec:intro}
Quantum computers would exploit quantum phenomena to easily perform certain calculations that existing computers struggle with. These could include 
large optimization problems \cite{Farhi2014Nov,Amaro2022Feb}, simulating quantum systems \cite{Bauer2020Nov,Cao2019Oct,Ma2020Jul}, cracking cryptography \cite{Shor2006Jul,Haner2020Apr}, etc. 
Yet, there is an apparent contradiction at the heart of quantum computing: It demands errors to be extremely rare, but it exploits quantum phenomena that are extremely sensitive to noise and imprecision.
This contradiction can be resolved by fault-tolerant quantum computing, which uses quantum error correction to make high-quality {\it logical qubits} out of very many lower-quality {\it physical qubits} \cite{Shor1996,KLZ1998,Preskill1998,AharonovBenOr2008} (see also \cite{Aliferis2005Apr,Fowler2012Sep,Terhal2015Apr,campbell_roads_2017,Roffe2019Jul}).
This requires a huge resource of physical qubits, leading to a search for ways to minimize such resources without impacting the fault tolerance.

The logical operations of a typical fault-tolerant quantum computer 
can be divided into two types that we call {\it magic} operations and {\it normal} operations. 
A magic operation is one whose fault-tolerant implementation requires the injection of a so-called magic state \cite{knill2005,bravyi2005}, which is a complicated state of additional qubits that must be prepared, verified, and then injected into the calculation.
In contrast, normal operations require no magic states. 
This normal versus magic dichotomy exists for most fault-tolerant schemes, including surface codes (see e.g., \cite{Fowler2012Sep,Terhal2015Apr}, and concatenated schemes \cite{Aliferis2005Apr}, such as the ones based on the 7-qubit code. 
In many schemes the normal operations include Clifford gates implemented transversally on the physical qubits.  The magic operations are non-Clifford gates, often $T$-gates, needed to complete the universal set of logical gates. 

The complexity of magic operations has led to the expectation that they require a lot of resources \cite{krishna_towards_2019,nikahd_low-overhead_2017,bravyi_magic_2012,webster_reducing_2015,liu2023magic},  vastly more than normal operations \cite{haah_codes_2018,hastings_distillation_2018,goto_step-by-step_2014}.
This was a motivation for works to reduce the number of magic operations in given algorithms \cite{Kissinger2020Aug}, and others to reduce their resource costs using new magic-state preparation schemes  \cite{Aliferis2005Apr,bravyi_magic_2012,fowler_surface_2013,Li2015Feb,Chamberland2017Feb,OGorman2017Mar,haah_magic_2017,campbell_magic_2018,haah_codes_2018,krishna_towards_2019,chamberland_fault-tolerant_2019,chamberland_very_2020} or replacements for magic-state injection \cite{nikahd_low-overhead_2017,chamberland_thresholds_2016,jochym-oconnor_using_2014,Chamberland2017Feb}. 
Yet, for surface codes, this expectation was challenged by Litinski \cite{Litinski2019Dec}, 
who found that magic operations were not so costly, and gave examples where they do not contribute significantly to the total resource cost of the quantum computer (see also \cite{Gidney2024Sep}). This implies that little is gained by further reducing resources required for magic operations in such surface-codes.
However for other fault-tolerant schemes, such as concatenated schemes,  it is unclear how the resources for magic and normal operation compare,
making it unclear if optimizing magic operations significantly impacts the total resource requirements.

While concatenated schemes (also called concatenated codes) were an early route to achieving fault-tolerance, it is worth studying their resource requirements, because recent works have renewed interest in them. In particular, it has been argued that their fault-tolerance thresholds could be made comparable to surface codes, while using fewer qubits \cite{yoshida2025concatenate}. 
This would require long-range operations (unlike surface codes), limiting it to platforms with this capacity (hopefully ion traps, neutral atoms, or platforms with flying qubits), but it might circumvent challenges \cite{yoshida2025concatenate} faced by high-rate quantum Low Density Parity Check (q-LDPC) codes \cite{Gottesman2014}, such as efficiently parallelizing logical gates.
In this context, Ref.~\cite{tamiya2025fault} recently suggested combining 
concatenated 7-qubit code with high-rate q-LDPC codes 
for resource-efficient fault-tolerant quantum computing.
Thus, concatenated schemes clearly remain of interest.

Yet, there are few tools to evaluate the resource requirements of concatenated schemes, making it hard to evaluate resources for even the oldest and best-understood schemes.
While a few works \cite{suchara2013estimating-techreport,goto_step-by-step_2014,chamberland_overhead_2017,Chamberland2019May} have evaluated resources for particular schemes, they do not provide a general recipe to do this. They typically use numerical calculations, which make it hard to see how modifying 
one component of the circuit will impact the overall resource requirements. 
In contrast, a method that provides algebraic expressions would make it much easier to understand how different circuit modifications will impact the overall resource requirements.

In this work, we present a method that gives algebraic expressions for the resource requirements for almost any concatenated scheme for fault-tolerant error correction,  including both normal and magic operations.
Our approach differs from that in earlier numerical resource evaluations 
\cite{suchara2013estimating-techreport,chamberland_overhead_2017,Chamberland2019May},
by treating the problem with a matrix approach that depends on only a few parameters of the concatenated scheme. This allows it to give closed-form algebraic expressions that remain simple for an arbitrary-sized algorithm with an arbitrary number of levels of concatenation.  As such, it becomes straightforward to evaluate the physical qubit resources required by any calculation with any concatenation scheme. 
Such algebraic expressions make it easier to understand how a change in one component of the circuit will impact the overall resource requirements. They can be used to reveal the parts of the scheme requiring the most resources, and so merit the most research to optimize them, as we show here by comparing the resources required for magic operations and normal operations.

\begin{figure}
\centering   
\includegraphics[width=0.9\columnwidth]{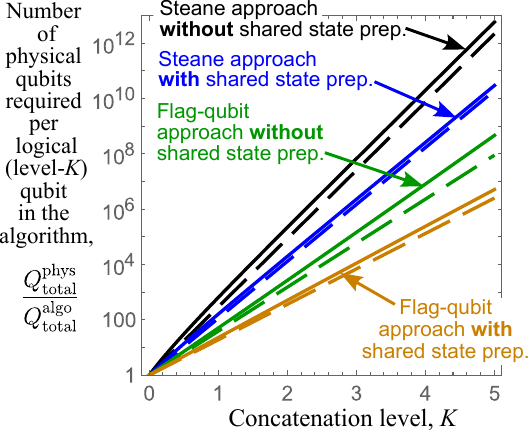}
\caption{Our matrix approach's algebraic expressions can give the physical qubit cost for an arbitrary-sized 
circuit with an arbitrary number of levels of concatenation $K$. Here we use those expressions
to calculate the physical qubit cost required for four different schemes of $7$-qubit codes  (four different colors). These costs are plotted as a function of the concatenation levels $K$. This is a fair comparison, because the four schemes are expected to have comparable error-rates for the logical gates at given $K$ \cite{11_footnote:see-appendix-thresholds}.
The four schemes are Steane or flag-qubit approaches, with or without shared resources for state-preparation. 
Solid curves are for a large-scale algorithm requiring thousands of logical qubits with 5\% of operations being magic operations (such as an algorithm to crack RSA-2048 encryption \cite{Gidney2021Apr}).  
Dashed curves are what one would get if magic operations cost the same as normal operations. On this scale, the solid and dashed curves for the same error-correction scheme are very close. In contrast, a much bigger reduction in costs comes from replacing a given scheme by a more optimal one. Switching to a scheme that share state preparation reduces the physical qubit cost by a factor of about 100 at $K=5$, switching from Steane to flag-qubits reduces this cost by a factor of about $10^4$. 
The curves for much smaller scale circuits (such as for a algorithm requiring only 20 logical qubits) would look similar \cite{10_footnote:smallscale_circuits}.
}
\label{fig:total-resources}
\end{figure}

We demonstrate this general matrix approach by evaluating the resources required by specific fault-tolerant schemes, looking at qubit resource costs of algorithms containing magic and normal operations. 
These are plotted in Fig.~\ref{fig:total-resources}, with solid curves corresponding to physical qubit costs for a typical large quantum algorithm (for example an algorithm requiring about six thousand logical qubits that could crack RSA-2048 encryption \cite{Gidney2021Apr}), while dashed curves of the same color correspond to what would happen if magic operations were no more costly than normal operations.  The four different colors in Fig.~\ref{fig:total-resources} correspond to four specific schemes for concatenated 7-qubit codes (Steane or flag-qubit approaches, with or without shared resources).
All four schemes are expected to 
have comparable error-rates for the logical gates, for 
$K$ levels of concatenation \cite{11_footnote:see-appendix-thresholds}, so it is fair to compare their resource costs at given $K$. These schemes' details are not important to understand our two main messages.  Our first main message is this work demonstrates that our matrix approach can treat such schemes, and that it gives algebraic formulas for the resources that generate the plots like Fig.~\ref{fig:total-resources}. Our second main message is that  Fig.~\ref{fig:total-resources} illustrates our conclusion that optimizing magic operations would only cause a modest reduction in physical qubit costs, corresponding to moving the solid curves towards the dashed curves of the same color. This reduction is small compared to reductions coming from  optimizations that affects all gate operations (not just magic operations), such as changing to more optimal error-correct, which allow orders-of-magnitude reductions in the physical qubit costs. Fig.~\ref{fig:total-resources} shows this occurring when 
moving from the black curve to curves of
other colors, by optimizing the error-correction through sharing state preparation, or switching to more resource-efficient error-correction gadgets (i.e., flag-qubit gadgets).

Our matrix approach can give results for any number of logical qubits. While Fig.~\ref{fig:total-resources} is for thousands of logical qubits, our approach reveals that Fig.~\ref{fig:total-resources}'s curves would be qualitatively similar whenever there are more than a few dozen logical qubits \cite{10_footnote:smallscale_circuits}. Hence, our main messages hold for any such number of logical qubits.

Our matrix approach provides a powerful tool to evaluate the qubit resources required by any concatenated schemes. To illustrate the usage of this tool in concrete situations, we give full details of examples involving 7-qubit codes. Our approach also reveals the importance of little-studied quantities that we call the time and failure multiplicities. 
For example, the failure multiplicity is responsible for the orders of magnitude difference between physical qubit costs with and without shared state-preparation in Fig.~\ref{fig:total-resources}. This makes it clear that these little-studied quantities may be critical for a reliable evaluation of the qubit resources in any fault-tolerant scheme.

Given the huge challenge of building large quantum computers, no optimization is too small to be ignored. However, here we reach the surprising conclusion that optimizing magic states is rarely the most important optimization. It is more important to first optimize aspects of the fault-tolerant scheme that affect all gates (such as optimizing the error-correction gadget), 
and only consider optimizing specific operations (such as those requiring magic states) afterward.

\subsection{A guide to reading this article}

Sec.~\ref{sec:summary} is a stand-alone six-page explanation of our matrix approach and our results. To keep this explanation clear yet short, it sketches our general approach and illustrates it with the simplest example, while postponing the justification of some assumptions and certain other technical details to later parts of the manuscript.
We emphasize that our matrix approach can be understood without the full circuit details of the underlying scheme, since its behavior is determined by only a few simple parameters of that underlying scheme. This is why this work's main messages can be understood from Sec.~\ref{sec:summary} alone.

However, the utility of our matrix approach is its applicability to any given concatenated fault-tolerant scheme.  To achieve this for a generic scheme, we must extract the few simple parameters needed for our matrix approach, and this requires careful definitions and detailed calculations, which Secs.~\ref{sec:Prelims}--\ref{sec:multiplicities} aim to explain.  More precisely, Sec.~\ref{sec:Prelims} sets up the definitions, and Sec.~\ref{sec:FTQC} explains how the calculations use the circuit details of any typical concatenated scheme to extract the few parameters required for our matrix approach. For this Sec.~\ref{sec:multiplicities} explains how to derive the time and failure multiplicity factors that are crucial to reliably compute these parameters. 
These sections consider a fairly generic case, but not the most general possible cases. More general cases typically obey the same underlying logic, but they require heavier bookkeeping that can obscure the main ideas without yielding additional insights.  Those sections also indicate simple adaptations of our derivations to more general cases.

Sec.~\ref{sec:FTQC7qb} then provides two worked examples for schemes based on 7-qubit code. 
This section's intention is to give step-by-step details for this well-known and well-understood code, so readers can follow the same steps for other concatenated schemes in the future.
We consider two versions of this 7-qubit code, the first version uses the standard Steane error-correction gadgets employed in the original scheme of Ref.~\cite{Aliferis2005Apr}, while the second replaces them with the more-recent (and less-costly) flag-qubit gadgets \cite{chao_quantum_2018,reichardt_fault-tolerant_2021,chamberland_fault-tolerant_2019}. 
Motivated by our examples, Sec.~\ref{sec:toy_model} contains a more general examination of the effect of reducing the size of the error correction gadget on the physical qubit costs. This Sec.~\ref{sec:toy_model} reinforces our qualitative conclusion, as it shows in yet other examples that optimizing the cost of magic-states brings marginal reduction compared to optimizing error correction gadgets. 
Finally, Sec.~\ref{sec:conc} gives our conclusions,  and recalls that the main  message of our work can be understood from Sec.~\ref{sec:summary} alone.

\section{Summary of our approach}
\label{sec:summary}
In this work, we introduce a matrix approach for evaluating the ``physical qubit cost'' of performing concatenated fault-tolerant quantum computation. Here ``physical qubit cost'' is the number of physical qubits needed for the computation, which determines the required size of the quantum computer. This section provides a summary of our methods and results. To provide the basic intuition, it presents the simplest application of our approach  
for an arbitrary-sized 
algorithm with an arbitrary number of levels of concatenation $K$. It then outline the generalization to more complicated settings, but leaves the full technical details to the remaining sections of this article.

\subsection{Problem setup and main method}
\label{sec:main-assumptions-definitions}

Consider an abstract circuit that carries out a specific computation of interest (or an algorithmic subroutine of that computation), expressed as a set of idealized error-free qubits undergoing a sequence of ideal gate operations. We choose a particular concatenated fault-tolerant quantum computing scheme to implement this circuit using real, potentially faulty, physical operations acting on real physical qubits. 
Through concatenation, each qubit in the abstract circuit is encoded into $n$ physical qubits, as prescribed by the fault-tolerant scheme, forming a ``logical qubit". Each operation---gate, state preparation, or measurement---in the abstract circuit is implemented as an encoded---or logical---operation on the logical qubit, accompanied by error correction to remove errors. This base level---``level-1"---of encoding provides protection against some maximum number of errors assured by the fault-tolerant scheme. To correct more errors 
we concatenate the circuit: Every physical qubit in the level-1 circuit is replaced by a logical qubit; every physical operation is replaced by the corresponding logical operation accompanied by error correction. This replacement increases the concatenation by one level, and we go recursively from level-0 (i.e., no encoding), to level-1, level-2, etc. 

Each time we increase the concatenation level, we gain the ability to correct more errors, until we  reach a level-$K$ that achieves our desired target computational accuracy. How large $K$ is, for a given target accuracy, depends on how the error-correction capability grows for the chosen fault-tolerant scheme. This increased accuracy comes at the cost of more physical qubits and operations (due to increasing the concatenation level). How this cost grows depends on the scheme used for concatenation. Our main task in estimating the physical qubit cost of implementing an abstract circuit is to understand how the number of physical qubits grows each time we increase the concatenation level. 

Of particular interest is the physical qubit cost of operations that require the injection of special states called ``magic states".
Magic states offer a well-studied route to complete the logical gate set in a fault-tolerant manner, but it is commonly felt that they incur high physical qubit cost. To facilitate our analysis of their costs, we categorize operations into ``normal" and ``magic". A magic operation is one whose fault-tolerant implementation requires the injection of ancillary qubits prepared in a magic state; normal operations are those that do not require such a state. Which operations are normal or magic is determined by the details of the fault-tolerant scheme; often, normal operations are those implemented transversally.

The abstract circuit that implements the algorithm of interest may demand varying numbers of physical qubits at different moments in the calculation. We refer to different timesteps of the level-$K$ circuit as ``layers", and focus on the ``bottleneck layer" $\Lmax$ that requires the largest number of physical qubits. $\Lmax$ comprises multiple lower-level timesteps, and includes every level-$k<K$ operation needed to implement the level-$K$ operations in $\Lmax$. The physical qubit cost of $\Lmax$ hence counts every physical qubit participating in these lower-level operations, and determines the size of the quantum computer needed for the abstract circuit. Often $\Lmax$ will be the layer with the most magic operations at level $K$, since magic operations require more qubits than normal operations. We assume that this is the case here (other possibilities and subtleties of defining $\Lmax$ are described in Sec.~\ref{sec:layer}).

To find the physical qubit cost of $\Lmax$, we count the resources required per level $k\leq K$ in $\Lmax$. For that, we assume the following, for $k=1,2,3,\ldots$ as the level of concatenation:
\begin{itemize}
\item {\bf Assumption 1.} 
A level-$k$ normal operation is built out of only normal operations at level-$(k-1)$. 
\item {\bf Assumption 2.}
A level-$k$ magic operation is built out of a combination of both normal and magic operations at level-$(k-1)$.
\end{itemize}
The level-$(k-1)$ operations include also the preparation and usage of all ancillary states needed in that level (for error-correction gadgets, magic-state injection, etc.) \cite{1_footnote:what-level-k-means}. 
Assumption 1 describes many known concatenated fault-tolerant schemes, where the desire is to limit the use of resource-expensive magic operations in the circuit design.

Next, let us turn to the qubits in the bottleneck layer, $\Lmax$, and classify each qubit at level-$k$ as ``magic" or ``normal''.
For this, we say that
a level-$k$ qubit is magic if it experiences \textit{at least one} level-$k$ magic operation in $\Lmax$; the level-$k$ qubit is normal otherwise \cite{1_footnote:what-level-k-means}.
We then have the following mental picture when evaluating the physical qubit cost: 
Each normal level-$k$ qubit is made up entirely of a certain number of level-$(k-1)$ normal qubits (with no magic qubits). In contrast, each magic level-$k$ qubit is made up of a certain number of magic level-$(k-1)$ qubits \textit{and} a certain number of normal level-$(k-1)$ qubits.
All these numbers can be determined for a given fault-tolerant scheme by looking at the details of the circuits that implement each gate in that scheme. They can then be concisely gathered into a matrix $\mathbf{M}_k$ that gives the number of normal and magic level-$(k-1)$ qubits needed for the level-$k$ normal and magic qubits. 
Our simplest example (relevant for the well-known 7-qubit scheme with Steane error-correction
gadgets) will have only one type of each, then the resulting $\mathbf{M}_k$ will be a two-by-two matrix. 

In more general settings, the matrix $\mathbf{M}_k$ may be larger than two-by-two, if there are multiple types of normal qubit or multiple types of magic qubit. For example, our Sec.~\ref{sec:example-flag-qubit_EC} treats an example for which $\mathbf{M}_k$ is three-by-three, because it has one type of normal qubit and two types of magic qubit.
Generalizations to such cases will be addressed later in this article (with  App.~\ref{sec:def_qubits_type} discussing how to identify the number of ``types'' of qubit for a general case), 
but they do not modify the logic of our method, nor the main intuition behind our results. Therefore, one can simply imagine a two-by-two matrix (as in Eq.~\ref{eq:recursion_T_Steane-1} below) to understand our method, and to grasp the general intuition underlying our quantitative results.

Once we have the matrix $\mathbf{M}_k$, we can develop a complete understanding of the physical qubit costs. 
For this, we begin by taking the algorithm of interest and counting the number of normal and magic qubits required at level-$K$ in the bottleneck layer, $\Lmax$. Then we use $\mathbf{M}_K$ to get the count of normal and magic qubits at level-$(K-1)$, which in turn allows us to use $\mathbf{M}_{K-1}$ to get this count at level-$(K-2)$, and so on, until we reach the physical qubits (level-0).
Thus, the product of these matrices, $\mathbf{M}_1\mathbf{M}_2 \cdots \mathbf{M}_K$, takes us from the number of level-$K$ qubits in $\Lmax$ to its physical (level-0) qubit cost. Then the sum of normal and magic level-0 qubits is the total number of physical qubits required by the computation.

The situation is particularly simple when the concatenated scheme has a {\it self-similar} recursive structure in which each level is implemented in the same manner, i.e., when all the circuit elements used for the concatenation (error-correction gadgets, magic-state gadgets, etc.) are the same at each level of concatenation.
In that case, the \emph{same} matrix $\mathbf{M}$ applies for all $k$, and the 
product $\mathbf{M}_1\mathbf{M}_2 \cdots \mathbf{M}_K$ reduces to $\mathbf{M}^K$.
Hence, the physical qubit cost is given by $\mathbf{M}^K$, the growth of which with $K$ can be found through an eigenanalysis of $\mathbf{M}$.

The problem of evaluating the physical qubit cost is thereby solved for any given $\mathbf{M}$ matrix.  Leaving only the issue of finding the entries of these $\mathbf{M}$ matrices for any given fault-tolerant computing scheme. These entries depend on the intricate circuit details of the fault-tolerant scheme, which Sec.~\ref{sec:FTQC} explains how to calculate for a generic concatenated fault-tolerant scheme. The analysis there highlights two complications that require careful treatment. The first is that many fault-tolerant schemes rely on probabilistic verification steps to assure fault-tolerant preparation of quantum states \footnote{The verification is probabilistic because it is a probabilistic process with a finite chance of failing verification.}. If verification fails, the state has to be re-prepared. To not hold up the computation, since any extra waiting time adds errors, one has to simultaneously prepare sufficiently many copies of the state, in the hopes that the necessary number of copies needed by the computation will pass the verification. We refer to this increase in the qubit cost from the verification steps as the {\it failure multiplicity}; see Sec.~\ref{sec:failure-multiplicity} for how to estimate this properly.

The second complication is that qubits can be re-used as they become available. Ancillary qubits participate in the computation for a certain number of timesteps, from their state preparation, interaction with the data qubits, to their measurement. Once they are measured, they are free to be reused as ancillary qubits in the next operation. Not accounting for this reuse possibility would give a gross overestimation of the true physical qubit cost. In contrast, treating ancillary qubits as always available (as in some past resource estimates
which neglected the fact that ancillary qubits can remain occupied for several timesteps and therefore cannot be immediately reassigned to subsequent operations) would give an underestimation of the true physical qubit cost. We refer to the effect of this aspect on the qubit cost as the {\it time multiplicity}. Section \ref{sec:time-multiplicity} explains how to estimate the time multiplicity by looking at the number of timesteps an ancillary qubit is in use, which in turn tells us how often it can be reused.


\subsection{Simplest example: our matrix approach}
\label{sec:simplest-case}
To illustrate our matrix approach, we look at the simplest situation. We assume that the concatenated scheme is self-similar (as explained above) so a single matrix $\mathbf{M}$ suffices to describe the problem for every level-$k$. Further, we assume that there is only a single type of normal qubit, and only a single type of magic qubit. 
Finally, we assume that all elements of $\mathbf{M}$ are known, having been carefully calculated in the manner discussed later in this work.
Below, we work through the matrix approach for this setting, and use it to examine the excess physical qubit cost of the magic operations.

To know if this simplest case describes a given fault-tolerant scheme, it can require a detailed analysis (see App.~\ref{sec:scaling-general-details}), but we see in Sec.~\ref{sec:FTQC7qb} that this simplest case suffices for the well-known 7-qubit scheme with Steane error-correction gadgets (though it is insufficient for the version with flag error-correction gadgets; see Sec.~\ref{sec:example-flag-qubit_EC}).

In this simplest case, we start by stating that
\begin{itemize}
\item
each level-$k$ normal qubit is built from $\lambda_{\opnormal}$ level-$(k-1)$ normal qubits;
\item
each level-$k$ magic qubit is built from $\lambda_{\opmagic}$ level-$(k-1)$ magic qubits and $b$ level-$(k-1)$ normal qubits.
\end{itemize}
The values of $\lambda_{\opnormal}$, $\lambda_{\opmagic}$ and $b$ are given by details of the concatenated scheme (they are $k$-independent because we have assumed self-similar concatenation). Once these parameters are known, finding the total resource cost for this simplest case requires minimal mathematics. Yet, we find it to also be qualitatively representative of more complicated cases. Thus, this simplest case provides an excellent basic intuition of our work, so we present it before discussing its limitations. 

Let $Q_\opnormal(k)$ and $Q_\opmagic(k)$ be the number of level-$k$ normal and magic qubits, respectively. Then, for this simplest case, we have
\begin{eqnarray}
\left( \begin{array}{c} Q_\opnormal(k-1) \\ Q_\opmagic(k-1) \end{array}\right) 
&=& \mathbf{M} \ 
\left(\! \begin{array}{c} Q_\opnormal(k) \\ Q_\opmagic(k) \end{array}\!\right), 
\nonumber \\
& & \hskip -18mm \mbox{with matrix }  
\mathbf{M} = 
\left( \begin{array}{cc} \lambda_\opnormal & b  \\ 0 & \lambda_\opmagic \end{array}\right). \qquad
\label{eq:recursion_T_Steane-1}
\end{eqnarray}
This equation determines the qubit resources for the single level of concatenation $k$.  However, the matrix  $\mathbf{M}$ is also the key ingredient for evaluating resources with multiple levels of concatenation. 
To get the physical qubit cost for all $K$ levels of concatenation, we iterate Eq.~(\ref{eq:recursion_T_Steane-1}) from $k=K$ down to $k=0$, giving $\mathbf{M}^K$ for relating the level-$K$ qubits to the physical level-0 qubits. 

At the highest level $k=K$, we have a total of $\Qalgo_\textrm{total}$ level-$K$ qubits that is equal to the number of logical qubits needed in the bottleneck layer, $\Lmax$, of the abstract circuit that implements the algorithm of interest. Of these, $\Qalgo_\opmagic=Q_\opmagic(K)$ of them participate in level-$K$ magic operations in $\Lmax$, while $\Qalgo_\opnormal\equiv\Qalgo_\textrm{total}-\Qalgo_\opmagic=Q_\opnormal(K)$ participate only in normal operations in $\Lmax$. The total number of normal physical qubits $\Qphys_\opnormal=Q_\opnormal(0)$, and total number of physical magic qubits, $\Qphys_\opmagic= Q_\opmagic(0)$. Hence, we have
\begin{eqnarray}
\left( \begin{array}{c} \Qphys_\opnormal \\ \Qphys_\opmagic \end{array}\right) 
&=& \mathbf{M}^K
\left( \begin{array}{c} \Qalgo_\opnormal \\ \Qalgo_\opmagic \end{array}\right),
\label{eq:recursion_T_Steane}
\end{eqnarray}
and the total number of physical qubits required is $\Qphys_\opnormal+\Qphys_\opmagic$.
To evaluate $\mathbf{M}^K$, we simply need the eigenvalues of $\mathbf{M}$. Since $\mathbf{M}$ is upper triangular, with the zero in its lower-left corner assured by Assumption 1, its eigenvalues are its diagonal entries, $\lambda_\opnormal$ and $\lambda_\opmagic$.
Expressing $\mathbf{M}^K$ in terms of its eigenvalues and eigenvectors, we find that
the total number of physical qubits required is
\begin{align}\label{Eq:Q-physical-total}
&~\quad\Qphys_{\rm total} \equiv  \Qphys_\opnormal \!+ \Qphys_\opmagic\\ 
&= \lambda_\opnormal^K \Qalgo_\opnormal + {\left[\lambda_\opmagic^K + b\frac{\lambda_\opnormal^K-\lambda_\opmagic^K}{\lambda_\opnormal-\lambda_\opmagic}\right]}\Qalgo_\opmagic. \nonumber 
\end{align}
We thus see that the physical qubit cost grows exponentially with $K$, as shown in Fig.~\ref{fig:total-resources}. This is, however, accompanied by a double-exponential increase in accuracy; see Eq.~\ref{eq:textbook-scaling_k-basics} below.

\subsection{Simplest example: Excess cost of magic operations}
\label{sec:simplest-cost-of-magic}

\begin{figure}
\centering   
\includegraphics[width=\columnwidth]{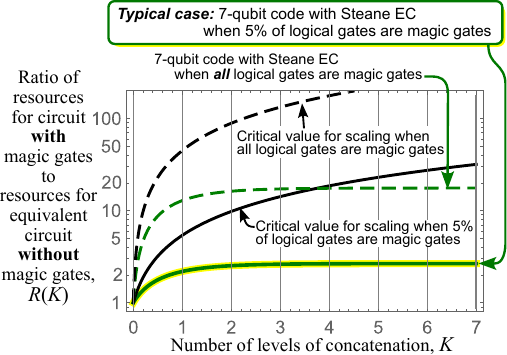}
\caption{Ratio $R(K)$ of the physical qubit cost for an abstract circuit of interest, $\cC_\mathrm{algo}$, divided by the cost for the equivalent circuit $\cC_\opnormal$
in which all level-$K$ magic operations are replaced by normal ones; see Eq.~(\ref{Eq:R-for-2-by-2}). The solid curves are for a $\cC_\mathrm{algo}$ where 5\% of the gates are magic operations
(comparable to some real algorithms in App.~\ref{sec:T-gate-proportion}). 
The dashed curves are for a $\cC_\mathrm{algo}$ with only magic operations (an unrealistic worst-case scenario). 
The parameters for the green curves correspond to a typical example ($\lambda_\opnormal=294, \lambda_\opmagic=84, b=3702$) for the 7-qubit code with Steane EC gadget; see Sec.~\ref{sec:example-Steane_EC}. 
The black curves also have $\lambda_\opmagic=84, b=3702$, 
but $\lambda_\opnormal$ is reduced to its critical value $\lambda_\opnormal=\lambda_\opmagic$.
We see that the 7-qubit code with Steane EC gadget is deep in the regime below criticality, which means that the proportion of resources for magic operations remains finite even as $K \to \infty$.
}
\label{Fig:phenomenological}
\end{figure}

\begin{figure}
\centering   
\includegraphics[width=\columnwidth]{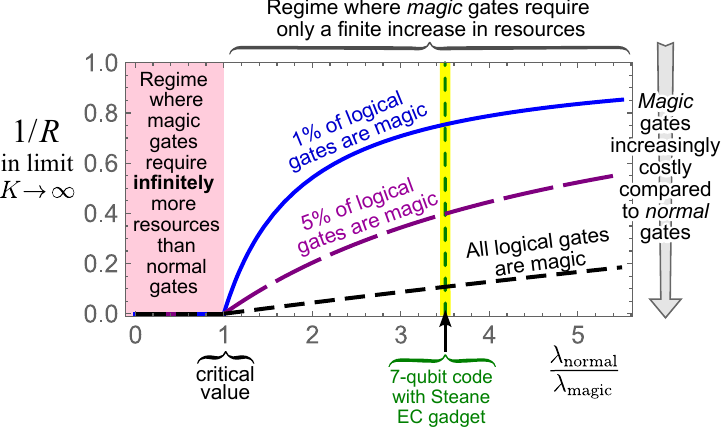}
\caption{
Analogy with a phase transition for the resource requirements in the limit of many levels of concatenation, $K\to \infty$. 
The role of the order parameter is played by $1/R$ on the vertical axis, with $R$ defined in Eq.~(\ref{Eq:def-R}). Hence, ``0.0'' on the vertical axis means magic operations require  
infinitely more resources than normal operations, while ``1.0'' means they require the same resources.
Each curve indicates the resources for a circuit with a given percentage of magic operations in the abstract  circuit, $\cC_\mathrm{algo}$, that implements the algorithm of interest. There are two regimes of behavior, with a transition between them at the critical value, 
$\lambda_\opmagic=\lambda_\opnormal$. 
} \label{Fig:phase-transition}
\end{figure}

To quantify the excess cost due to the magic operations, when we have $K$-levels of concatenation, we define a ratio $R(K)$. 
For this, we consider the bottleneck layer, $\Lmax$, of the abstract circuit that implements the algorithm of interest, which generally contains both magic operations and normal operations.
We then define $R(K)$ as the ratio of the physical qubit cost of the layer $\Lmax$, to the physical qubit cost of layer $\Lmax$ if
all the magic operations in the abstract circuit were replaced by normal operations (i.e., all level-$K$ magic qubits were replaced by level-$K$ normal qubits).

Skipping some minor subtleties in defining this ratio (see Sec.~\ref{sec:quantify-magic}), 
the numerator of $R(K)$ is Eq.~(\ref{Eq:Q-physical-total}). The denominator of $R(K)$ is obtained by setting $\Qalgo_\opmagic$ to zero and replacing $\Qalgo_\opnormal$ by $\Qalgo_\total$ in Eq.~(\ref{Eq:Q-physical-total}), so this denominator is
$\lambda_\opnormal^K \Qalgo_\total$.
Hence $R(K)$ reduces to 
\begin{eqnarray}
\label{Eq:R-for-2-by-2}
R(K) &=&
\frac{\Qalgo_\opnormal}{\Qalgo_\total}+\frac{b}{\lambda_\opnormal-\lambda_\opmagic}\frac{\Qalgo_\opmagic}{\Qalgo_\total}\\
&&+ {\left(1-\frac{b}{\lambda_\opnormal\!-\!\lambda_\opmagic}\right)}\frac{\Qalgo_\opmagic}{\Qalgo_\total}{\left(\frac{\lambda_\opmagic}{\lambda_\opnormal}\right)}^K\!.
\nonumber
\end{eqnarray}
Equations \eqref{Eq:Q-physical-total} and \eqref{Eq:R-for-2-by-2} fully describe the relative cost of magic operations for a given concatenated scheme with the above properties, and they are determined solely by the three circuit parameters $\lambda_\opnormal$, $\lambda_\opmagic$ and $b$. 

Equation \eqref{Eq:R-for-2-by-2} tells us that $R$ grows monotonically with $K$ (see footnote \footnote{Monotonic growth with $K$ is guaranteed by two observations. Firstly, $\lambda_\opnormal$, $\lambda_\opmagic$, and $b$ are positive numbers. Secondly, for the regime where  $\lambda_\opnormal > \lambda_\opmagic$, we use $b >\lambda_\opnormal$, which holds for the reasons explained below Eq.~(\ref{eq:b-formula}).}); Fig.~\ref{Fig:phenomenological} shows this behavior. This monotonic behavior motivates looking at $R$ in the infinite-$K$ limit for the upper bound at any finite $K$. In this limit, the large-$K$ behavior depends on the ratio $\lambda_\opmagic/\lambda_\opnormal$,
\begin{subequations}
\label{Eq:asymptotics_of_R}
\begin{eqnarray}
\lim_{K\to\infty} R(K) &=& \left\{ \begin{array}{ccc} C & \phantom{\Big|}\mbox{for} & \lambda_\opnormal > \lambda_\opmagic
\\
\infty & \phantom{\Big|}\mbox{for} & \lambda_\opnormal < \lambda_\opmagic
\end{array}\right., \label{eq:upper_bound}
\\
& & \hskip -16mm \mbox{with \ }
C\equiv \frac{\Qalgo_\opnormal}{\Qalgo_\total}+\frac{b}{\lambda_\opnormal-\lambda_\opmagic}\frac{\Qalgo_\opmagic}{\Qalgo_\total}. \qquad
\end{eqnarray}
\end{subequations}
For $\lambda_\opnormal > \lambda_\opmagic$, the physical qubit cost for magic operations grows with $K$ but remains a finite proportion of the overall physical qubit cost. In contrast, if $\lambda_\opnormal < \lambda_\opmagic$, the physical qubit cost for magic operations overwhelms that for normal operations. Hence, to know if magic states will dominate the resource requirements for a given scheme at large $K$, one only needs to compare $\lambda_\opnormal$ with $\lambda_\opmagic$.

We note that the behavior for $K\to \infty$ is reminiscent of a phase transition (with $K\to \infty$ resembling the thermodynamic limit), with an order parameter $1/R(K\to \infty)$; see Fig.~\ref{Fig:phase-transition}.
When $\lambda_\opmagic/\lambda_\opnormal$ is below the critical value of 1, we have $1/R(K\to \infty)=0$; when $\lambda_\opmagic/\lambda_\opnormal$ is above the critical value, $1/R(K\to \infty)$ is finite. 

Having $\lambda_\opnormal \geq \lambda_\opmagic$ means more level-$(k-1)$ normal qubits are needed for a level-$k$ normal qubit than level-$(k-1)$ magic qubits are needed for a level-$k$ magic qubit. This is common in many schemes, including the example of the 7-qubit code with Steane error correction discussed in Sec.~\ref{sec:example-Steane_EC}. 
This may seem surprising, when we know that magic operations (i.e., operations that must be implemented using magic states) typically require significantly more qubits than normal operations. The reason is that a level-$k$ magic qubit is typically built using numerous level-$(k-1)$ normal qubits, while involving relatively few level-$(k-1)$ magic qubits. So, while $b+\lambda_\opmagic>\lambda_\opnormal$ because of a large $b$, we can have $\lambda_\opmagic$ smaller than $\lambda_\opnormal$. 

Crucially, when $\lambda_\opnormal \geq \lambda_\opmagic$, then Eq.~(\ref{Eq:asymptotics_of_R}) tells us that the physical qubit cost of level-$K$ magic operations, is upper bounded by the cost of normal operations multiplied by a constant $C$ that is \textit{independent} of $K$. This is a key result for all fault-tolerant schemes with  $\lambda_\opnormal \geq \lambda_\opmagic$, such as the 7-qubit code with Steane error correction in Sec.~\ref{sec:example-Steane_EC}. It tells us that the number of physical qubits involved in a level-$K$ magic operation does not blow up with increasing $K$, compared to that for a level-$K$ normal operation. Given that many algorithms contain only a low density of magic operations --since often less than $5 \%$ of the operations in the algorithm require magic states (see App.~\ref{sec:T-gate-proportion})---
the constant factor $C$ is often fairly small. For example,  $C$ is 2.66 or less for the 7-qubit code with Steane error correction (see discussion below 
Eq.~(\ref{eq:Steane-C})), and one can see that this is fairly small from Fig.~\ref{fig:total-resources}. In that figure, the factor $C$ represents the fairly small separation of a dashed and solid curve of the same color. This is a modest difference in physical qubit cost compared to other changes shown in Fig.~\ref{fig:total-resources}. If one develops methods to make magic operations less costly, it will move the solid curve toward the dashed curve of the same color. Yet, this will have a much smaller effect than improving other aspects of the circuit, which take us from one color curve to another in Fig.~\ref{fig:total-resources} and often result in orders-of-magnitude reductions in the total physical qubit cost.

In contrast, when $\lambda_\opmagic \geq \lambda_\opnormal$, Eq.~(\ref{Eq:asymptotics_of_R}) shows that there is a divergence in
the cost of those level-$K$ magic operations when $K\to\infty$. However, if $\lambda_\opmagic$ is only slightly larger than $\lambda_\opnormal$, the proportion of the physical qubit cost coming from such magic operations only diverges slowly with $K$. We have an example of this with the flag-qubit code in Sec.~\ref{sec:example-flag-qubit_EC}
(though, the relevant $\mathbf{M}$ is $3\times 3$; see careful treatment in that section). The slowness of this divergence with $K$ is significant, since we expect $K\leq 5$ at least in the near- to middle-term devices, as larger $K$ would require such a large number of physical qubits that it will be inaccessible in the near future. Hence, the cost of magic operations also does not blow up for situations of practical interest with $K\leq 5$. 
Again the fact that the typical cost of operations requiring magic states remains modest can be seen by comparing dashed and solid curve of the same color for those marked  ``flag-qubit'' in Fig.~\ref{fig:total-resources}.

\subsection{Beyond the simplest example}
\label{sec:beyond_simplest_example}
Above, we focused on conveying the basic logic of our work through intuitive (if at times less precise) definitions, and illustrated our matrix approach with the simplest example of a recursive scheme characterized by a single $k$-independent 2$\times$2  matrix $\mathbf{M}$. The remainder of our paper provides a more careful and complete treatment, with more precise definitions, and going beyond the simplest $2\times 2$ example.
 
To explain why a more careful treatment is necessary, as well as provide a reading guide for key parts of the remaining text, we list here some of the complications that can arise and point to the relevant sections below. 
\begin{itemize}
\item 
The problem of correctly defining all quantities for the general case is treated in Sec.~\ref{sec:Prelims}.
In particular, this requires unambiguously defining what a layer $\Lmax$ looks like at lower levels (levels with $k< K$), so we know when a level-$k$ qubit is part of layer $\Lmax$. This can be subtle because at lower levels, different parts of $\Lmax$ are spread over different timesteps (see Sec.~\ref{sec:layer}).

\item Our matrix treatment above assumes the various parameters (e.g., $\lambda_\opnormal$) that enter the matrix $\mathbf{M}$ are given. We need to further understand how to extract the values of these parameters given the circuits for a particular concatenated scheme. This is described in Sec.~\ref{sec:FTQC}.

\item The treatment of the time and failure multiplicities, mentioned earlier, is a critical aspect that is required for accurate resource estimates, as explained in Sec.~\ref{sec:multiplicities}. In particular, Sec.~\ref{sec:shared-failure-multiplicity} reveals that a huge reduction in qubit resources is achieved by reducing the failure multiplicity through shared state preparation.

\item Complete and more elaborate examples illustrating our approach, as well as how various complications that arise can be properly handled, are in Sec.~\ref{sec:FTQC7qb}. There, we discuss well-studied concatenated schemes based on the 7-qubit code, first with standard Steane error-correction gadgets, and then with flag-qubit gadgets.
\end{itemize}

For better logical flow in the remaining text, some of the more peripheral technical details are relegated to the Appendix sections. We mention two aspects addressed there that may nevertheless be of interest to the reader:
\begin{itemize}
\item The proportion of magic to normal operations in the abstract circuit for typical algorithms is discussed in App.~\ref{sec:T-gate-proportion}.
\item Our simplest example above assumed there is only one type of normal qubit and one type of magic qubit. The treatment of the general situation of multiple types of normal and/or magic qubits (when $\mathbf{M}$ becomes a larger matrix) is in App.~\ref{sec:formalism_general_case}. It is worth noting that a single type of magic operation at level $k$ may require multiple types of magic qubits at level $(k-1)$; an example can be found in Sec.~\ref{sec:example-flag-qubit_EC}. The same may be true of normal operations.
\end{itemize}

\subsection{Main message}

We conclude our summary with the main message of our matrix approach: The growth of physical qubit costs, as one increases the number of concatenation levels $K$, depends only on a small number of parameters. In the simplest case above, these parameters are $\lambda_\opnormal$, $\lambda_\opmagic$, and $b$. These parameters fully determine the physical qubit costs for a given concatenated fault-tolerant scheme, and allow one to write relatively simple closed-form algebraic expressions for those costs, e.g., Eq.~(\ref{Eq:Q-physical-total}). 
These expressions apply for an arbitrary-sized 
algorithm with an arbitrary number of levels of concatenation $K$.
They allow us to identify the relative cost of magic operations (these being gate operations that must be implemented using magic states) to normal operations (these being gate operations that do not require magic states). 

This leaves only the tedious work of extracting concrete values of these parameters for a given fault-tolerant scheme. A careful study of the circuit gadgets that implement the concatenated scheme is necessary to properly count the number of qubits needed when going from level-$k$ to $(k-1)$. 
The remainder of the paper elaborates on this for a common class of concatenated schemes, and discusses the specific examples of the 7-qubit concatenated scheme with the Steane and the flag-qubit error correction approaches.

In such examples, we find that the cost of magic operations is not that different from normal gate operations.
These examples provide evidence for our more general claim: that we believe that it is more important to modify things that reduce resources required by all operations than modify things that only reduce resources required by magic operations.


\newpage

\section{Defining the problem for a generic concatenated scheme} 
\label{sec:Prelims}

Here, we start going beyond the summary of our work in Sec.~\ref{sec:summary} above. For this we are more precise in formalizing the assumption we make about the calculation than before.
We assume that the algorithm of interest is represented by an abstract circuit $\cC_\mathrm{algo}$ (recalling that an abstract circuit is one written in terms of idealized error-free qubits undergoing fault-free operations).
We choose a particular concatenated fault-tolerant scheme to carry out the computation to some desired computational accuracy using real (faulty) operations. The target accuracy determines the minimum concatenation level $K$ we need for the fault-tolerant implementation. 
This abstract circuit is assumed now to be written in terms of the (level-$K$) logical operations (state preparations, gates, and measurements) implementable in the fault-tolerant scheme that employs both normal and magic operations (with magic operations being those that must be implemented using magic states). 

\subsection{The abstract circuit that implements the algorithm of interest}
\label{sec:C_algo}

We start by defining the logical level of the abstract circuit, $\cC_\mathrm{algo}$, that implements the algorithm of interest. It is made of operations (preparations, gate-operations and measurements) acting on level-$K$ qubits. 
We label operations at level $K$ as ``magic'' if their fault-tolerant implementation will require injection of a magic-state, and label all other operations as ``normal''. 
Importantly, in our convention, the resources required for level-$k$ magic state injection only starts being counted at level $k-1$ and below: see footnote \cite{1_footnote:what-level-k-means}.

We now make assumptions about the abstract circuit, $\cC_\mathrm{algo}$, that are sometimes called the ``standard circuit model''.
Firstly, we assume that the circuit $\cC_\mathrm{algo}$ requires that all  (level-$K$) qubits are initialized in $\ket{0}$ and are  measured in their Pauli $Z$ basis.
Secondly, we assume that such level-$K$ preparations and measurements are normal operations (rather than magic operations). 
Thirdly, we assume that the algorithm has a rectangular shape, namely, that all level-$K$ qubits are prepared at the same initial timestep, and measured at the same final timestep, with logical gate operations (but no measurements) in between.

We emphasize that we only make these three assumptions at the level of the abstract circuit (level-$K$), since we know that they do not hold at level-$k$ for $k<K$.  
For example, to perform level-$K$ gates we often have to prepare and measure lower level states throughout the calculation, and some of these are not prepared in the state $\ket{0}$. For example, some schemes have magic gates at level $K$ \cite{1_footnote:what-level-k-means} that must be implemented at level $(K-1)$ using ancillary level-$(K-1)$ qubits prepared in a complicated state (for example the state $\ket{H}$ in Fig.~\ref{fig:sketch-flag-qubit}) that require specific preparation at level $(K-2)$.
Thus, neither of the first two assumptions hold at lower levels.
Similarly, at lower levels
the third assumption does not hold because the circuit ceases to be rectangular, and contains many measurements throughout the algorithm (in both EC gadgets and in magic operations).

In principle, our matrix approach does not require these ``standard circuit model'' assumptions for the abstract circuit being implemented. However, they simplify its application.  Thus, we assume them throughout, mentioning the places where they allow us to simplify the analysis.

\subsection{Formal definition of a layer}
\label{sec:layer}

To find the physical qubit cost of a given algorithm, one notes that the circuit implementing that algorithm $\cC_\mathrm{algo}$ (described in Sec. \ref{sec:C_algo}) requires different resources at different moments in the calculation. Hence, we cut $\cC_\mathrm{algo}$ into what we call {\it layers},  each of which is a single time step of $\cC_\mathrm{algo}$. 
The physical qubit cost of an algorithm is then given by the layer that requires the most physical qubits; we call this the bottleneck layer, $\Lmax$. 
To evaluate the physical qubit costs, we need to extend the definition of such layers to lower levels, all the way down to the level of physical qubits (level-0).  We explain this in the next paragraph.

To detail the definition of a given layer at level-$k$, 
we start with the above definition of that layer at the algorithm level (level-$K$), and iteratively define that layer at lower levels. 
For any $k \in [0,K]$, the layer at level-\mbox{$(k-1)$} is defined to include all the operations necessary to perform that layer, at level-$k$. Thus, it includes the full level-$(k-1)$ circuit that prepares and measures all ancillary states that are used in all operations occurring in the layer at level-$k$. 
To avoid any risk of under-estimating costs,
we make two further inclusions in the layer.
Firstly, when going from level-$k$ to level-$(k-1)$ we include the ancillary states used to prepare those level-$k$ qubits, even if that preparation is at an earlier time in the level-$k$ circuit. 
Secondly, as fault-tolerant schemes often require multiple timesteps to prepare and measure ancillary states, we include in the layer the full circuits for all ancillary states used in preceding or subsequent layers, if they are being prepared or measured at the same time as the layer of interest. These two inclusions allow us to be sure that the level-0 layer has enough physical qubits for the computer to implement that layer at the level of the algorithm $\cC_\mathrm{algo}$.  
After that the physical qubit cost for the {\it whole algorithm} is given by the layer requiring the most physical qubits, which we call the bottleneck layer, $\Lmax$. 
In all our examples, we assume that $\Lmax$ is a level-$K$ layer containing the largest number of magic operations, since we are considering magic operations that cost more than normal operations. This is typically the case when the algorithm requires the same number of level-$K$ qubits in each layer \cite{5_footnote-Lmax-general}. 
We next need to make an assumption for the ancillary states being prepared and measured during $\Lmax$ for previous or subsequent layers. For simplicity of our calculations, and to ensure our cost evaluations apply for any algorithm $\cC_\mathrm{algo}$, we assume that all layers preceding and following $\Lmax$ require the same ancillary states as $\Lmax$ \footnote{An algorithm implementing $\Lmax$ at every step, requires more qubits than an algorithm implementing $\Lmax$ only once. This is because the latter would not need to be preparing ancillary states for future layers (or reading ancillary states from preceding layers) at the same time as carrying out $\Lmax$. The extra resources required to implement $\Lmax$ at every step will be quantified by the ``time multiplicity'' introduced  in Sec.~\ref{sec:resources_for_basic_components}.}. In other words, we assume that every layer in the algorithm is as costly as the most-costly layer,  $\Lmax$.

\subsection{Comparing costs of normal and magic operations}
\label{sec:quantify-magic}

With the full circuit details for the chosen fault-tolerant quantum computing scheme, and going from level $K$ down to the physical level (level 0), we can estimate --- as we will do below --- the total physical qubit cost, $\Qphys_\mathrm{total}$, i.e., the total number of physical qubits needed to implement the bottleneck layer, $\Lmax$, of the abstract circuit $\cC_\mathrm{algo}$, when using a chosen fault-tolerant scheme with a desired target accuracy (which determines the desired $K$).

We then want to understand how much of the physical qubit cost $\Qphys_\mathrm{total}$ is due to the magic operations in the circuit $\cC_\mathrm{algo}$. To do that, we consider a fictitious abstract circuit $\cC_\opnormal$, whose bottleneck layer is the same as the bottleneck layer of $\cC_\mathrm{algo}$, except that we replace every level-$K$ magic operation in $\cC_\mathrm{algo}$ by a level-$K$ normal operation. 
We then evaluate the physical qubit cost $\Qphys_\mathrm{total}$ for the bottleneck layer $\Lmax$ of circuit $\cC_\opnormal$, in the same manner as for $\Lmax$ of $\cC_\mathrm{algo}$.  The difference between the $\Qphys_\mathrm{total}$ values for $\cC_\mathrm{algo}$ and $\cC_\opnormal$ can be attributed to the physical qubit cost of doing the magic operations in $\cC_\mathrm{algo}$. 

Note that the fictitious circuit $\cC_\opnormal$ is not intended to be one that 
does any useful calculation, it is simply there to compare with the original circuit.
Also $\cC_\opnormal$ is not necessarily a uniquely defined circuit: There are usually many different normal gate operations, and we can choose which one we use to replace the magic operations in $\cC_\mathrm{algo}$ to get $\cC_\opnormal$. 
Nevertheless, in the examples we will study, magic operations are single-qubit gates and all single-qubit normal gates have the same qubit cost. Thus, any replacement of magic operations by single-qubit normal gates will yield the same $\Qphys_\mathrm{total}$ for $\cC_\opnormal$. In Appendix~\ref{sec:formalism_general_case}, we explain how to treat more general cases. In addition, we will also assume that the chosen fault-tolerant scheme is such that a level-$k$  normal operation is composed only of level-$(k-1)$  normal operations (see next section), irrespective of whether that operation is a gate operation, a state preparation or a measurement. Hence, since $\cC_\opnormal$ contains only normal operations at level-$K$, then there are no magic operations at any level in $\cC_\opnormal$. 

To quantify the excess cost of the magic operations, we define $R(K)$ as the ratio between $\Qphys_\mathrm{total}$ for the two fault-tolerant circuits (both with $K$ levels of concatenation); in other words
\begin{eqnarray}
R(K) = 
\frac{\hbox{$\Qphys_{\rm total}$ for  $\Lmax$ of ${\cal C}_{\rm algo}$}}
{\hbox{$\Qphys_{\rm total}$ for $\Lmax$ of ${\cal C}_\opnormal$}}\,.
\label{Eq:def-R}
\end{eqnarray}
When this ratio is large, $R(K)\gg 1$, it means magic operations require significantly more physical qubits than normal operations, while $R(K)=1$ means that magic and normal operations have the same physical qubit costs.

Of crucial interest for the scalability of a fault-tolerant scheme is the behavior of $R(K)$ as $K$ grows. An $R(K)$ that grows unchecked suggests a limit to scalability, and a need to reduce the cost of magic operations. $R(K)$ is also an important ingredient in estimating any resource (such as power or energy) in a full-stack quantum computing model. Such estimations are simpler when $R(K)\sim 1$, because one can neglect differences in qubit costs between magic and normal operations, as in Ref.~\cite{Fellous-Asiani2023Oct}.

In Sec.~\ref{sec:FTQC7qb}, we explicitly evaluate $\Qphys_{\rm total}$ and $R$, by applying this  matrix approach to concatenated schemes based on the 7-qubit code.

\section{Qubit resource costs for 
a generic concatenated scheme}
\label{sec:FTQC}

We are now ready to explain the parameters entering the matrix $\mathbf{M}$ in more detail than in the summary in Sec.~\ref{sec:summary} above.
In particular, we show how $\lambda_\opnormal$, $\lambda_\opmagic$, and $b$ are determined from all the microscopic technical details required for a fault-tolerant scheme.

Here, we focus on concatenated schemes that are recursively defined using the same code at every level, a general class that includes some of the most commonly discussed examples in the literature of concatenated fault-tolerant quantum computing. For such schemes, we need only specify the circuit details for how to go from level-0 to level-1 concatenation; further increase in the concatenation level is carried out by recursively replacing level-0 components by the level-1 versions. 
 
This is not the most general possible concatenated construction, but it is fairly generic in the sense that more general constructions will follow a similar logic.
Focusing on this construction allows us to present this logic, while avoids the heavy bookkeeping required for more general cases that would obscure the main ideas without yielding additional conceptual insights.   In addition, to keep our logic as clear as possible, this section
(along with Sec.~\ref{sec:multiplicities}) will initially make some extra simplifications (such as assuming only a single type of normal qubit and a single type of  magic qubit), and then later explain how this is generalize the logic to more complicated cases.

\subsection{Basics of concatenated fault-tolerant schemes}
\label{sec:basics-of-concatenation}
Before we discuss the circuit designs that implement fault-tolerant error correction, 
we first recall the basics of concatenated schemes.
As explained earlier, in a fault-tolerant quantum computation, a given $\cC_\mathrm{algo}$, specifying a sequence of logical operations that carries out an algorithm, is built from potentially faulty physical (i.e., level-0) qubits and operations following the fault-tolerance prescription. As we increase the concatenation level $k$, the error probability per logical operation after $k$ levels scales according to the formula \cite{Aliferis2005Apr},
\begin{eqnarray}
p_{k}=\frac{1}{B^{1/t}} (B^{1/t} p_0)^{(t+1)^k}\,,
\label{eq:textbook-scaling_k-basics}
\end{eqnarray}
where $t$ is the number of errors the underlying code corrects. 
Here, $p_0$ bounds the error probability of a physical operation, and $B$ is a numerical constant that captures the increased complexity of the physical circuit due to the error correction (for each level of concatenation). $B$ depends on the number of fault locations in an ``exRec" (extended rectangle) for the scheme (see Ref.~\cite{Aliferis2005Apr} for details) and determines the quantum accuracy threshold $p_\mathrm{thres}\equiv B^{-1/t}$ below which $p_k$ shrinks as $k$ grows. Using this formula, we can determine the minimum concatenation level $K$ needed to achieve a target computational accuracy, provided we are below the threshold. 
Our goal here is to estimate, for a given value of $K$, an upper bound --- hopefully, a close-to-tight one --- on the number of physical qubits $\Qphys_{\rm total}(K)$ needed to implement $\cC_\mathrm{algo}$ to a desired target accuracy.

Below, we begin with a review of the circuit designs for recursive concatenated schemes. We then explain how to extract the parameters of interest for characterizing the resource scaling, and specialize in the case of the 7-qubit concatenated scheme for concrete numbers.
Of course, circuit designs are specific to a particular fault-tolerant scheme with its choice of underlying code. Nevertheless, there are common features, at least in the known examples so far, that allow us to come up with concrete formulas for our resource estimates below.

\subsection{Circuit designs}

We begin with the physical operations that can be performed. We assume we can do a set of physical gates $\Gphys$, prepare a physical qubit in some (usually small) set of ``default" physical states $\Sphys$, and also do measurements on the physical qubits for some (again, usually small) set of observables $\Mphys$. Each of these physical operations (gates, preparations, and measurements) is assumed to take time $\tauP$. In practice, of course, the different operations may take different times. For example, in many implementations, a two-qubit gate takes a longer time than single-qubit ones; measurement times can be significantly longer than gate times; preparation of different states may also take a different number of timesteps. Here, we ignore these device-dependent variabilities, and opt for the simplicity of a single time parameter $\tauP$ for all physical operations.

\begin{figure}
\begin{quantikz}[column sep=0.25cm]
\qw&\gate{G_\mathrm{L}}&\qw&\equiv
\end{quantikz}
\hspace*{-0.1cm}
\begin{quantikz}[row sep=0.1cm, column sep=0.25cm]
\qw&\gate{G_1}&\qw\\
\qw&\gate{G_2}&\qw\\
&\vdots&\\
&&\\
\qw&\gate{G_n}&\qw
\end{quantikz}
\caption{\label{fig:TransG} A transversal logical gate $G_\mathrm{L}$, with $G_1,G_2,\ldots\in\Gphys$ as physical single-qubit gates. Each horizontal line on the right-hand side represents a single physical qubit; the horizontal lines on the left-hand side represents a single logical qubit.}
\end{figure}
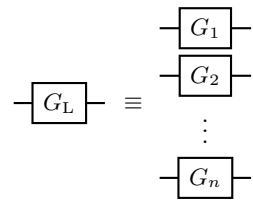

Next, we discuss logical gates. 
Below, we indicate logical operations and states with a subscript ``L".
Logical gates can be separated into the normal and magic classes as discussed earlier. 
We assume each normal gate takes a fixed time $\tauG\equiv g\tauP$ to complete, with $g$ being a positive integer, assumed for simplicity to be the same for all normal gates. The prototypical examples of normal gates are logical gates that can be done transversally, i.e., the logical gate is achieved by applying a physical gate to each physical qubit in the code block; see Fig.~\ref{fig:TransG}. In this transversal case, $g=1$. 

We assume the magic operations we need are single-logical-qubit gates, as is often the case for lower implementation demands. The normal and magic operations that we can implement fault-tolerantly together form the set of logical gates $\GL$, which we assume to be computationally universal. Before we go on to describe how these magic operations are implemented, we need a few additional ingredients, namely, fault-tolerant measurements and preparations, as well as a circuit for the error-correction operations. 

We assume that a set of logical measurements (e.g., measurement of the $\overline Z$ operator) $\ML$ can be implemented as normal operations (requiring no magic states): A logical measurement in $\ML$ requires no ancillary qubits, and can be done in a fixed time $\tauMeas\equiv m\tauP$, with $m$ a positive integer. Again, a prototypical example for an element of $\ML$ is a logical measurement performed by doing transversal physical measurements, i.e., measurements from $\Mphys$ on each individual physical data qubit.

\begin{figure}
\begin{quantikz}[font=\small, row sep=0.1cm]
\gate{~~\psi\textrm{-FT-Prep}~~}
\end{quantikz}
$=$\!\!\!
\begin{quantikz}[font=\small, row sep=0.1cm]
\gate{~~\psi\textrm{-Prep}~~}&\gate[2]{G_\psi}&\qw&\qw\\
\gate{\ket{A_\psi}\!\textrm{-Prep}}&\qwbundle{\hspace*{-0.15cm}v_\psi}&\meter{}&\cw\rstick{pass}
\end{quantikz}
\caption{\label{fig:FT_prep}
The method to prepare an arbitrary state $\psi\in{\cal S}_L$ in a fault-tolerant manner (indicated as $\psi$-FT-Prep).
One first prepares the qubits carrying the (yet-to-be-verified) $\psi$ state (indicated as $\psi$-Prep). To ensure fault-tolerance this state must then be verified via the verification step, $G_\psi$. This verification step requires an input made of $v_\psi$ ancillary (physical) qubits prepared in the state $\ket{A_\psi}$ (indicated as $\ket{A_\psi}$-Prep). These ancillary qubits are then measured after $G_\psi$. If the measurement result gives a ``pass", the state is fault-tolerantly prepared, and can be used as a $\psi$-FT-Prep state in the calculation. In contrast, if a ``fail" result is obtained, the state is useless. Thus, one should be running multiple such preparation circuits in parallel to ensure a very high chance that at least one of them passes verification and so generates the $\psi$-FT-Prep state to be used in the calculation. 
Note that this preparation circuit includes the circuits that give $\psi$-Prep and $\ket{A_\psi}$-Prep, which can themselves require ancillary qubits. In many cases, the states being prepared are logical states (like $\ket{0}_\mathrm{L}$ or $\ket{+}_\mathrm{L}$); in such cases 
we call them $\ket{\psi}_\mathrm{L}$ rather than $\psi$-FT-Prep.  Thus, $\ket{\psi}_\mathrm{L}$ in Figs.~\ref{fig:EC_gadget} and \ref{fig:hard_gate} was prepared in the manner shown here.
}
\end{figure}
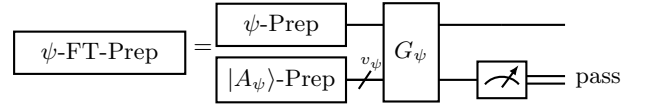

Next, we assume that a set of states $\SL$ can be fault-tolerantly prepared, but in a manner that requires the use of ancillary physical qubits to verify that the state is correctly (i.e., fault-tolerantly) prepared.  Here we differentiate between physical qubits that carry the logical information --- referred to as ``data qubits'' --- and extra physical qubits --- referred to as ``ancillary qubits'' --- needed to help carry out the operations.
In the verification step, the ancillary qubits interact with the data qubits carrying the (yet-to-be-verified) prepared state, and those ancillary qubits are measured to check for possible errors; see Fig.~\ref{fig:FT_prep}. The state is considered fault-tolerantly prepared only if the ``pass" measurement outcome is obtained. If a ``fail" outcome occurs, that copy is discarded. Of course, such a repeat-until-success strategy that may require indefinite waiting times is often difficult to incorporate; long waiting times entail more noise in the qubits waiting to interact with the ancillary state. Instead, a more practical approach is to have a parallel procedure where multiple copies of the needed state are prepared at once, so that there is a high chance that at least one copy will pass the verification. This requires extra qubits, an increase we will quantify by a ``failure multiplicity" factor (see Sec.~\ref{sec:multiplicities}).

$\SL$ can contain standard logical states like $\ket{0}_\mathrm{L}$ or $\ket{+}_\mathrm{L}$, but also multi-physical-qubit states needed in the fault tolerance scheme. These multi-qubit states need not be logical states, and the subscript L in $\SL$ simply indicates that it contains states needed in the fault-tolerance scheme. Each round (pass or fail) of preparation of $\ket\psi\in\SL$ requires $v_\psi$ ancillary qubits, takes total time $\tau_\psi$ and has probability $p_\psi$ of getting the ``pass" outcome.

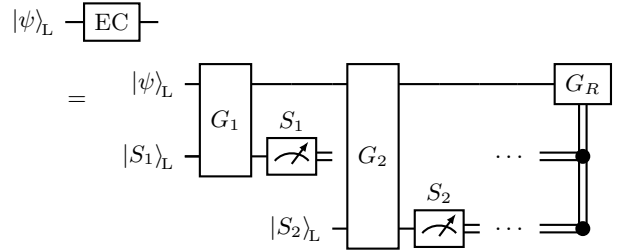
\begin{figure}
\hspace*{-1.5cm}
\begin{quantikz}[column sep=0.25cm]
\lstick{$\psiL$}\qw&\gate{\textrm{EC}}&\qw
\end{quantikz}
\hskip 5cm \ 

\vskip 0.6cm
\hskip 0.5cm $=$ \hskip 7cm \ 

\vskip -0.6cm\hskip 1cm
\begin{quantikz}[font=\small, row sep=0.1cm, column sep=0.2cm]
\lstick{$\psiL$}&\gate[2]{G_1}&\qw&\qw&\gate[3,nwires={2}]{G_2}&\qw&\qw&\qw&\gate{G_R}\\
\lstick{$\ket{S_1}_{\!\mathrm{L}}$} &\qw  &\meter{$S_1$}&\cw&&&&\ldots~~&\cwbend{-1}\\
&&&\lstick{$\ket{S_2}_{\!\mathrm{L}}$} &&\meter{$S_2$}&\cw&\ldots~~&\cwbend{-1}\\
\end{quantikz}
\caption{\label{fig:EC_gadget} The circuit for a generic EC gadget acting on a state $\ket{\psi}_\mathrm{L}$. The gadget involves $I_S$ syndrome operators $S_i$, measured sequentially, using fault-tolerantly prepared ancillary states $\ket{S_i}_\mathrm{L}\in\SL$. Each $G_i$ gate is a sequence of $d_i$ normal gates, and the measurement of $S_i$ is assumed to be in the set of easily-implemented measurements, $\ML$. The measurement outcomes are used to deduce what errors occurred and hence what recovery to apply. The recovery operation, denoted simply as $G_R$ above, is a sequence of $d_R$ normal gates on the physical qubits in the code block, controlled on the $I_S$ measurement outcomes.}
\end{figure}

We also need the circuit design for the fault-tolerant implementation of the EC gadget, i.e., the circuit that implements the syndrome measurement and recovery to remove errors. Here, we assume a generic structure for the EC gadget; see Fig.~\ref{fig:EC_gadget}. $v_i$ ancillary qubits, fault-tolerantly prepared in a multi-qubit state $\ket{S_i}\in\SL$, are used to measure syndrome operator $S_i$, for $i=1,\ldots, I_S$. We will refer to the $\ket{S_i}$s as the ``syndrome states". The $I_S$ syndrome operators are measured sequentially as shown in the figure, with each $G_i$ comprising a sequence of $d_i$ normal gates, and thus taking time $d_i\tauG=d_ig\tauP$ to complete. The recovery operation $G_R$ is then performed, using normal gates only, depending on the measurement outcomes. Note that the syndrome operators $S_i$ need not be the check operators for the code; instead, in some EC-gadget designs, a single measured syndrome operator can yield the results of multiple check operators at once. Note also that, since the syndrome states come from $\SL$, their fault-tolerant preparation is nondeterministic, and one requires again multiple preparation attempts at the same time, to be assured of available syndrome states when they are needed for the EC gadget. 

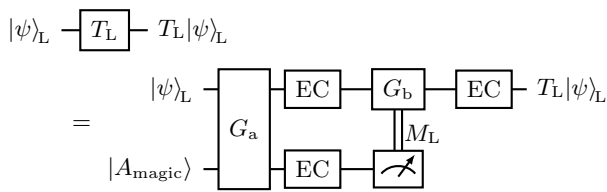
\begin{figure}
\begin{quantikz}[column sep=0.25cm]
\lstick{$\psiL$}\qw&\gate{T_\mathrm{L}}&\qw\rstick{$T_\mathrm{L}\!\psiL$} 
\end{quantikz}
\hskip 5cm \ 

\hskip 1cm $=$
\begin{quantikz}[row sep=0.2cm, column sep=0.2cm]
\lstick{$\psiL$}&\gate[wires=2]{G_{\rm a}}&\gate{\mathrm{EC}}&\gate{G_{\rm b}}& \gate{\mathrm{EC}}&\qw\rstick{$T_\mathrm{L}\!\psiL$} \\
\lstick{\ket{A_\opmagic}}&&\gate{\mathrm{EC}}&\meter{~~\quad~$M_\mathrm{L}$}\vcw{-1}&
\end{quantikz}
\caption{\label{fig:hard_gate}
Magic-state implementation of a magic operation $T_\mathrm{L}$ on the input logical state $\psiL$ via the use of a magic state $\ket{A_\opmagic}\in\SL$. $G_{\rm a}$ and $G_{\rm b}$ are normal gates, and $M_\mathrm{L}\in\ML$. The $G_{\rm b}$ gate is applied controlled on the measurement result of $M_\mathrm{L}$. Each horizontal line here is a logical qubit, and the boxes marked ``EC" are the EC gadgets of Fig.~\ref{fig:EC_gadget}.
}
\end{figure}

Now, we can return to the magic operations. We assume, as is the case in known schemes, a magic operation is implemented using a magic-state approach; see Fig.~\ref{fig:hard_gate}. In this approach, a magic operation $T_\mathrm{L}$ --- a single-logical-qubit gate --- is implemented using only normal gates and measurements from $\ML$ (interspersed with error correction to maintain fault tolerance), but the process consumes a resource state $\ket{A_\opmagic}\in\SL$ carried by ancillary qubits, a so-called ``magic state" appropriate for that magic operation. In general, $\ket{A_\opmagic}$ can be carried by some $n_A$ physical qubits and need not be a logical state for the underlying code; for simplicity, however, we will assume that $\ket{A_\opmagic}$ is a logical state, carried by $n$ qubits, on which we can perform logical gates as well as error correction using same EC gadget as for the data qubits. In Fig.~\ref{fig:7qb-Tgate}, we show the situation where $G_{\rm a}$ and $G_{\rm b}$ are each a single normal gate; this is what we will assume in our resource estimates below. More generally, $G_{\rm a}$ and $G_{\rm b}$ can be composed from a sequence of normal gates interspersed with the EC gadgets (after every normal gate).

We need an additional feature to ensure fault tolerance of the resulting circuit. As mentioned earlier, we increase the circuit concatenation by one level by replacing every physical component --- a preparation, a gate, or a measurement --- by its level-1 version. This  level-1 version is a fault-tolerantly implementable logical counterpart of the corresponding gate (with this logical counterpart of magic operations being implemented using magic-state injection). 
The level-1 version is not just the corresponding level-1 version of the logical operation; it includes also the error-correction circuit for each code block involved in that operation.  We use the terminology of Ref.~\cite{Aliferis2005Apr} and refer to this level-1 version (including the error correction) as a ``1-Rec".
This applies for gates, preparation and measurement. Every gate in $\Gphys$ is replaced by a ``1-gate-Rec", comprising the circuit that performs the corresponding level-1 logical gate, followed by the EC gadget. Every state preparation in $\Sphys$ is replaced by a ``1-prep-Rec", comprising the fault-tolerant preparation of those states, followed by the EC gadget. Finally, every measurement in $\Mphys$ is replaced by a ``1-meas-Rec", which is simply the measurement of the level-1 logical operator.

Ref.~\cite{Aliferis2005Apr} showed 
that we can then repeat the above fault-tolerant construction to go from level-1 to level-2, thereby concatenating up to a level-$K$ of fault-tolerant error correction.
Such a ``recursive simulation" gives a fault-tolerant construction at every level. To make our notation clear, we will reserve the subscript L for level-1 elements (so the earlier figures for the circuit designs are understood to be for level-1 concatenation); we will specify the concatenation level when talking about higher concatenation levels. 

Finally, we need one more critical assumption about the circuit designs for fault-tolerant gate implementation: Logical normal operations are assumed to be implementable using only physical normal operations. In contrast, logical magic operations can in general also make use of physical magic operations, for example, in the preparation of the magic states. This, together with the recursive manner of concatenation, assures that Assumptions 1 and 2 
of Sec.~\ref{sec:main-assumptions-definitions} are valid.

\subsection{Qubit costs for basic components}
\label{sec:resources_for_basic_components}

We start by evaluating the qubit costs for the basic components of the fault-tolerance scheme introduced in the previous section. The normal gates and measurements in $\ML$ require no additional ancillary qubits, and hence require no additional qubit costs. 
However, extra ancillary qubits are required for state preparation of elements in $\SL$, the EC gadget and the magic operations. Let us treat these one by one, for the first level of concatenation ($k=1$); we extend this to multiple levels of concatenation in subsequent subsections.

When evaluating the qubit costs for a given component in a concrete fault-tolerant scheme, the first step is to take that component's circuit diagram and simply count the number of qubits (qubits carrying data plus qubits involved in ancillary states).
For simplicity, we first explain this for the simple case which gave the $2\times2$ scaling in Sec.~\ref{sec:simplest-case} (the case in which there is only one type of normal qubit, and one type of magic qubit); we describe the generalization to other cases at the end of this section.
The second step is then to account for two multiplicity factors that are rarely shown in circuit diagrams, but have a crucial effect on the qubit costs.  These two multiplicity factors are the following. 
\begin{itemize}
\item {\bf Time multiplicity:} 
Ancillary states take multiple timesteps to prepare, but must be ready when needed. Thus, at a given timestep in the calculation, one may be preparing the ancillary states for multiple subsequent timesteps. 
In short, if a certain type of ancillary state takes a time $\tau_{\rm anc}$ to be prepared, but is required by the calculation at time-intervals of $\tau_{\rm L}$, 
then we assume that we
have $r=\lceil\tauAnc/\tau_\mathrm{L}\rceil$ ancillary states at the same time \cite{ceiling}, 
with one such state used in the current timestep and the others being prepared for the subsequent $(r-1)$ timesteps. We call $r$  the ancillary state's {\it time multiplicity}.
It takes a different value for each type of ancillary state, and its calculation is explained in Sec.~\ref{sec:time-multiplicity}. 
\item {\bf Failure multiplicity:}
The preparation of quantum states usually requires a verification procedure to guarantee that the state has been prepared in a fault-tolerant manner before it is used. Usually, the verification of each of these $\mu$ copies requires ancillary qubits that are prepared in a specific way, interact with the qubits encoding the state, and are then measured to tell us if the state is verified or not. The states that fail verification should not be injected into the calculation, and are thrown away. Thus, whenever the circuit diagram indicates a verification of such a state, we must prepare $\mu$ copies of that state in parallel (along with the ancillary states used to verify them) to account for the fact that there is a chance that each one will fail verification. 
We call $\mu$ the state preparation {\it failure multiplicity}. It is different for each type of quantum state, because each one has a different probability of failing verification. The larger the failure probability, the larger $\mu$ must be to ensure a high enough probability for a successful verification of the  state.  We quantify what we mean by ``high enough probability" in Sec.~\ref{sec:failure-multiplicity}, where we also calculate $\mu$ for different states.
\end{itemize}
Such time and failure multiplicities appear in the qubit costs required by quantum states used in state preparation, EC gadgets, and magic states. 
Our strategy in determining their values is to ensure that data qubits {\it never} need to wait mid-calculation for a verified ancillary state to become available, to avoid accumulation of errors when the data qubits are waiting.
In principle, one could reduce the time multiplicity by making the data qubits wait longer between uses of ancillary states (increasing $\tau_{\rm L}$ to reduce $r$). 
Similarly, one could eliminate the failure multiplicity by preparing only one quantum state for each usage,
then if it fails verification, one starts preparing a new quantum state, making the calculation wait for a verified state to become available.  
However, such waiting times come at the cost of many more errors, and hence a worse fault-tolerance threshold (for instance in the case of ancillary states required by the EC-gadgets, or magic states). Thus, we do not consider any reductions of multiplicities that come from making data qubits wait.

The first component to consider in any fault-tolerant scheme is the state preparation shown in Fig.~\ref{fig:FT_prep}. The fault-tolerant preparation of a state $\ket{\psi}\in \SL$ carried by $n$ physical qubits requires a verification with an additional $v_\psi$ ancillary qubits.
The number of physical qubits needed for the fault-tolerant preparation of the state $\ket{\psi}\in\SL$ is thus given by
\begin{equation}\label{eq:StateCost}
\lambda_\psi\equiv \mu_\psi(n+v_\psi).
\end{equation} 
where the failure multiplicity that appears here $\mu_\psi$ corresponds to preparing multiple states in parallel to account for the fact that we need to verify the state $\ket{\psi}$ before use, and each one has a chance of failing verification. In contrast, there is no time multiplicity (in other words $r_\psi=1$), because one prepares each state only once.

The next component to consider is the EC gadget shown in Fig.~\ref{fig:EC_gadget}. We see that ancillary qubits are required to carry the syndrome states $\ket{S_i}\in\SL$, for $i=1,2,\ldots,I_S$. 
The total number of physical qubits needed to prepare and carry the syndrome states for an EC gadget is thus given by
\begin{equation}\label{eq:lambda_EC}
\lambda_\textrm{EC}=\sum_{i=1}^{I_S}r_i \mu_i(n_i+v_i).
\end{equation}
Here, $\mu_i$, $n_i$, and $v_i$ are, respectively, the failure multiplicity, number of physical qubits carrying the state $\ket{S_i}$, and the associated number of verification ancillary qubits for the state preparation. This counting assumes that it is impossible to re-use any ancillary qubits within the same EC gadget, e.g., none of the qubits employed in preparing $\ket{S_1}$ can be reused for preparing $\ket{S_i}$, for some $i>1$. Such a re-use would allow one to reduce the qubit costs
in the EC gadgets, but it is not possible in the examples that we consider in this work.

With this, we can already obtain an expression for $\lambda_\opnormal$, the number of physical qubits needed for a level-1 normal qubit. A normal qubit, by definition, never goes through a magic operation, and it requires ancillary physical qubits only for two things: its initial fault-tolerant state preparation, and in the subsequent EC gadgets. We thus have
\begin{eqnarray}
\lambda_\opnormal &=& \lambda_\psi+ \lambda_\textrm{EC}
\nonumber \\
&=& \mu_\psi(n+v_\psi) +\sum_{i=1}^{I_S}r_i \mu_i(n_i+v_i),
\label{eq:lambda-easy-formula}
\end{eqnarray}
with the time multiplicity $r_i$ accounting for the qubit costs needed to run consecutive EC gadgets on the same level-1 qubit. Note that Eq.~(\ref{eq:lambda-easy-formula}) is an overestimate, since, while $\mu_\psi(n+v_\psi)$ qubits are used in the preparation 
of the initial state, only $n$ of them carry the state through to subsequent timesteps, leaving $\mu_\psi(n+v_\psi)-n$ ancillary qubits for re-use in later EC gadgets. However, when there are multiple levels of concatenation, the possibility for such re-use may be limited and anyway complicated to keep track of. We thus prefer the simplicity of an overestimate.

Finally, we consider a single magic gate applied on a single qubit. At level-1 concatenation, the qubit is replaced by a logical qubit carried by $n$ physical qubits, and the physical magic gate is promoted to a 1-gate-Rec, with a logical magic gate (implemented using magic-state injection) followed by the EC gadget.
We have already discussed the qubit costs for the EC gadget, so let us focus on the logical magic gate.

The magic gate requires a magic state $\ket{A_\opmagic}$ for fault-tolerant implementation. Here $\ket{A_\opmagic}$ is a logical state carried by $n$ physical qubits and inside $\SL$. Its preparation requires $v_{A;{\rm prep}}$ ancillary qubits, followed by an EC gadget consisting of $\lambda_{EC}$ ancillary qubits. The magic state must be verified, with a chance of failing verification, so one must add a failure multiplicity factor $\mu_A$. The total qubit cost for $\ket{A_\opmagic}$ is thus
\begin{equation}
\lambda_A\equiv\mu_A(n+v_A).
\end{equation}
where $v_A= v_{A;{\rm prep}}+\lambda_{EC}$. Note that including the ancillary qubits required by the EC gadget, $\lambda_{EC}$, inside $v_A$ means that we will also throw away these qubits whenever the magic state fails verification. While it may seem an excessive over-counting of resources, in practice, an EC gadget is often used inside the actual fault-tolerant magic-state preparation circuit. This is at least the case in the schemes that we study (see Figs.~\ref{fig:A_pis4_prep}(a) and \ref{fig:H_FT_Prep_flag}). It is thus natural to directly include the qubit cost for the EC gadget applied on the magic state in $v_A$ itself. In addition, there is the qubit cost for the state carrying the data $\ket{\psi}_\mathrm{L}$ associated with the EC gadget that must be applied after the gate operation between it and $\ket{A_\opmagic}$. The qubit cost for $\ket{\psi}_\mathrm{L}$ is thus the same as for a normal gate and given by $\lambda_\opnormal$.
Hence, the total qubit cost for the magic gate is
\begin{equation}\label{eq:NDCost}
\lambda_{\opmagic,\mathrm{total}} =\lambda_\opnormal+r_A\lambda_A.
\end{equation}
The term $\lambda_\opnormal$ counts the physical qubits needed to prepare, carry, and protect the logical data qubit. The remaining term counts the qubits needed for the magic state, with an added time multiplicity $r_A$ to account for the possibility of multiple magic gates applied on the same logical qubit close in time.

Now we are ready to divide the total qubit cost in Eq.~(\ref{eq:NDCost}) into two categories: normal and magic. There are only normal gates in the EC gadget, and there are only normal gates applied between the magic-state qubits and the data qubits. Hence, the only qubits in Eq.~(\ref{eq:NDCost}) that could be magic qubits are the $r_A \lambda_A$ qubits involved in the magic-state preparation. 
Looking at the magic-state preparation circuit allows us to identify the number of magic qubits in it, $\lambda_\opmagic$, with the remaining $b=\lambda_{\opmagic,\mathrm{total}}-\lambda_\opmagic$ qubits being normal.
These are the remaining two numbers,  $\lambda_\opmagic$ and $b$, needed for the matrix approach in Sec.~\ref{sec:simplest-case}.

For a concrete example of finding $\lambda_\opmagic$ and $b$, consider the 7-qubit scheme that we analyze in Sec.~\ref{sec:FTQC7qb}. There only the $n$ qubits carrying the magic state undergo magic gates, all other qubits undergo only normal gates. Thus, the number of magic qubits in a magic gate is 
\begin{eqnarray}
\lambda_\opmagic&=&r_A\mu_A n,
\label{eq:lambda-hard-formula}
\end{eqnarray}
and the number of normal qubits in a magic gate is
\begin{eqnarray}
b&\equiv& \lambda_{\opmagic,\mathrm{total}}-\lambda_\opmagic
\nonumber \\
&=& \lambda_\opnormal+r_A \mu_A v_A.
\label{eq:b-formula}
\end{eqnarray}
Other schemes may give different numbers, but the principle used to identify  $\lambda_\opmagic$ and $b$ remains the same.

It is worth recalling that below Eq.~(\ref{Eq:R-for-2-by-2}), we used the fact $b > \lambda_\opnormal$ to show that $R(K)$ grows monotonically with $K$, specifically in the regime $ \lambda_\opnormal > \lambda_\opmagic$. 
Now we have all the ingredients to see why $b \geq \lambda_\opnormal$. 
A magic gate requires us to perform normal gates on the qubits that carry the logical state, in addition to the preparation of the magic state 
$\ket{A_\opmagic}$. Hence, $b$ (the number of normal qubits in this magic gate) must equal the number of normal qubits in a normal gate, $\lambda_\opnormal$, plus the number of normal qubits in the preparation of the magic state
$\ket{A_\opmagic}$. This means that $b\geq \lambda_\opnormal$.
Further, we assumed here that the preparation of the magic state involves an EC gadget (which requires $\lambda_{\rm EC}$ normal qubits), which means that $b \geq \lambda_\opnormal+\lambda_{\rm EC}$. Generically, $b$ will be even larger than this because of the factors of time and failure multiplicities that appear in the preparation of the magic state. 
Thus, it is no surprise that our examples have $b$ larger than $\lambda_\opnormal$.  This has a interesting consequence for the results of the matrix approach in Sec.~\ref{sec:simplest-cost-of-magic}; it means that generically 
$b/(\lambda_\opnormal-\lambda_\opmagic)$ in Eq.~(\ref{Eq:asymptotics_of_R}) is notably larger than one, as we will see in our examples.

Note that to keep the logic clear here, we assumed that there is only one type of normal qubit and one type of magic qubit. More generally, there can be multiple types of both that follow different scaling rules, for example if different gates have different qubit costs, or if some ancillary states prepared to carry out a given gate have higher qubit costs than others.
Then one might have $j_\opnormal$ types of level-$k$ normal qubits, and $j_\opmagic$ types of level-$k$ magic qubits. The scaling analysis for such a situation is described in Appendix~\ref{sec:scaling-general-details}; 
it is more complicated than the simple case presented in Sec.~\ref{sec:simplest-case}, but the logic is the same, so the simple case remains representative of more general cases.
As a concrete example, we will see that there is one type of magic gate, but two types of magic qubits in the flag-qubit example of Sec.~\ref{sec:example-flag-qubit_EC}. With multiple types of magic qubits, the $b$ and $\lambda_\opmagic$ parameters have multiple components.  In such a case, one can calculate each component of $\lambda_\opmagic$ and $b$ by following the same recipe as above. First, one identifies the total number of qubits necessary for a given type of magic qubit, with terms coming from preparation, error correction, and verification of its magic state. Some of these terms are likely to be different for different types of magic qubits, each with a different $\lambda_{\rm \opmagic,total}$. Second, one separates this total into the different types of qubits, identifies the number of each type of normal qubits (these are the components of $b$), 
and the number of each type of magic qubits (these are the components of $\lambda_\opmagic$). We will discuss this more carefully in the flag-qubit example; see Sec.~\ref{sec:example-flag-qubit_EC}.

\subsection{Resources for ``normal'' circuits with multiple levels of concatenation}
\label{sec:DOnly}

Having understood the resource costs of the basic components, we now turn to logical circuits formed from these components, beginning with ``normal'' circuits, namely, those that employ only normal operations. Once again, we assume all normal qubits are the same --- the simple case in Sec.~\ref{sec:simplest-case} --- while noting that the extensions to more general cases are straightforward but tedious (using the method in App.~\ref{sec:scaling-general-details}). In this case, as the circuit has no magic operations, we have $\Qalgo_\opmagic=0$ and $\Qalgo_\mathrm{total}=\Qalgo_\opnormal$.
We start with $K=1$. We encode each logical qubit as a level-$1$ logical qubit carried by $n$ physical qubits. Each logical qubit is prepared at the start of the computation in state $\ket{\psi_0}\in\SL$ with a 1-prep-Rec, namely, the fault-tolerant preparation of $\ket{\psi_0}$ followed by the EC gadget. Each logical gate is promoted to a 1-gate-Rec, comprising a normal gate implementable in time $\tauG$ with no need for any ancillary qubits, followed by the EC gadget.

The physical qubit costs are then estimated
using Eq.~(\ref{eq:lambda-easy-formula}), we have
\begin{equation}
\label{eq:roughLambda1}
\Qphys_\mathrm{total}\  \equiv \ \Qphys_\mathrm{total}(K=1) \ = \ \lambda_\opnormal \Qalgo_\mathrm{total}
\end{equation}
for this $K=1$ situation.

Now, for $K>1$, we simply treat the \emph{physical} circuit for the above $K=1$ case as if it is the new logical circuit, set $\Qalgo_\mathrm{total}$ in the formula above to now be $\Qphys_\mathrm{total}(K=1)$, and then reapply the above logic. We need only check that the physical circuit after each additional level of concatenation remains a normal circuit. This is the case since logical normal operations are implemented in a fault-tolerant manner using only physical normal operations, as per our earlier assumption; the EC gadget also only employs normal operations. Every logical qubit then again requires $\lambda_\opnormal$ physical qubits, and we pick up one additional factor of $\lambda_\opnormal$ for every additional level of concatenation. Thus, for $K$  levels of concatenation, we have
\begin{equation}\label{eq:roughLambdak}
\Qphys_\mathrm{total}(K) \,=\, \lambda_\opnormal^K \Qalgo_\mathrm{total},
\end{equation}
where $\Qalgo_\mathrm{total}$ here still refers to the number of logical qubits in the original 
level-$K$ logical circuit. An implicit assumption made here is that the failure and time multiplicity factors are the same for all levels $k\leq K$ of concatenation (so that $\lambda_\opnormal$ is a $k$-independent constant). This will be the case in all the concrete examples we will treat.

\subsection{Resources for general computational circuits with multiple levels of concatenation}

In the previous section, a logical circuit that contains only normal operations gives circuits that contain only normal operations at all levels of concatenation. In that case, ancillary qubits are needed only for the EC gadget, but not the gate operations themselves. For general algorithmic circuits, we will also need ancillary qubits for the magic-state preparation for any magic operations. Our resource analysis thus needs to account for those extra ancillary qubits.

Here, our goal is to give a more rigorous explanation of the formulas in Sec.~\ref{sec:summary}, in terms of the details of the recursive schemes introduced above.
We consider a general algorithmic circuit $\cC_\mathrm{algo}$ of the type described in Sec.~\ref{sec:C_algo}, that employs $\Qalgo_\mathrm{total}$ qubits. 
We consider the bottleneck layer $\Lmax$ in $\cC_\mathrm{algo}$, which is the step of the abstract circuit that requires the most physical qubits.
Under the assumptions explained in detail in Sec.~\ref{sec:C_algo} and Sec.~\ref{sec:layer}, this is the timestep of the logical (level-$K$) circuit that has the most magic gates. 
As in Sec.~\ref{sec:simplest-case}, we denote  $\Qalgo_{\opmagic}$ as the number of level-$K$ qubits on which those magic gates are applied in layer $\Lmax$, and set $\Qalgo_{\opnormal}\equiv \Qalgo_\mathrm{total}-\Qalgo_{\opmagic}$. 
For simplicity, we assume that all the magic gates have the same qubit cost given by Eqs.~\eqref{eq:lambda-hard-formula} and \eqref{eq:b-formula}.  This corresponds to the simple case in Sec.~\ref{sec:simplest-case}, while again noting that the extension to the general case is straightforward but tedious (see App.~\ref{sec:scaling-general-details}).

We now begin with a physical circuit (this can be the unencoded algorithmic circuit, or the physical circuit at some intermediate level of concatenation) comprising a certain number of normal and magic operations, applied on a number of physical qubits. We note that a physical qubit that is acted upon only by normal operations, when promoted to a logical qubit upon concatenation, will be carried by physical qubits that undergo only normal operations (as per our earlier assumption).
However, a physical qubit that undergoes a physical magic operation, at the next level of concatenation, will require $\lambda_{\opmagic,\mathrm{total}}$ physical qubits from Eq.~\eqref{eq:NDCost} above, but this can be split into $\lambda_\opmagic$ qubits that will undergo at least one physical magic operation (see Eq.~\eqref{eq:lambda-hard-formula}), and the remaining $b\equiv \lambda_{\opmagic,\mathrm{total}}-\lambda_\opmagic$ qubits that are acted upon by normal operations only (see Eq.~\eqref{eq:b-formula}). Then a single step in the recursion is given by the matrix equation in Eq.~(\ref{eq:recursion_T_Steane-1}).
Thus, the recursion for $K$ levels of concatenation can be concisely written as 
in Eq.~(\ref{eq:recursion_T_Steane}).

As explained in Sec.~\ref{sec:simplest-case}, we see that the number of physical qubits grow exponentially with $K$, at a speed dictated by the two eigenvalues of $\mathbf{M}$, 
namely, $\lambda_\opnormal$ and $\lambda_\opmagic$. 
In our examples below, we will explore the relative sizes of the two eigenvalues, which will then determine the dominant scaling behavior, quantified by how $R$ grows with $K$ [see Eq.~\eqref{Eq:R-for-2-by-2}].

\subsection{This matrix approach gives an overestimate on physical qubit costs}

Our application of the matrix approach is actually computing an overestimate of the qubit requirements. This is because there are a few places where we make simplifying assumptions that overestimate the physical qubit costs of a calculation. Some of these assumptions could be improved upon within our matrix approach (at the cost of making the calculations more complicated), thereby reducing the estimate that it predicts. 

As a first example of our overestimating of the qubit cost of magic-gates, we assume that a single magic gate at level-$K$ in the layer $\Lmax$ has the same physical qubit cost as doing a magic-gate at every step at level-$K$ during $\Lmax$.  While this may sometimes be an overestimate of the cost of magic-gates, it is not a bad estimate in our examples, for which this overestimate (solid curves in Fig.~\ref{fig:total-resources}) is within an order of magnitude of an under-estimate (dashed curves in Fig.~\ref{fig:total-resources}) in which we assume the cost od a magic qubit cost is no more than a normal qubit.

A second example is that we overestimate the physical qubit costs when we assume that each normal qubit always requires $\lambda_\opnormal$ qubits, assigning the same time multiplicity to all of them. (as outlined above, and calculated in detail below in Sec.~\ref{sec:time-multiplicity}). Yet, a few of the ancillary qubits are measured earlier than others, making them available for re-use earlier.
If the circuit can be designed to reassign and re-use them as soon as they become available (which would probably require a complicated numerical optimization of the full circuit), then we would require less qubits than our matrix approach indicates. 
However, such optimizations are likely to be numerically heavy and highly algorithm-dependent. So our feeling is that such algorithm-specific numerical optimization should only be considered after we have applied the matrix approach presented here to identify more generic ways of reducing the number of qubits, such as those identified in this article and shown in Fig.~\ref{fig:total-resources}.


\section{Multiplicity factors}
\label{sec:multiplicities}

This section provides a detailed evaluation of the multiplicity factors that were mentioned as important in our summary of our work (Sec.~\ref{sec:summary} above), and then introduced in Sec.~\ref{sec:resources_for_basic_components} above.

For a given fault-tolerant quantum computing scheme, most of the quantities that determine $\lambda_\opnormal$,$\lambda_\opmagic$, and $b$ (e.g., $n_A$, $n$, $v_i$, etc.) come directly from the scheme's circuit designs; we show how to write them down for the 7-qubit-code scheme in Sec.~\ref{sec:FTQC7qb}. 
The time and failure multiplicity factors, $r_i$s and $\mu_i$s, however, go beyond circuit designs and warrant a separate explanation of how we can determine them. 
We note that the considerations discussed in this section are often neglected in standard fault-tolerance analysis. As we will see below, however, they can play a significant role in properly accounting for the physical qubit cost of a computation, and hence warrant a careful treatment.

\subsection{Time multiplicity}
\label{sec:time-multiplicity}

In our formulas for $\lambda_\opnormal$ and $\lambda_{\opmagic, \mathrm{total}}$, we assume that once ancillary qubits are measured, they become immediately available to be reused to prepare whatever new ancillary state is needed at that moment. We noted earlier, however, that the preparation time for these ancillary states 
(such as syndrome states necessary for EC gadgets or magic states necessary for magic gates) is often longer than the time between two consecutive uses of those ancillary states in the same code block. The time multiplicity for a given type of
ancillary state accounts for the additional ancillary qubits needed to prepare multiple copies of that state at the same time, to have enough of them ready for all upcoming usages. 
To establish the upcoming usage of ancillary states, let us recall that we want the physical qubit resources for the bottleneck layer of the algorithm, $\Lmax$ (the moment when the algorithm requires the most physical qubits) defined in Sec.~\ref{sec:layer}. As 
we want to allow for any algorithm, we assume that all subsequent layers
use the same number of ancillary states as the bottleneck layer, $\Lmax$ (see Sec.~\ref{sec:layer} for details).  This assumption makes the required number of ancillary states unvarying in time, greatly simplifying the calculation of the time multiplicity.

Here, we explain in detail how to get an expression for time multiplicity $r$ for syndrome states used in EC gadgets.  The expression for the time multiplicity for other states --- such as the time multiplicity for a magic state, $r_A$ --- follows the same logic.

Let us consider the preparation of syndrome states that are used in EC gadgets.
We begin as usual with $k=1$. We first count how often the EC gadget is employed for each code block. This is simply the time $\tauL$ for the 1-gate-Rec, assumed to be for a normal gate. $\tauL$ is given by a sum of the time $\tauG$ for the normal gate itself and the time of the EC gadget $\sum_{i=1}^{I_S}d_i\tauG+d_R\tauG$, where $d_i$ is the number of gates in the sequence that performs $G_i$ in the EC gadget, and $d_R$ is the number of gates in the sequence implementing $G_R$ (see Fig.~\ref{fig:EC_gadget}). The measurements of $S_i$s are assumed to be done in the same timestep as the subsequent gates on the data qubits within the EC gadget and hence take no extra computational time. This thus gives
\begin{equation}\label{eq:tauL}
\tauL= {\left(1+\sum_{i=1}^{I_S}d_i+d_R\right)}\tauG
\end{equation}
as the time between two consecutive applications of the EC gadget. 
$\tauL$ tells us how often fault-tolerantly prepared syndrome states are needed.

We want to compare $\tauL$ with the total time $\tauAnc$ the ancillary qubits needed in the EC gadget are in use, from the initial state preparation to their final measurement. After that time, the qubits can be reused. For simplicity, we compute $\tauAnc$ by considering the longest time among all ancillary qubits used in the EC gadgets (note that not all ancillary qubits carry information for the same amount of time).
From Fig.~\ref{fig:EC_gadget}, we find
\begin{align}\label{eq:tauAnc}
\tauAnc&=\max_{i=1,2,\ldots,I_S}(\tau_i+d_i\tauG+\tau_\mathrm{M}).
\end{align}
Here, $\tau_i\equiv \tau_{S_i}$ is the time taken for 
the fault-tolerant preparation of the ancillary state $\ket{S_i}$; the second term $d_i\tauG$, as before, gives the time taken for the $G_i$ gate in the EC gadget; $\tau_\textrm{M}$ is the time for the measurement of $S_i$ in the EC gadget. 

Comparing $\tauAnc$ and $\tau_\mathrm{L}$, we find that often $\tauAnc >\tau_\mathrm{L}$.  Then
for the computation to continue uninterrupted (under the above assumption that the required number of ancillary states is unvarying in time), we need at least 
\begin{equation}
r\equiv \lceil\tauAnc/\tau_\mathrm{L}\rceil
\label{eq:r_i-definition}
\end{equation}
sets of ancillary qubits at any one time \cite{ceiling}.
Put differently, at any given moment in the computation, we need to be preparing the syndrome states for $r-1$ future EC gadgets, as well as using syndrome states in the current EC gadget.

A similar argument gives $r_A$, which is the time multiplicity for the ancillary qubits carrying the magic states, and depends on the depth of the circuit that prepares the magic-state. We explain how to obtain $r_A$ in Sec.~\ref{sec:FTQC7qb}, using the specific example of the 7-qubit-code scheme.

Now, the above discussion was for one level of concatenation. In principle, $r$ can depend on $k$. For simplicity, we remove this dependence by redefining $r\equiv\max_{k \geq 1} \lceil \tauAnc(k)/\tau_\mathrm{L}(k) \rceil$, a maximum that can be found for a given fault-tolerance scheme together with its full circuit designs \footnote{In the case where $\tauAnc(k)/\tau_\mathrm{L}(k)$ is unbounded when $k$ grows (which never occurs in the examples we treat later on), one can simply redefine $r\equiv\max_{k \in [1,K]} \lceil \tauAnc(k)/\tau_\mathrm{L}(k) \rceil$ so that the maximum is well-defined over the targeted concatenation levels that the experimentalist wishes to implement.}.
We also assume we do not mix ancillary qubits, in other words we assume certain ancillary qubits are reserved for EC gadgets throughout the calculation, while others are reserved for magic-states. One can imagine lowering the physical qubit cost by using qubits wherever and whenever they are needed, but such a complicated orchestration can be devised only for a given scheme and a given computational algorithm, and is beyond the scope of the current paper.

\subsection{Failure multiplicity}
\label{sec:failure-multiplicity}

Let us now discuss how to compute the failure-multiplicity factors $\mu_i$s and $\mu_A$. Recall that the failure multiplicity accommodates the finite probability of failure of the verification test during the preparation of a state in $\SL$ (i.e., the initial computational state, the syndrome states, or the magic states). To avoid holding up the computation, we must prepare more copies than actually required, so that we have a good chance that at least one copy passes the verification to be used in the computation at the time of need. 
For this, we imagine preparing $\mu$ copies of the state every time the circuit requires a single verified copy, i.e., a copy that passes the verification. Of course, it might happen that all $\mu$ copies fail verification, and so are considered useless, and we are forced to inject one of these useless states into the circuit, possibly causing an error.  However, we face a contradiction: On the one hand, we want a $\mu$ large enough that this failure case happens only very rarely; on the other hand, it is prohibitively costly to take an extremely large $\mu$. We hence want to find the smallest $\mu$ that is ``good enough''.

To resolve this contradiction, and identify when $\mu$ is ``good enough'', we note that the injection of a useless state (because all $\mu$ states failed verification) gives a potential preparation fault in the circuit, occurring with a probability given by the likelihood of all $\mu$ states failing verification. This is an additional fault location in the circuit, not counted in the standard case when $\mu$ is taken as infinite so that we never inject a useless state. This increases the total number of fault locations that determines the threshold $p_\mathrm{thres}$ [as explained below Eq.~\eqref{eq:textbook-scaling_k-basics}], hence reducing the value of $p_\mathrm{thres}$. This interplay between the code's threshold, and the number of states prepared for verification, $\mu$, has often been ignored. Some works explicitly make the assumption that there is always a prepared state that successfully passed verification when needed \cite{Aliferis2005Apr,Chamberland2017Feb}. Other studies do acknowledge the extra resources verification failure implies under simplifying assumptions, but ignore how these events modify the fault-tolerance threshold. They typically compute an average number of qubits required per verified state, relying on the law of large numbers \cite{Chamberland2017Feb,Chamberland2019May}.
In contrast, we show below that even very large quantum circuits are not in the regime governed by the law of large numbers, and its use under-estimates the physical qubit cost.

Now, we want the smallest $\mu$ for which the probability that all $\mu$ copies fail verification is small enough to not {\it significantly} reduce the fault-tolerance threshold. While we analyze the general case below, when we want precise numbers for $\mu$, we decide that $\mu$ should take the smallest value large enough that the threshold is only reduced by $1\%$.

\begin{table*}[t]
\begin{tabular}{|l|c|c|c|c|c||c|c|}
\hline
\parbox[t]{2.1cm}{Fault-tolerant scheme}&\parbox[t]{1.9cm}{State being prepared}&\parbox[t]{1.2cm}{Circuit diagram}&\parbox[t]{1.9cm}{N$^{\circ}$ of fault locations per verification, \\ $N$}
&\parbox[t][1.6cm]{1.9cm}{Verification failure probability $\leq Np_{\rm thres}$}
&\parbox[t]{1.9cm}{N$^{\circ}$ of verifications per exRec, $M$} 
&\parbox[t]{2.5cm}{Failure multiplicity \textit{\textbf{without}} shared prep.} 
& \parbox[t]{2.2cm}{Failure multiplicity \\
\textit{\textbf{with}} shared \phantom{p}prep.\phantom{p}} 
\\
\hline \hline
\multirow{3}{*}{\parbox[t]{2cm}{Steane approach}} & 
$\ket{0}$ or $\ket{+}$ & Fig.~\ref{fig:FT_prep} & 50 & $\leq 0.10\%$ & 8 & 
\phantom{$\Big|$}%
$ 
\muCl = 3$ \ & 
\, $\muCl = 1.16$ \ 
\\
 & 
$\ket{A_{\pi/4}}$ & 
Fig.~\ref{fig:A_pis4_prep} & 521 & $\leq 1.0\%$ & 8 &
\phantom{$\Big|$}$
\ \, 
\mu_\text{A} = 4$ & \ \ \ \ $\mu_\text{A} = 1.30$ \ 
\\
 & 
 $\ket{\text{Rep}}$ & Fig.~\ref{fig:A_pis4_prep}c & 17 &
 $\leq 0.034\%$ & 8 &
\phantom{$\Big|$}$
\mu_\text{Rep} = 2$ \ &
 \ $\mu_\text{Rep} = 1.13$ \ 
\\
\hline 
\multirow{2}{*}{\parbox[t]{2cm}{Flag-qubit approach}} & 
$\ket{0}$ or $\ket{+}$ & Fig.~\ref{fig:+_prep_flag} & 
36 & 
$\leq 0.072\%$ & 1 &
\phantom{$\Big|$}$
\muClFlag = 3$\ &
\phantom{$\Big|$}$
\muClFlag = 1.19$\ 
\\ 
& $\ket{\text{H}}$ & Fig.~\ref{fig:H_FT_Prep_flag} & 
\ $\approx \sqrt{\frac{2}{p_{\text{thres}}}}=316 $ \ & $\leq 0.63\%$  & 1 &
\phantom{$\Big|$}$
\mu^{\text{flag}}_H = 4$ \ & 
\phantom{$\Big|$}$
\mu^{\text{flag}}_H = 1.31$
\\
\hline
\end{tabular}
\caption{A list of the parameters needed to evaluate failure probability and failure multiplicity for the 
7-qubit code examples considered in Sec.~\ref{sec:FTQC7qb}. 
In all cases we assume the physical error probability per gate to be $p_0< p_{\text{thres}}=2 \times 10^{-5}$.
We identify $M$ as the number of such verified states needed in whichever exRec requires the most such verified states. For Steane EC gadgets, the exRec with the maximum number is that of the $\cnot$, with $M=8$. 
For flag-qubits, the only verification is a single one needed in the preparation of $\ket{H}$ or $\ket{0/+}$, so $M=1$.
The probability that any exRec lacks a verified state is then upper bounded by $Mp_{\rm lack}$; see below Eq.~\eqref{eq:plack_cond}. 
For the 7-qubit code with the Steane approach, $|0\rangle$ or $|+\rangle$ are required for all gates (normal and magic), while $|A_{\pi/4}\rangle$ and $|\text{Rep}\rangle$ are only required for magic gates. The failure probabilities (upper bounded by $Np_{\rm thres}$) are relatively small (1\% or less). Yet they are much too big for a successful fault-tolerant quantum computation. This is why we need to prepare multiple copies of a state to ensure that the probability $p_{\rm Lack}$ of all $\mu$ copies failing is many orders of magnitude smaller.
For the 7-qubit code with the flag-qubit approach, $|0\rangle$ or $|+\rangle$ are required for all gates (normal and magic), while  $|\text{H}\rangle$ is only needed for magic gates. The failure multiplicity without shared state-preparation (labeled ``without state prep.'') is the smallest integer satisfying Eq.~(\ref{eq:rob-m-inequality-simpler}) for $\kappa=1\%$.
The last column uses the approach in Sec.~\ref{sec:shared-failure-multiplicity} that shares state preparation for a group of 100 gates. In this case, the failure multiplicity drops to fairly close to one.
}
\label{table:failure-multiplicity}
\end{table*}

We start with the probability that a copy fails verification, and use the fact that this is upper-bounded by $Np$, where $p$ is the probability of a fault per gate, and $N$ is an upper bound on the number of locations inside the preparation and verification circuit where a single fault can lead to 
failed verification. Typical values of $N$ are shown in Table~\ref{table:failure-multiplicity} (for the examples discussed in Sec.~\ref{sec:FTQC7qb}). 
If we prepare $\mu$ copies every time the circuit requires a single {\it verified} copy of a state, the probability we lack a verified state is the probability 
that all $\mu$ copies fail verification. This is upper bounded by \footnote{The upper-bound of $Np$ does not require independent errors. However, when errors are independent, such a probability takes the form $1-(1-p)^N$, for which the upper bound $Np$ is quite tight for the concrete values in Table~\ref{table:failure-multiplicity}.} 
\begin{eqnarray}
p_{\text{Lack}}=(Np)^\mu.
    \label{eq:plack_cond}
\end{eqnarray}
The exRec --- see below  Eq.~\eqref{eq:textbook-scaling_k-basics} --- will require $M$ such verified states\footnote{If there are multiple exRecs with different $M$s, we simplify our analysis (while getting an upper bound) by replacing all these $M$s by whichever $M$ is largest.}. Then the probability of that exRec lacking verified states is upper bounded by $Mp_{\text{Lack}}$.

If all $\mu$ do fail, then one is forced to use a possibly badly prepared state in the circuit, potentially introducing an error in the quantum computation. Otherwise, with probability $>1-p_{\rm Lack}$, one has a verified state  to use in the circuit, and one can throw away the other $(\mu-1)$ copies.
Throwing away the other $(\mu-1)$ copies is wasteful, since many may have passed verification, but it keeps the algebra simple; we will be less wasteful in Sec.~\ref{sec:shared-failure-multiplicity}.

Verification failure is a source of errors usually ignored in standard treatments of fault-tolerant quantum computing, which assume, for simplicity, that the circuit {\it never} lacks verified states, implicitly assuming that $\mu\to \infty$. To incorporate verification failure in the description, it is useful to recall how Eq.~\eqref{eq:textbook-scaling_k-basics} is derived. We do this here for the situation where the code underlying the fault-tolerant quantum computing scheme corrects $t=1$ error; it is straightforward to generalize the analysis below to $t>1$. We begin with $p_k$, which is the  upper bound on the probability of error for a logical gate after $k$ levels of concatenation. For $\mu\to \infty$ (and $t=1$), we have
\begin{eqnarray}
p_{k+1} &=& B p_{k}^2\, , 
\label{eq:textbook-scaling_kp1_fct_k}
\end{eqnarray}
The right-hand side of Eq.~\eqref{eq:textbook-scaling_kp1_fct_k} is the probability 
that there is an error that the circuit cannot correct. Such an uncorrectable error comes from at least a pair of faults in the exRec of interest (namely, the largest one, which hence determines the threshold), and $B$ is primarily the number of such pairs that are malignant (i.e., pairs that cause an uncorrectable error in the end), corrected to account also for uncorrectable errors from more than two faults in the circuit. The error probability per logical gate operation after $k$ levels of concatenation is then
\begin{eqnarray}
p_{k}=\frac{1}{B}(B p_0)^{2^k}\, ,
\label{eq:textbook-scaling_k}
\end{eqnarray}
which corresponds to Eq.~(\ref{eq:textbook-scaling_k-basics}) for the case considered here with $t=1$. This gives the usual fault-tolerance threshold that requires that $p<p_{\rm thres} \equiv 1/B$, for $p_k$ to decrease with increasing $k$. 

The above analysis was for $\mu\to\infty$. For finite $\mu$, Eqs.~\eqref{eq:textbook-scaling_kp1_fct_k} and \eqref{eq:textbook-scaling_k} must be modified to account for errors that come from the lack of verified states to inject at the relevant moment. This is another source of uncorrected errors that must be added to the right-hand side of Eq.~(\ref{eq:textbook-scaling_kp1_fct_k}), so that we have
\begin{eqnarray}\label{eq:modifiedThres}
p_{k+1} = B p_{k}^2 + M p_{\text{Lack}}\,.
\end{eqnarray}
We want the value of $\mu$ for which the additional contribution from the $M p_{\text{Lack}}$ term is {\it no worse} than a small reduction of the threshold from $1/B$ to $1/[(1+\kappa)B]$ for 
small $\kappa$ (below we choose $\kappa=0.01$).
In other words, we want 
\begin{eqnarray}
B p_{k}^2 + M p_{\text{Lack}} \ \leq\  (1+\kappa) B p_{k}^2\,.
\label{Eq:inequality_for_pLack}
\end{eqnarray}
This  would guarantee that $p_{k+1} \leq (1+\kappa) B p_{k}^2$, meaning that the error per logical gate operation after $k$ levels of concatenation would obey
\begin{eqnarray}
p_k \leq \frac{ \big(B(1+\kappa) p_0\big)^{2^k}}{B(1+\kappa)}. 
\label{eq:pk_new_threshold}
\end{eqnarray}
In this case, the logical error rate after $k$ levels would be about the same as if one never lacked verified states but the inverse of the threshold, $B$ is larger by a factor of $\kappa$.
If we take $\kappa=1\%$, then this can be interpreted in an approximate way as follows:
A quantum computer with $\mu$ satisfying Eq.~(\ref{eq:rob-m-inequality}) is about as reliable as a quantum computer with infinite $\mu$ (so it never lacks verified states) but with a physical error probability about 1\% higher. So, if one can reduce the physical error probability by about 1\%, one has compensated for $\mu$ not being infinite.

To find the value of $\mu$ that ensures Eq.~(\ref{Eq:inequality_for_pLack}) holds for a chosen $\kappa$, we substitute in Eq.~(\ref{eq:plack_cond}) and rearrange it to get the inequality 
\begin{eqnarray}
\mu \,\geq\ \ \frac{2\ln[1/p_k] +\ln[1/\kappa] - \ln[B]+\ln[M]}
{\ln[1/p_k] - \ln[N]}\, .
\label{eq:rob-m-inequality}
\end{eqnarray}
In the regime of interest ($p_k < p_0 < p_{\rm thres} \equiv 1/B$),  the right-hand side of Eq.~(\ref{eq:rob-m-inequality}) decays monotonically as we reduce errors, so will never exceed the value given by replacing $p_k$ by $p_{\text{thres}}$. Hence, to be sure  Eq.~(\ref{eq:rob-m-inequality}) is satisfied for any $p_k$ of interest, it is sufficient to satisfy it when $p_k$  is replaced by $p_{\text{thres}}$.
Then, Eq.~(\ref{eq:rob-m-inequality}) becomes 
\begin{eqnarray}
\mu \ \geq\   \frac{\ln[M B/\kappa]}
{\ln[B/N]} \,.
\label{eq:rob-m-inequality-simpler}
\end{eqnarray}
Of course, $\mu$ must be an integer, and we will take it to be the smallest integer satisfying this inequality.

As the right-hand side of Eq.~(\ref{eq:rob-m-inequality-simpler}) only depends logarithmically on $B$, $\kappa$, $M$, and $N$, it only depends weakly on the exact choice of these parameters.
To get a feeling for the typical values of $\mu$ given by Eq.~(\ref{eq:rob-m-inequality-simpler}), we can look at
the column of Table~\ref{table:failure-multiplicity} labeled ``Failure multiplicity without shared prep.''. It gives $\mu$ for the $N$ and $M$ typical of states in the two 
fault-tolerant schemes considered in Sec.~\ref{sec:FTQC7qb}, assuming we want $\kappa=1\%$.
The important thing we learn from this is that these failure multiplicities are typically fairly small integers (typically two, three or four). So one can expect that each time one requires a single verified state, one must prepare between two and four copies of that state for verification. In principle, $\mu$ can depend on $k$, but we neglect that here. To be rigorous, we can say that we take the largest $\mu$ over all $k$ values, and use that value for all $k$, thereby overestimating this failure multiplicity $\mu$.

\subsection{Lowering the failure multiplicity by sharing state preparation}
\label{sec:shared-failure-multiplicity}
In the previous section, we assume that each time we need a verified copy of a state at a given place in the circuit, we prepare $\mu$ copies of that state, but in the end, we discard all but one copy for use in the actual circuit.
This is extremely wasteful, because the majority of those discarded states also successfully passed verification. 
We will show here that it is vastly less wasteful if we created a shared reservoir of each type of state that needs to be prepared, do verification on each of them, and then distribute those that successfully passed verification to where they are needed (while throwing away those that fail verification).
If the probability that each state is verified is of order $(1-Np_{\text{thres}})$, then the ``law of large numbers'' means that you might immediately guess that when preparing a very large number of copies, one only needs to prepare a bit more than $1/(1-Np_{\text{thres}})$ unverified copies for every verified state needed, implying a failure multiplicity $\sim 1/(1-Np_{\text{thres}})$,
which is clearly much smaller than without sharing. Unfortunately, these sorts of ``large-number'' arguments are too simplistic for finite numbers (e.g., it cannot tell us what happens if we need only 100 verified states, 
thus we treat the problem differently here.
However, as we explain below, we do reach the conclusion that sharing the preparation of verified states in this manner greatly reduces the failure multiplicity. 

To see how much it can reduce the failure multiplicity in practical situations, one can compare the last two columns of Table~\ref{table:failure-multiplicity}, and note that sharing state preparation can reduce failure multiplicities from 4 to less than 1.3 \footnote{These values are still significantly larger than that given by the law of large numbers, $1/(1-Np_{\text{thres}})$. This shows that even very large computations will not be large enough to be approximated by the law of large numbers.}. As we will see, the reduction is based on a simple argument that overestimates the failure multiplicity for shared resources, so one can hope it could be even smaller. However, it cannot be smaller than one, so this already gives us a good idea of its order of magnitude.

To be more precise, we note that in a typical computation, multiple verified copies of the \emph{same} state will be needed at different parts of the circuit at the same time. For example, every single level-$k$ qubit experiences an EC gadget at the same moment, so a vast number of verified ancillary states will be used at the same time for the EC gadgets. To a lesser extent, in a large quantum computer, a large number of qubits will be undergoing a magic gate at the same time, so many qubits will require verified magic states at the same time.
It is thus natural to consider sharing resources between all those gates that require the same type of verified states in the same timestep.

We imagine that copies of a state are produced and verified centrally in a state-preparation factory, often called an {\it ancilla factory}. This factory prepares and verifies as many copies as necessary (throwing away those that fail verification) to ensure that it produces enough {\it verified} states to distribute to all parts of the circuit in need of such a state at that timestep. This, of course, assumes the possibility of long-range gates or transport for distributing the states across the circuit, but long-range gates are usually already a requirement for implementing concatenated schemes, and this also fits well with architectures that have different zones for different computational functions (e.g., in the recent neutral-atom experiments \cite{Bluvstein2024Feb}).

Let us try to estimate the resource savings arising from such a shared state-preparation factory. Suppose we need $V$ verified copies of a state in a given timestep. For that, we prepare $V+S$ copies, each of which has a probability $Np$ of failing the verification. We define $p^\mathrm{Share}_\mathrm{Lack}$ as the probability that we are short of copies that pass the verification, and are hence forced to use at least one failed copy, possibly introducing an error into our computation. We want to understand how large $S$ has to be, to assure a small enough $p^\mathrm{Share}_\mathrm{Lack}$. If the EC gadgets of two logical gates (gates $i$ and $j$) use states coming from the same shared ancilla factory, if the logical gate $i$ lacks a verified state, it is more likely that the logical gate $j$ will also lack a verified state: the events are correlated, making the gate failure correlated. These correlations complicate the analysis.

Here, we prefer to simplify the analysis by making simple overestimates. These overestimates will be enough to show that sharing resources reduces the failure multiplicity to fairly close to one for reasonable values of $V$, for practical values of the maximum concatenation level $K$ (we assume that $K$ does not exceed 5 in practical situations).
Our simplification is to decide that it is enough to find the value of $S$ that ensures that $p^\mathrm{Share}_\mathrm{Lack}$ is so extremely small that there is a probability close to one that the whole calculation occurs without any operation (at any level of $k$) in that calculation lacking a verified state. 
A lack of verified states only occurs if $S+1$ of the $V+S$ copies fail verification. 
Hence, $p^\mathrm{Share}_\mathrm{Lack}$ is the probability
that more than $S$ of the $V+S$ copies of the state fail verification.
We take the probability that each copy fails verification as  $Np$. Then we can see, using the binomial distribution, that
\begin{align}
   p^\mathrm{Share}_\mathrm{Lack} (S,p) \ =\sum_{l=S+1}^{V+S} \binom{V\!+\!S}{l} (Np)^l(1\!-\!Np)^{V+S-l}.
    \label{eq:def_pLack-shared}
\end{align}
This is in fact a slight overestimate of  $p^\mathrm{Share}_\mathrm{Lack} (S,p)$, because $Np$ is a slight  overestimate of the probability that a copy fails verification (just as in Sec.~\ref{sec:failure-multiplicity} above). We now make another overestimate: We replace $p$ by $p_{\rm thres}$ in Eq.~(\ref{eq:def_pLack-shared}), and work with $p^\mathrm{Share}_\mathrm{Lack} (S,p_{\rm thres})$ from now on.
Since $p\leq p_{\rm thres}$, we know that $p^\mathrm{Share}_\mathrm{Lack} (S,p_{\rm thres})$ is a significant overestimate of the probability of lacking a verified state. This means that when we next impose an upper bound on $p^\mathrm{Share}_\mathrm{Lack} (S,p_{\rm thres})$, we are safe in the knowledge that the true probability to lack a verified copy of the state is smaller.

We now require that $p^{\text{Share}}_{\text{Lack}}$ is so small
that there is only a small chance $\kappa$ of lacking a verified state of a given type in the whole calculation, when $p=p_{\rm thres}\equiv 1/B$. 
The probability to lack a verified state (of any type) in the whole computation is upper bounded by $\alpha\,\kappa$, where $\alpha$ is the number of different types of verified states needed in the computation (e.g., $\alpha=4$ for the example discussed in Sec.~\ref{sec:example-Steane_EC}, with the four types of verified states being $|0\rangle$, $|+\rangle$, $|A_{\pi/4}\rangle$, and $|{\rm Rep}\rangle$). To ensure that $p^{\text{Share}}_{\text{Lack}}$ is small enough for a state of type $\nu$, we need to choose the number of spares for that type of states, $S_\nu$, to be large enough to ensure that the probability to lack type $\nu$ states $p^{\mathrm{Share}}_{\mathrm{Lack};\nu} (S_\nu,p_{\rm thres})$ obeys
\begin{eqnarray}
p^{\mathrm{Share}}_{\mathrm{Lack};\nu}(S_\nu,p_{\rm thres}) \ \leq\  \frac{\kappa}{N^{(\nu)}_{\rm states}},
\label{eq:p_Lack_mut-1}
\end{eqnarray}
where $N^{(\nu)}_{\rm states}$ is the number of type $\nu$ states required in the whole calculation.

Now, we note that the following inequality holds:
\begin{eqnarray}\label{eq:Nnustates}
    N^{(\nu)}_{\rm states} < K M\, N_{\rm gates}^{\text{level-1}},
\end{eqnarray}  
where $N_{\rm gates}^{\text{level-1}}$ is the total number of gates in the entire algorithm at level 1. This inequality is based on two observations. The first observation is that the number of verified states needed at level-1 concatenation is larger than at any other levels. This is because the number of verified states needed scales like the number of gates, and so 
grows exponentially as one goes down the concatenation levels towards the physical level; yet the physical level (level 0) itself requires no verified states.
So the total number of verified states required must be much fewer than $K$ times the number required at level 1. The second observation is that the number of verified states required at level 1 must be fewer than $M\, N_{\rm gates}^{\text{level-1}}$. 
Inequality \eqref{eq:Nnustates} overestimates $N^{(\nu)}_{\rm states}$, but it allows for a simple sufficient bound: We assume we choose the smallest $S_\nu$ that ensures that
\begin{eqnarray}
p^{\mathrm{Share}}_{\mathrm{Lack};\nu} (S_\nu,p_{\rm thres}) \ \leq  \ \frac{\kappa}{K M  N_{\rm gates}^{\text{level-1}}},
\label{eq:p_Lack_mut-2}
\end{eqnarray}
since any $S_\nu$ that satisfies this will easily satisfy Eq.~(\ref{eq:p_Lack_mut-1}).
We then use this $S_\nu$ to get the failure multiplicity for states of type $\nu$ as 
\begin{eqnarray}
\mu_\nu = \frac{V+ S_\nu}{V}, 
\label{eq:mu_mut}
\end{eqnarray}
where we recall that $V$ is the number of verified states we want to produce at a given timestep.

For a given fault-tolerant quantum computing scheme, it is fairly easy to get reasonable estimates of $V$, $M$, $K$ and $N_{\rm gates}^{\text{level-1}}$. We do not need rigorous values; we simply need to show that there is a clear advantage in sharing resources and how it can significantly reduce the value of $\mu$ to something fairly close to $1$. We give a worked example for the case of the 7-qubit code scheme with Steane EC gadgets in Sec.~\ref{sec:steane-shared}.
It is more involved to estimate the number of level-1 gates, $N_{\rm gates}^{\text{level-1}}$, in a circuit with $K$ levels of concatenation, however the concatenated structure means that it should be well approximated by the number of physical gates (level-0 gates) in a circuit with $(K-1)$ levels of concatenation, and this number scales, at worst, like $Q^{\text{phys}}_{\text{total}}(K-1) \times D^{K-1} \times 2 D_L$.
Here $Q^{\text{phys}}_{\text{total}}(K-1)$ is the number of physical qubits for $(K-1)$ concatenations, and $D^{K-1} \times 2 D_L$ 
is the physical circuit depth for $(K-1)$ concatenations,
where $D_L$ is the algorithm's depth
and $D$ is how the physical circuit depth increases with each concatenation level. The factor of $2$ in front of $D_L$ acknowledges the fact that a magic gate in $\cC_\mathrm{algo}$ takes two logical-step to be implemented, as shown in Fig.~\ref{fig:hard_gate} (we implement two normal gates $G_a$ and $G_b$ to perform state injection) \cite{6_footnote-DepthSharing}.
These quantities ($Q^{\text{phys}}_{\text{total}}$, $D$ and $D_L$) are all accessible for a given calculation with a given fault-tolerant scheme, as we will show in Sec.~\ref{sec:FTQC7qb}, allowing us to find $\mu_\nu$ in such cases.

Note that, formally, as $K$ increases, the
right-hand side of Eq.~(\ref{eq:p_Lack_mut-2}) vanishes. Hence, satisfying the inequality requires increasingly large $S_\nu$, as $K$ increases. However, we consider $K=5$ (or fewer) for conceivable middle-term computers \cite{2_footnote-K=5}, and we will see in the examples that $S_\nu$ remains small enough for $K=5$ that $\mu_\nu$ is fairly close to one.

We conclude this section on sharing resources by recalling that it contains big overestimates of both the likelihood of lacking verified states, and the consequential danger for the results of a calculation.
Despite these overestimates, we find that when sharing resources for reasonable parameters in the specific examples that we consider, the failure multiplicity drops to fairly close to one (see examples in the last column of Table~\ref{table:failure-multiplicity}). As these multiplicity factors cannot be below one, our overestimates nevertheless provide fairly tight bounds on the true failure multiplicity values for shared resources.

\section{Example: The 7-qubit code scheme}
\label{sec:FTQC7qb}

This section provides a concrete example of the application of our matrix approach to a well-known fault-tolerant scheme. 
This example provides the results that were shown in Fig.~\ref{fig:total-resources}. However, this section's main goal is to give a full worked example of our matrix approach for a very well-known fault-tolerant scheme, so that readers can see how to apply the matrix approach to any concatenated scheme of interest to them. 

As a concrete example, we look at the well-studied concatenated fault-tolerance scheme built upon the 7-qubit code; see Ref.~\cite{Aliferis2005Apr} for the full description of the scheme.
The underlying 7-qubit code is a Calderbank-Shor-Steane (CSS) stabilizer $[[n,k,d]]=[[7,1,3]]$ ($n$ physical qubits encoding $k$ logical qubits, with distance $d\equiv 2t+1$, and hence corrects $t$ errors) code. The two-dimensional code space --- the logical qubit --- is the $+1$ joint eigenspace of the stabilizer group generated by the set of $X$-type operators $\Xgen\equiv \{IIIXXXX,IXXIXXI,XIXIXIX\}$ and the $Z$-type version $\Zgen\equiv \Xgen(X\rightarrow Z)$. The logical $X$ and $Z$ operators can be taken to be $XXXXXXX$ and $ZZZZZZZ$, respectively. The code corrects arbitrary errors on a single one of the seven physical qubits carrying the logical qubit: Pauli $X$ errors are diagnosed by the $X$-type operators and $Z$ errors by the $Z$-type operators; $Y$ errors are considered as simultaneous $X$ and $Z$ errors; any single-qubit error can be written as a linear combination of $X$, $Y$, and $Z$ errors and are hence correctable by the code. The error diagnosis involves the nondestructive (through the use of ancillary qubits) measurement of the $\Xgen$ and $\Zgen$ operators. The measurement results (the syndromes) uniquely identify the qubit with the error and the type of error ($X$, $Y$, or $Z$). The recovery for this code can be performed ``virtually", via what is known as a Pauli-frame rotation, so that no actual quantum gates need to be carried out. This means that $d_R=0$ (see Fig.~\ref{fig:EC_gadget}) for this code. 

\vskip 3mm

\subsection{Circuit designs}

To analyze the physical qubit cost for this scheme, we need the detailed circuit designs for the 7-qubit code scheme. Here, we simply reproduce the designs devised in Ref.~\cite{Aliferis2005Apr}, leaving the reader to refer to the original paper for the reasoning behind those designs.

We begin with the physical operations that can be performed. We make the common assumption that we can do six physical gates: the single-qubit Hadamard gate, $H\equiv \frac{1}{\sqrt 2}(\ketbra{+}{0}+\ketbra{-}{1})$; the phase gate, $S\equiv\exp{\left(-\upi\frac{\pi}{4}Z\right)}$; its Hermitian conjugate $S^\dagger$; the two-qubit controlled-$X$ gate, $\cnot\equiv\ketbra{0}{0}\otimes \id+\ketbra{1}{1}\otimes X$; the $T$-gate, $T\equiv\exp{\left(-\upi\frac{\pi}{8}Z\right)}$, and its Hermitian conjugate $T^\dagger$.  These form the set $\Gphys$. $H$, $S$, and $\cnot$ together generate the Clifford group of operations; adding the $T$-gate (and $T^\dagger$) gives full universality over qubit operations. In addition, we assume we can prepare a physical qubit in the states $\ket0$ and $\ket+$ --- these form the set $\Sphys$ --- and that we can do a physical $Z$ measurement or a physical $X$ measurement, giving $\Mphys$. Each of these physical operations is assumed to take time $\tauP$ to complete.

Many logical operations for the 7-qubit code can be done transversally, automatically ensuring their fault tolerance. The code admits transversal constructions of the logical $H$ and $S$, i.e., the logical gate is achieved by applying physical gates to individual physical qubits. The logical $\cnot$ is also transversal, composed from physical $\cnot$ gates acting on corresponding pairs of physical qubits in the two code blocks. $H$, $S$, and $\cnot$ are hence all normal gates, and since they generate the whole Clifford group, any Clifford operation is a normal operation. The measurement of the logical $Z$ operator is transversal as well: It can be achieved by measuring the physical $Z$ operator on each of (a subset of) the physical qubits (e.g., qubits 1, 2, and 3) and then classically computing the parity of the measurement outcomes. The same is true of the logical $X$ measurement. 

\begin{figure}
\raisebox{7mm}{(a)}\hskip -2mm
\begin{quantikz}[font=\small, row sep=0.1cm]
\gate{\ket0_{\!{\rm L}}\!\textrm{-FT-Prep}}
\end{quantikz}
\ $=$
\begin{quantikz}[font=\small, row sep=0.1cm]
\gate{\ket0_{\!{\rm L}}\!\textrm{-Prep}}\slice{}&\ctrl{1}\slice{}&\qw\slice{}&\qw\\
\gate{\ket0_{\!{\rm L}}\!\textrm{-Prep}}\slice{}&\targ{}\slice{}&\meter{$Z_L$}&\cw\rstick{pass}
\end{quantikz}

\vspace{0.2cm}

\raisebox{10mm}{(b)} \hskip -2mm
\begin{quantikz}[font=\small, row sep=0.1cm]
\gate{\ket0_{\!{\rm L}}\!\textrm{-Prep}}
\end{quantikz}
\ $=$
\begin{quantikz}[row sep=0.2cm, column sep=0.1cm]
\lstick{\ket+}\slice{}&\ctrl{2}&\qw\slice{}&\qw&\ctrl{4}&\qw&\qw\slice{}&\ctrl{6}&\qw&\qw\slice{}&\qw&\qw\\
\lstick{\ket+}&\qw&\ctrl{5}&\ctrl{4}&\qw&\qw&\qw&\qw&\ctrl{1}&\qw&\qw&\qw\\
\lstick{\ket0}&\targ{}&\qw&\qw&\qw&\gate{I}&\qw&\qw&\targ{}&\qw&\qw&\qw\\
\lstick{\ket+}&\ctrl{2}&\qw&\qw&\qw&\ctrl{3}&\qw&\qw&\ctrl{1}&\qw&\qw&\qw\\
&&&\lstick{\colorbox{white}{\ket0}}\qw&\targ{}&\qw&\qw&\qw&\targ{}&\qw&\qw&\qw\\
\lstick{\ket0}&\targ{}&\qw&\targ{}&\qw&\qw&\qw&\qw&\gate{I}&\qw&\qw&\qw\\
\lstick{\ket0}&\qw&\targ{}&\qw&\qw&\targ{}&\qw&\targ{}&\qw&\qw&\qw&\qw
\end{quantikz} \hskip 10mm \ 

\caption{\label{fig:7qb-FT_prep} Fault-tolerant preparation of the logical (level-1) $\ket0$ state 
\cite{Aliferis2005Apr}; the logical $\ket+$ state is prepared using the same circuits, except for an $X\leftrightarrow Z$ transformation (see main text). (a) A verification step, involving two (unverified) copies of $\ket0_{\!{\rm L}}$ (indicated in the figure as $\ket0_{\!{\rm L}}$-Prep), is carried out to assure fault tolerance. A logical $\ket0$ (indicated as $\ket0_{\!{\rm L}}$-FT-Prep) is successfully prepared if the measurement result gives a ``pass"; if a ``fail" result is obtained, the entire preparation procedure is repeated. (b) The circuit to prepare $\ket0_{\!{\rm L}}$-Prep, with $\ket0$ and $\ket+$ as physical states.}
\end{figure}
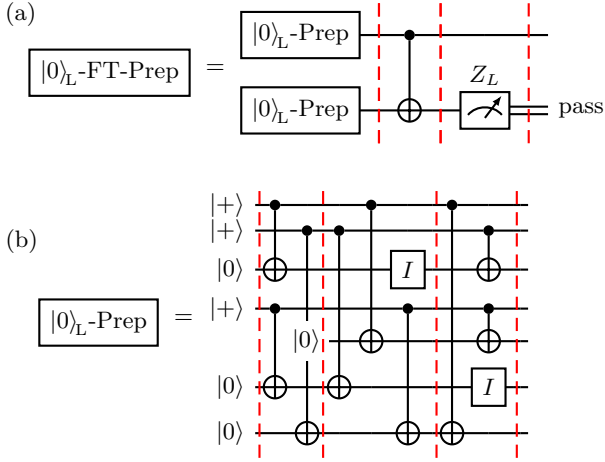

The fault-tolerant procedure for preparing the logical $\ket 0$ state is given in Fig.~\ref{fig:7qb-FT_prep}. The circuit that verifies this state (necessary for a fault-tolerant procedure) is shown in Fig.~\ref{fig:7qb-FT_prep}(a); it takes as input two copies of the unverified state $\ket{0}_{\!{\rm L}}\!\textrm{-Prep}$ prepared by the circuit in Fig.~\ref{fig:7qb-FT_prep}(b). One can fault-tolerantly prepare a logical $\ket+$ with the same procedure, except that we need an $X\leftrightarrow Z$ transformation; this simply swaps the roles of the target and control for every $\cnot$ gate, every $Z$ measurement becomes an $X$ measurement, and 
the preparation of each physical $\ket{+}$ becomes a preparation of a physical $\ket{0}$ (and vice-versa) \cite{Aliferis2005Apr}. Each attempt at preparing the logical $\ket0$ or $\ket+$ state thus takes exactly the same number of qubits, gates, measurements, and timesteps. 

\begin{figure}
\begin{quantikz}[font=\small, row sep=0.1cm]
&\gate{\textrm{EC gadget}}&\qw
\end{quantikz}
$=$
\begin{quantikz}[font=\small, row sep=0.1cm]
\qw\slice{}&\targ{}\slice{}&\ctrl{2}&\qw\slice{}&\qw\slice{}&\qw\\
\gate{\ket0_{\!{\rm L}}\!\textrm{-FT-Prep}}&\ctrl{-1}&\qw&\meter{$Z_L$}&\cw\\
&\lstick{\framebox{$\ket+_{\!{\rm L}}\!\textrm{-Prep}$}}&\targ{}&\qw&\meter{$X_L$}&\cw
\end{quantikz}
\caption{\label{fig:Steane_EC_gadget} The Steane error-correction (EC) gadget for the 7-qubit code. Two logical ancillary states are fault-tolerantly prepared in $\ket0_L$ and $\ket+_L$ (see Fig.~\ref{fig:7qb-FT_prep} for the preparation procedure). They interact with the data block (top line) via logical $\cnot$s (which are transversal) and are measured in the $Z_L$ and $X_L$ bases (again, transversally), respectively. The measurement outcomes are the syndrome results used for error correction.}
\end{figure}
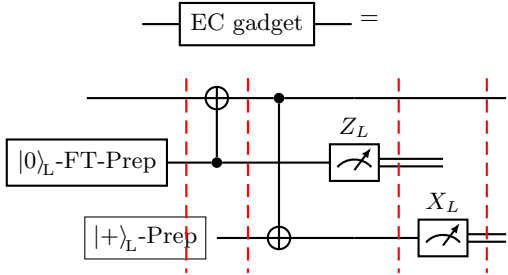

We also need the circuit design for the fault-tolerant implementation of the error-correction gadget. Here, we consider two different circuit designs: the Steane EC gadget in Fig.~\ref{fig:Steane_EC_gadget} from Ref.~\cite{Aliferis2005Apr}, and the flag-qubit EC circuit in Fig.~\ref{fig:EC_gadget_flag} \cite{reichardt_fault-tolerant_2021}. The Steane EC gadget requires the syndrome states $\ket0_L$ and $\ket+_L$, used to detect $X$ and $Z$ errors, respectively, in the data-carrying block. The $\cnot$ gates followed by the $X$ and $Z$ measurements together measure the stabilizer generators of the 7-qubit code, which allow for diagnosis of the errors in the data. As mentioned earlier, the recovery operation is done only virtually. Comparing Fig.~\ref{fig:Steane_EC_gadget} and the earlier general Fig.~\ref{fig:EC_gadget}, we see that $I_S=2$, with $S_1=X_\mathrm{L}$, $S_2=Z_\mathrm{L}$, $d_1=d_2=1$, and $d_R=0$.

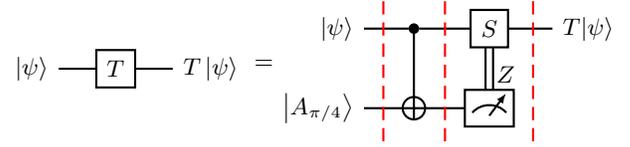
\begin{figure}
\begin{quantikz}
\lstick{\ket\psi}\qw&\gate{T}&\qw\rstick{$T\ket\psi$}
\end{quantikz}
~$=$\hspace*{-0.1cm}
\begin{quantikz}[row sep=0.25cm]
\lstick{\ket\psi}\slice{}&\ctrl{1}\slice{} &\gate{S}\slice{}&\qw\rstick{$T$\ket\psi} \\
\lstick{\ket{A_{\pi/4}}} &\targ{}&\meter{\quad~$Z$}\vcw{-1}&
\end{quantikz}
\caption{\label{fig:7qb-Tgate}Implementation of a $T$-gate on the single-qubit input $\ket\psi$ via the use of the magic state $\ket{A_{\pi/4}}$.}
\end{figure}

\begin{figure*}
\centering   
\includegraphics[width=0.95\textwidth]{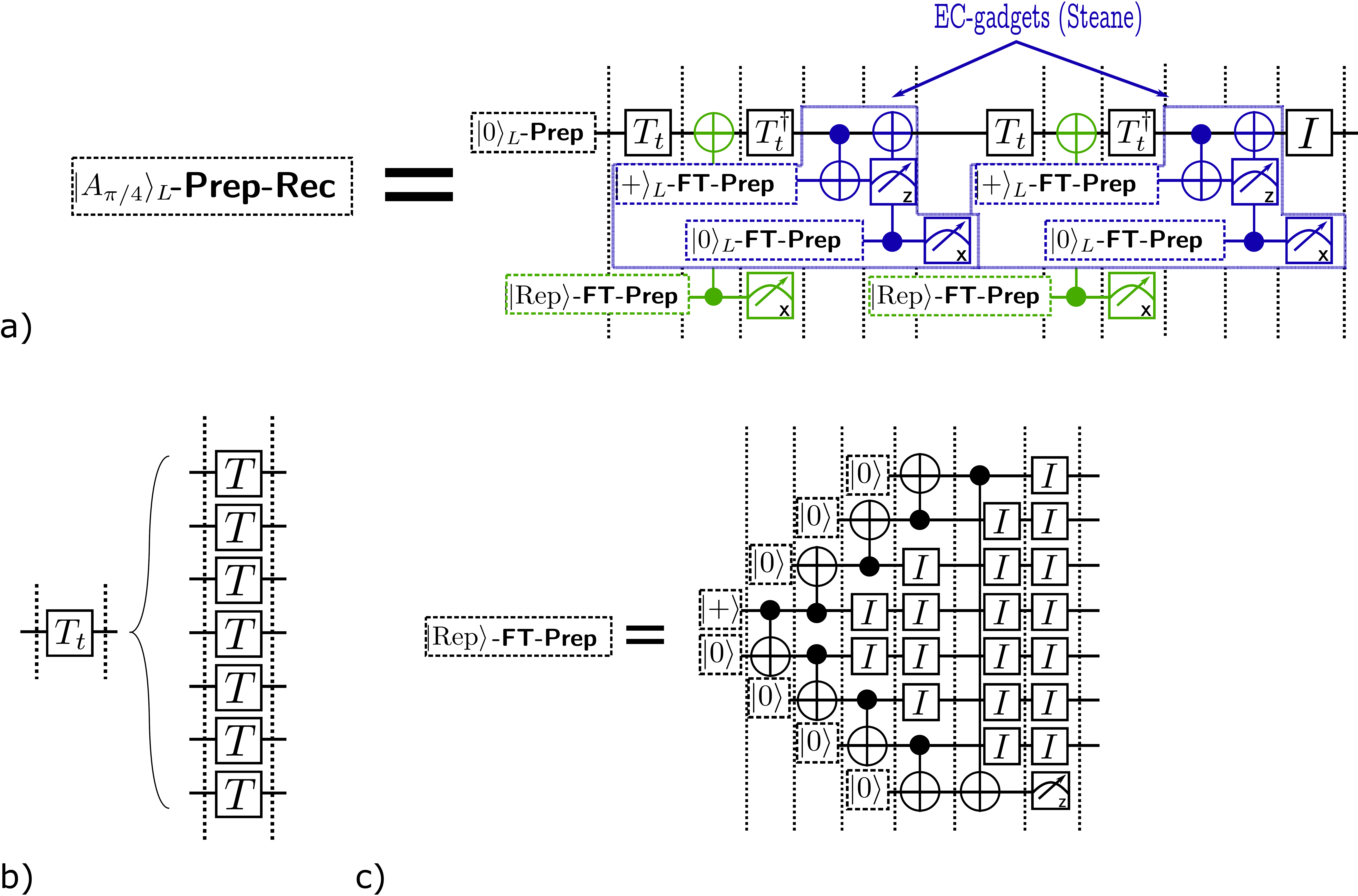}
\caption{Magic-state preparation circuits for the 7-qubit code with Steane EC gadgets. (a) The 1-prep-Rec of the state $\ket{A_{\pi/4}}$ from Ref.~\cite{Aliferis2005Apr} (indicated as $\ket{A_{\pi/4}}_L$-Prep-Rec). Each horizontal line represents a logical qubit encoded in the 7-qubit code, with the exception of the two green lines connected to the $\ket{\text{Rep}}\mathrm{-FT-Prep}$ boxes: Those represent logical qubits in the state $\ket{\text{Rep}}\equiv(\ket{0}^{\otimes 7}+\ket{1}^{\otimes 7})/\sqrt{2}$, the logical-$+$ state of the $7$-qubit repetition code (sometimes called the cat state \cite{Aliferis2005Apr}). The gadgets drawn in green measure the logical observable $TXT^{\dagger}$ on the top qubit (which eventually carries the  magic state) through the use of the transversally defined $T_t$ and $T_t^\dagger$ gates [see (b)]. The first measurement projects the top qubit into the $TXT\dagger$ eigenstates $\ket{A_{\pi/4}}$ or $Z \ket{A_{\pi/4}}$; the latter can also be used for the $T$-gate \cite{3_footnote-ZApis4}). The two Steane EC gadgets (in blue) and a second green gadget provide verification for a fault-tolerant preparation: The magic state passes the verification only if trivial syndromes are detected by the two EC gadgets, and if both measurements of $TXT^{\dagger}$ give the same outcome (see Ref.~\cite{Aliferis2005Apr}). (b) The definition of the gate $T_t$ used in circuit in (a). (c) Fault-tolerant preparation of the state $\ket{\text{Rep}}$. In addition to the $7$ qubits carrying $\ket{\text{Rep}}$, an extra qubit is used for verification: The state passes verification only if the $Z$ measurement on that eight qubit gives $+1$. 
\label{fig:A_pis4_prep}
}
\end{figure*}

The logical universal gate-set for the 7-qubit code is completed by a logical $T$-gate. We will find it useful to also be able to implement the logical $T^\dagger$-gate. Both can be done via the magic-state approach, in fact, with the same magic state $\ket{A_\opmagic}$. The $T$ and $T^\dagger$ gates are hence the magic gates in this scheme. We first recall the magic-state implementation of a \emph{physical} $T$ gate \cite{Aliferis2005Apr}
employs only Clifford operations ($S$, $\cnot$, and a Pauli-$Z$ measurement) but consumes the magic state $\ket{A_{\pi/4}}\equiv T\ket+$, where $\ket+\equiv\frac{1}{\sqrt 2}(\ket0+\ket1)$; see Fig.~\ref{fig:7qb-Tgate}. To promote this to an implementation of a logical $T$-gate, we replace every physical qubit in Fig.~\ref{fig:7qb-Tgate} by a logical qubit, every physical operation is replaced by its 1-Rec version, and the physical magic state is replaced by a fault-tolerantly prepared logical state, $\ket{A_{\pi/4}}_{\!{\rm L}}\!\textrm{-FT-Prep}$. The procedure for preparing  $\ket{A_{\pi/4}}_{\!{\rm L}}\!\textrm{-FT-Prep}$ is given in Fig.~\ref{fig:A_pis4_prep}, which involves verification steps that employ verified ``cat" states. Using the fact $T^\dagger=S^{\dagger} T$, the $T^{\dagger}$ gate can be implemented with the same circuit as Fig.~\ref{fig:7qb-Tgate} with an extra $S^{\dagger}$ applied on the top qubit at the end of the circuit. This $S^{\dagger}$ can in practice be incorporated in the classically controlled operation so that if the magic state collapses to $\ket{0}$, $S^{\dagger}$ is implemented and if it collapses to $\ket{1}$, no gate is applied. Importantly, $S^{\dagger}$ can then be implemented transversally in the 7-qubit code and is hence a normal gate.

\subsection{Qubit cost with the Steane EC gadget}
\label{sec:example-Steane_EC}

\begin{table}[t]
\begin{tabular}{|l||c|c|}
\hline
\parbox[t]{2.8cm}{Fault-tolerant scheme} &\parbox[t]{2cm}{State being prepared}& \parbox[t]{2.2cm}{Time multiplicity} \\
\hline \hline
\multirow{3}{*}{\parbox[t]{2.8cm}{Steane approach}} & \ syndrome \ & $r_{\rm s} = 3$\\
& $|A_{\pi/4}\rangle$ & $r_A=3$ \\
& $|\text{Rep}\rangle$ & $r_{\rm Rep}=2$ \\
\hline
\multirow{2}{*}{\parbox[t]{2.8cm}{Flag-qubit appoarch}} 
& syndrome & $r^{\rm flag}_s = 2$\\
& $|H\rangle$ & $r^{\rm flag}_H= 2$\\
\hline
\end{tabular}
\caption{The time multiplicity for the various states discussed in the examples in this work. In Sec.~\ref{sec:example-Steane_EC} [around Eq.~\eqref{eq:tauA_anc}], the details on how $r_A$ can be determined is provided as an example.
}
\label{table:time-multiplicity}
\end{table}

\begin{figure*}
\centering   
\includegraphics[width=0.95\textwidth]{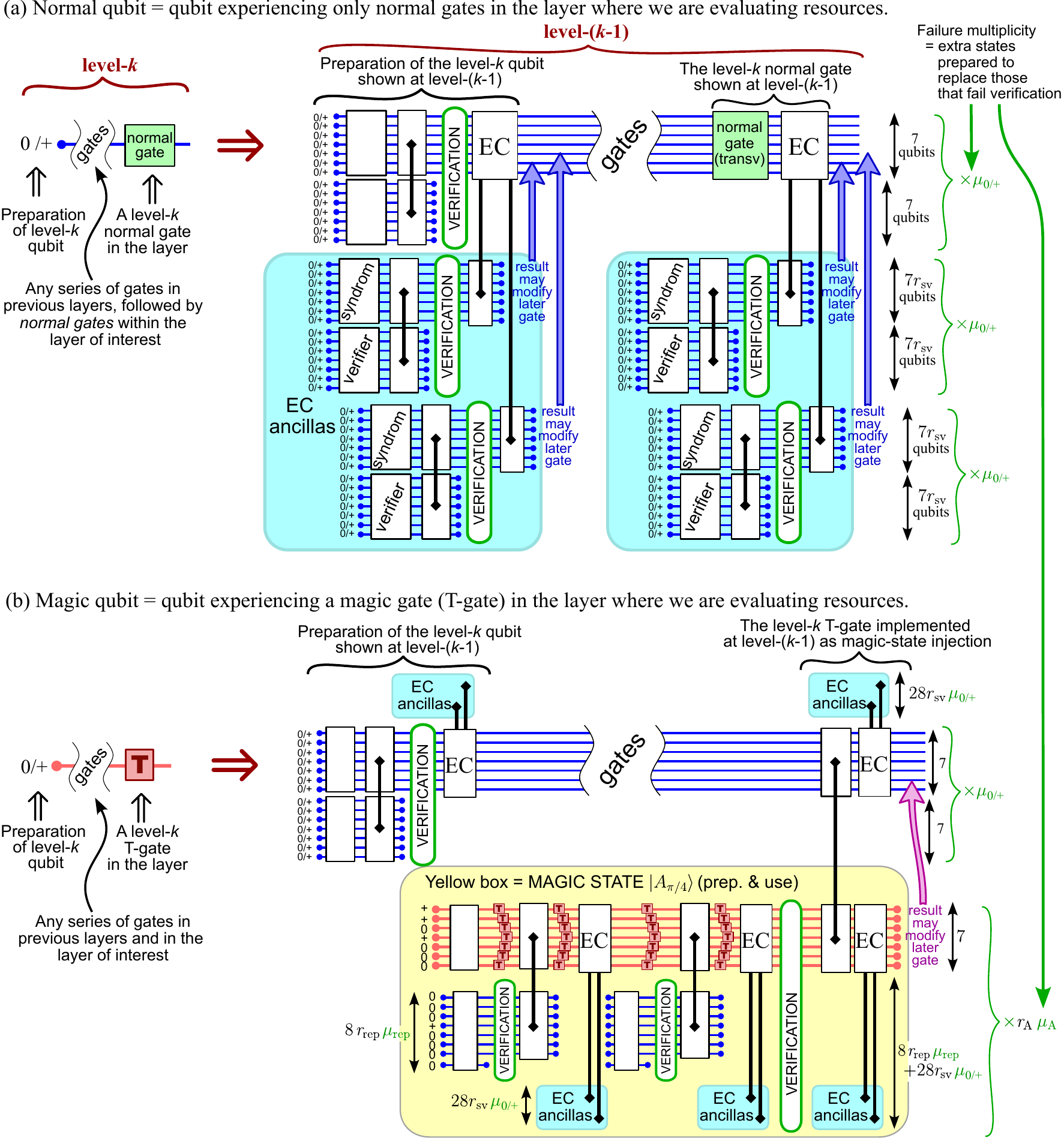}
\caption{
Simplified sketch of the error-correction circuitry for 7-qubit code with the Steane approach, showing the aspects necessary to evaluate the qubit cost. The states at level $k$ (to the left of the red arrows) are shown in terms of states at level $k-1$ (to the right of the red arrows). Rectangles are blocks of gate-operations, whose detailed circuits are in other figures in this work. 
The horizontal red lines are qubits that start in $|0\rangle$ or $|+\rangle$ and experience a magic gate ($T$-gate). The blue horizontal lines are qubits that start in $|0\rangle$ or $|+\rangle$ which only experience normal gates.  Dots at the right-hand of horizontal lines indicate measurements, while dots at the left-hand ends of horizontal lines indicate $|0\rangle$ or 
$|+\rangle$
For simplicity, we indicate $|0\rangle$ and $|+\rangle$ with ``0/+'', and we indicate $T$ and $T^\dagger$ gates with ``T'' (the detailed circuits show which are which); this simplification is fine when evaluating resources, because 0 and + require the same resources, as do $T$ and $T^\dagger$. 
The blue arrows indicate that any errors detected by the measurement of the EC gadget are corrected by modifying the gate-operations at a following timestep.
For compactness in (b), we do not show internal details of the ``EC ancilla'' blocks (nor the arrows indicating modification of later gates); these details are the same as in the light-blue blocks in (a).
}
\label{fig:sketch-Steane}
\end{figure*}

With the circuit designs in place, we can now apply our earlier formulas for qubit cost to this 7-qubit example. We begin with the case where the EC gadget is done using the Steane EC circuit.
To proceed, we take the circuits in Figs.~\ref{fig:7qb-FT_prep}-\ref{fig:A_pis4_prep} for the 7-qubit code with Steane EC gadgets, and combine them all into the simplified
sketches in Fig.~\ref{fig:sketch-Steane}, where all circuit details irrelevant to the evaluation of qubit cost have been removed. Next to each part of the circuit, we indicate the number of qubits needed, allowing us to simply read off the qubit costs for the normal and magic qubits.

We now need Eqs.~\eqref{eq:StateCost}--\eqref{eq:lambda-easy-formula}, \eqref{eq:lambda-hard-formula}, and \eqref{eq:b-formula}, which we reproduce here as
\begin{eqnarray}
\lambda_\mathrm{EC}&=&\sum_{i=1}^{I_S} r_i \mu_i(n_i+v_i)\nonumber\\
\lambda_\psi&=&\mu_\psi(n+v_\psi)\nonumber\\
\lambda_\opnormal &=& \lambda_\psi +\lambda_\mathrm{EC},\nonumber\\
\lambda_\opmagic&=&r_A\mu_A n,\nonumber\\
b&=&\lambda_\opnormal+r_A\mu_Av_A
\end{eqnarray}
For the 7-qubit code, we have $n=7$, and $v_\psi=7$. 
$I_S=2$ in $\lambda_{\rm EC}$ and the terms in its sum have $n_1=n_2=v_1=v_2=7$.
The four states we need to prepare are listed in Table \ref{table:failure-multiplicity}; our $\psi$ labels in the formulas above thus take the values 0, 1, Rep, or $A$ (short for $A_{\pi/4}$).
Each one has a corresponding $\mu$ in Table \ref{table:failure-multiplicity}, with $\muCl$ being the appropriate one for states $|0\rangle$ and $|+\rangle$, needed in $\lambda_\opnormal$.

Next, we need the time multiplicity factors for the preparation of the different states. Those for the ancillary states needed in the EC gadget are calculated using Eq.~(\ref{eq:r_i-definition}); we find $r\equiv r_{\rm s}=3$ as listed in Table~\ref{table:time-multiplicity}. The calculation of $r_A$ follows the same logic, but is more involved, so we explain it here. Let us initially assume a single concatenation level ($K=1$). We apply the same logic already used for other ancillary states to derive Eq.~(\ref{eq:r_i-definition}). This gives
$r_A \equiv \lceil \tau_{A,\text{anc}}/\tau_{\scriptscriptstyle T\to T} \rceil$, where $\tau_{\scriptscriptstyle T\to T}$ is the minimum time separating two consecutive $T$-gates, and $\tau_{A,\text{anc}}$ is the time to prepare, use, and measure the qubits carrying the magic state (or those qubits used to verify it). 
To evaluate $\tau_{T\to T}$, we note that $T$-gates are always separated by a Clifford gate (since $T^2=S$ is a Clifford gate), and each $T$-gate lasts for $2 \tau_L$ because of the state injection procedure (see fig.~\ref{fig:7qb-Tgate}). Hence, $\tau_T=3 \tau_L=3 D \tau_P$ where $D=3$ is the number of timesteps to implement a Clifford gate. Here, $\tau_L$ is the duration of a level-1 logical gate, formally the duration of a 1-Rec including the syndrome extraction steps, and $\tau_P$ is the time for a single timestep for physical (level-0) qubits.

Turning to $\tau_{A,\text{anc}}$, Figs.~\ref{fig:7qb-Tgate} and \ref{fig:A_pis4_prep} tell us that
\begin{align}
    \tau_{A,\text{anc}}=\tau_{A,\text{prep}}+\tau_L+\tau_P,
    \label{eq:tauA_anc}
\end{align}
where $\tau_{A,\text{prep}}$ is the duration to prepare the magic state, and $\tau_L+\tau_P$ is the duration of the state-injection procedure ($\cnot$ lasting for $\tau_L$, and the magic-state measurement lasting for $\tau_P$ in Fig.~\ref{fig:7qb-Tgate}). We get $\tau_{A,\text{prep}}$ from Fig.~\ref{fig:A_pis4_prep}; the magic-state preparation starts with the preparation of the first repetition-code state, and ends when the last syndrome measurement is performed. Hence, we have: $\tau_{A,\text{prep}}=\tau_{\text{Rep},\text{prep}}+10 \tau_P$. The duration to prepare the repetition-code state is $\tau_{\text{Rep},\text{prep}}=7 \tau_P$; see Fig.~\ref{fig:A_pis4_prep}c (one step to initialize the fourth qubit from the top in $\ket{+}$, 5 steps of $\cnot$s, and one measurement, with each of these steps lasting for $\tau_P$). We thus have $\tau_{A,\text{anc}}=21 \tau_P$, and hence, $r_A=\lceil 21 \tau_P/(9 \tau_P) \rceil=3$.

This reasoning for $r_A$ generalizes to multiple concatenation levels, if one is careful with the following two facts. Firstly, while the preparation of a physical state takes $\tau_L(k=0)=\tau_P$ (and a measurement lasts for $\tau_P$ for all $k$), the fault-tolerant preparation of a level-$k$ state takes longer than $\tau_L(k)$. 
Secondly, while the preparation of the $k=1$ magic state started with the preparation of the repetition-code state, it is not so clear for $k>1$. 

There, one must check which of the following tasks takes the longest time: (i) the preparation of the repetition-code state; (ii) the initialization in $\ket{0}$ of the qubits that carry the magic state; or (iii) the state injection necessary to implement $T$-gates at level-($k-1)$ for $(k-1)>0$. 
Despite these facts, one can check numerically that $r_A=3$ is sufficient for general $k$.

In the calculation of the failure multiplicities we need to know the number of verifications per exRec, $M$. For Steane EC gadgets, we take the same $M$ for all states, with that being the worst-case $M=8$.  This worst-case $M$ is the number of verified syndrome states required in a $\cnot$ exRec; each EC gadget requires two syndrome states, and the $\cnot$ exRec contains four EC gadgets (each qubit in the $\cnot$ has one before and one after the $\cnot$). This is clearly an overestimate of $M$ for most states, so in fact this shows that $M$ is in the interval $1<M\leq 8$. However, the failure multiplicity only depends logarithmically on $M$, so we take $M=8$ for simplicity, knowing that the failure multiplicity would barely change if we had taken a different value of $M$ in this interval.

Now we have all necessary parameters, we find that for normal gates,
\begin{eqnarray}
\lambda_{0/\!+}&=&14\muCl \, , \nonumber\\
\lambda_\mathrm{EC}&=&28 r_{\rm sv} \muCl = 84 \muCl\, .
\end{eqnarray}
From this we get
\begin{eqnarray}
\lambda_\opnormal 
&=&98\muCl\nonumber\\
&=&{\left\{\begin{array}{ll}
294&\text{without shared prep.,} \\
114&\text{with shared prep.} 
\end{array}\right.} 
\label{eq:lambdaEasy_Steane}
\end{eqnarray}
where we use the values of $\muCl$ in Table~\ref{table:failure-multiplicity}, and round to the nearest integer. Although, finding the value of $\muCl$ for shared state preparation (shared prep.) is postponed to Sec.~\ref{sec:steane-shared}.

Next, we turn to the parameters for magic gates, $\lambda_\opmagic$ and $b$, which involve the parameters associated with the magic-state preparation. We begin with $v_A$, the number of ancillary qubits needed to prepare the magic state $\ket{A_{\pi/4}}$. From Fig.~\ref{fig:A_pis4_prep}, we see that the preparation of $\ket{A_{\pi/4}}$ requires an ancillary state $\ket{\textrm{Rep}}$ that needs to be fault-tolerantly prepared, employing 8 physical qubits in all, with one qubit used as a verification qubit. This $\ket{\textrm{Rep}}$ state preparation can fail, meaning that we require a failure multiplicity $\mu_\mathrm{Rep}$ for it, and its multiple-timestep preparation circuit needs a time multiplicity $r_\mathrm{Rep}$. Thus, the number of ancillary qubits needed for the $\ket{\textrm{Rep}}$ alone is $8r_\mathrm{Rep}\mu_\mathrm{Rep}$. In addition, the $\ket{A_{\pi/4}}$ state preparation circuit contains the Steane EC gadget. Altogether then, we have 
\begin{eqnarray}
v_A=\lambda_\mathrm{EC}+8r_\mathrm{Rep}\mu_\mathrm{Rep} \ = \ 84\muCl+16\mu_\mathrm{Rep}\, .
\end{eqnarray} 
Putting these into $\lambda_\opmagic$ and $b$, using the multiplicity values from Tables \ref{table:failure-multiplicity} and \ref{table:time-multiplicity}, and rounding to the nearest integer, we get 
\begin{eqnarray}
\lambda_\opmagic&=&7r_A\mu_A \ = \ 21\mu_A\nonumber\\
&=&{\left\{\begin{array}{ll}
84&\textrm{without shared prep.,}\\
27&\textrm{with shared prep.} 
\end{array}\right.}
\label{eq:lambdaHard_Steane}
\end{eqnarray}
and finally, for $b$, we have
\begin{eqnarray}
b
&=&98\muCl+ 3\mu_A(84 \muCl+16\mu_\mathrm{Rep})
\nonumber\\
&=&{\left\{\begin{array}{ll}
3702&\textrm{without shared prep.,}\\
564 &\textrm{with shared prep.} 
\end{array}\right.}
\label{eq:b_Steane}
\end{eqnarray}

As we take more and more levels of concatenation, the proportion of resources necessary for a given circuit compared to an equivalent one with only normal gates saturates at a finite value given by Eq.~(\ref{Eq:asymptotics_of_R}), which becomes\begin{eqnarray}
R(K) &\leq& C =  
\big(\Qalgo_\opnormal + A_0\,\Qalgo_\opmagic\big)\big/\Qalgo_{\rm total}
\nonumber \\
\hbox{ with } A_0&\equiv&\frac{b}{\lambda_\opnormal-\lambda_\opmagic}\nonumber\\
&=& \left\{\begin{array}{lc}
17.6 &
{\hbox{without shared prep.,}}
\\
6.48 &
{\hbox{with shared prep.}}
\end{array}\right. \qquad
\label{eq:Steane-C}
\end{eqnarray}
Thus, we find that even in the worst-case scenario where the algorithm has a timestep where all gates are magic gates \big(hence $\Qalgo_\opmagic=\Qalgo_{\rm total}$\big), that algorithm  will never require more than 17.6 times the physical qubit cost of an equivalent-sized algorithm made entirely of normal gates. The factor of 17.6 is reduced to 6.48 if state preparation is shared. 

However, as we discuss in Appendix~\ref{sec:T-gate-proportion}, magic gates are fairly rare in typical algorithms, and it is reasonable to assume only about 5\% of gates are magic gates. To account for this we now make a simplifying assumption to avoid treating the time multiplicity at the logical level (level-$K$) differently from other levels in the concatenation.
We recall that the time multiplicity for magic gates was calculated knowing that a qubit only experiences (at most) a magic operation ($T$-gate) every two timesteps. This slightly complicates the estimation of the percentage of magic qubits at the logical level (level-$K$) in $\Lmax$. In short, when we take 5\% of magic-operations in each level-$K$ layer, we actually assume 10\% of magic-operations occur every two level-$K$ layers (and no magic-operations occur in between these two level-$K$ layers). It guarantees that the computer has enough resources to implement 5\% of magic-operations at each level-$K$ layer. We do this because it allows us to use the same time multiplicity at the logical level (level-$K$) as at other levels. This is what is done to get the 5\% values in Figs.~\ref{fig:total-resources}, \ref{Fig:phenomenological} and \ref{Fig:phase-transition}, and 
corresponds to taking $\Qalgo_\opmagic\big/\Qalgo_{\rm total}\simeq 0.1$. Then the physical qubit cost for such a circuit is only $2.66$ times that for an equivalent-sized logical circuit made entirely of normal gates.  This factor of $2.66$ is reduced to $1.55$, if state preparation is shared.

In other words, in typical cases, the presence of gates that require magic states (the $T$-gates in this case), does not even triple the total physical qubit cost.  Furthermore, it only adds about 50\% to the total physical qubit cost when state preparation is shared. While this is clearly not a negligible contribution to the total resources, it is much less than implied in the literature, which insists on $T$-gates being extremely expensive in terms of resources \cite{krishna_towards_2019,nikahd_low-overhead_2017,bravyi_magic_2012,webster_reducing_2015,haah_codes_2018,hastings_distillation_2018,goto_step-by-step_2014}.
This tells us that while optimizing magic-states (or optimizing circuits to minimize the number of magic-states) will reduce the  qubits resources  required by the quantum computer, any proposed (or hypothetical) optimization is unlikely to reduce the qubits resources by very much (probably much less than a factor of three). 
Of course, any such optimization is worthwhile, but it is extremely modest compared to other specific optimizations like sharing state preparation (which can reduce qubit costs by a couple of orders of magnitude),
or replacing Steane EC gadgets by more resource efficient flag-qubit EC gadgets (which can reduce resources by four orders of magnitude). All this is shown in Fig.~\ref{fig:total-resources}. 

\subsection{Shared state preparation for the Steane EC gadget}
\label{sec:steane-shared}

Sec.~\ref{sec:shared-failure-multiplicity} explained the basic logic of evaluating the failure multiplicity factors for shared state preparation. Here, we explain how this logic applies for our current example of the 7-qubit code scheme with the Steane EC gadget, and show how we obtained the failure multiplicities given in the last column of Table~\ref{table:failure-multiplicity}, which were used above in Eqs.~(\ref{eq:lambdaEasy_Steane}-\ref{eq:Steane-C}).
Hence, our goal is to get rough estimates of $V$, $M$, $K$ and $N_{\rm gates}^{\text{level-1}}$ for Steane EC gadgets and magic-states. As explained in Sec.~\ref{sec:shared-failure-multiplicity}, we are not looking for a rigorous estimate here; we simply want to show with a simple argument the clear advantage in sharing state preparation and how it can significantly reduce the value of $\mu$ to something fairly close to $1$.
Here $K=5$ is likely large enough for conceivable middle-term computers \cite{2_footnote-K=5}, and $M=8$.
This leaves us with estimating $N_{\rm gates}^{\text{level-1}}$.
For this, we start by counting the number of qubits at level 1 in a code with $K$ levels of concatenation. We see it is equal to the total number of physical qubits in a code with $(K-1)$ levels of concatenation, $Q^{\text{phys}}_{\text{total}}(K-1)$. We multiply this number by the number of level-1 logical timesteps in the algorithm. For $K$ total levels of concatenation, we would have about $D^{K-1} \times 2 D_L$ level-1 timesteps (see Sec.~\ref{sec:shared-failure-multiplicity} and footnote \cite{6_footnote-DepthSharing} for more details), where $D$ is the number of timesteps required to implement a normal gate, and $D_L$ is the depth of the algorithm implemented on the universal gateset (in this case Clifford gates plus $T$-gates);
the factor of 2 is because a logical $T$-gate takes two logical steps to be implemented. 

Hence, we have:
\begin{align}
    N_{\rm gates}^{\text{level-1}} \ \approx \ Q^{\text{phys}}_{\text{total}}(K-1) \times D^{K-1} \times 2 D_L\, .
    \label{eq:Ngate-level-1}
\end{align}
To show that sharing state preparation can significantly reduce the overheads, we will reason with a pessimistic scenario: for type $\nu$ states, we will find a value of $S_\nu$ satisfying Eq.~(\ref{eq:p_Lack_mut-2}) with Eqs.~(\ref{eq:def_pLack-shared},\ref{eq:Ngate-level-1}), and thus extract a value for its failure multiplicity $\mu_\nu$.

As an example, we assume the quantum computer is large enough to perform Shor's algorithm to crack RSA-2048 \cite{Gidney2021Apr}, which requires $Q^{\text{algo}}_{\text{total}}\sim10^4$ qubits and a depth
$D^{\rm algo}\sim10^{10}$ at the level of the abstract circuit (level-$K$ circuit) implementing Shor's algorithm.
We take $\kappa=0.01$ and $K=5$ \cite{2_footnote-K=5}, while taking the worst case scenario in which there is a timestep in the level-$K$ circuit where all level-$k$ qubits  experience $T$-gates: $Q^{\text{algo}}_\opmagic=Q^{\text{algo}}_{\text{total}}$.
To get a lower bound on the right-hand side of Eq.~(\ref{eq:p_Lack_mut-2}), we take an overestimate of $Q^{\text{phys}}_{\text{total}}(K-1)$ for $K=5$ by taking its value when state preparation is {\it not} shared from Fig.~\ref{fig:total-resources} (this is a clear overestimate, because sharing the state preparation can only reduce this number); this gives $Q^{\text{phys}}_{\text{total}}(K-1) = 10^{15}$ \footnote{For this, we start with the dashed curve for Steane code without shared state preparation in Fig.~\ref{fig:total-resources}, which corresponds to a circuit without magic gates. We then multiply it by 17.6 (taken from Eq.~\ref{eq:Steane-C}) to arrive at the case where all gates are magic gates, and round to the nearest order of magnitude.}. In this case, the ratio of physical depth to logical depth, $D=3$. Using the fact $M=8$, we know that Eq.~(\ref{eq:p_Lack_mut-2}) will be satisfied if we find the smallest $S$ that ensures we take 
\begin{eqnarray}
p^{\mathrm{Share}}_{\mathrm{Lack};\nu} (S_\nu,p_{\rm thres}) \ \leq  \ 10^{-31},
\label{eq:p_Lack_mut-3}
\end{eqnarray}
If we take $V=100$, then the smallest $S$ that satisfies the inequality 
Eq.~\eqref{eq:p_Lack_mut-3} with Eqs.~(\ref{eq:def_pLack-shared},\ref{eq:Ngate-level-1}) depends on the number of fault-locations, $N$, in the type of  state being considered since $N$ appears in the expression of $p^{\mathrm{Share}}_{\mathrm{Lack};\nu} (S_\nu,p_{\rm thres})$).
The failure multiplicity for that type of  state is then given by $(V+S)/V$
with the $S$ for that type of state.
This gives the values of $\mu$s in the last column of Table~\ref{table:failure-multiplicity}. 

This raises the question of why we take $V=100$ in our calculation of failure multiplicities, $\mu_\nu$ where $\nu$  is ``$0/+$'', ``A'' or ``Rep''. 
Firstly, we note that $\mu_\nu$ only depends weakly on $V$ for the large $V$ that is of interest to us, so any plausible choice of large $V$ will give a similar $\mu_\nu$. Secondly, we note that $\mu_\nu$
decays (slowly) with increasing $V$. Thus, to get a worst case estimate of $\mu_\nu$ with shared state preparation, we should take the smallest $V$ likely to occur in the large-scale algorithm of interest to us.
We now argue that $V=100$ is a plausible such lower bound on $V$. To do this we assume the total number of qubits at the algorithmic level, $Q_L$, is at least $2000$ (for comparison, $Q_L\sim10^4$ for cracking  RSA-2048 encryption \cite{Gidney2021Apr}), 
and assume about 5\% of the gates are magic gates (as explained below, our conclusion also holds for less than 5\% of magic gates). 
Then we can expect at least 100 magic gates in parallel at the highest concatenation level (with more at lower concatenation levels), each of which will require a verified magic state. Hence, $V=100$ is the smallest number of verified magic states an ancilla factory will have to produce in parallel, giving an upper bound on $\mu_\nu$. We thus use $V=100$ for all the failure multiplicity in the last column of table~\ref{table:failure-multiplicity}. 
In the case of ancillary states for EC gadgets, the smallest $V$ is likely to be much larger
(of order $Q_L$), which would reduce failure multiplicities below those in table~\ref{table:failure-multiplicity}. 
However, those failure multiplicity is already fairly close to 1 with $V=100$ 
(and it will not go below 1 for $V\to \infty$), so there is  little practical gain in taking larger $V$.

Now turning to similar-sized circuits with less than 5\% of magic gates, we note that the above value of $V$ would be too large, so the value of $\mu_\nu$ would be a bit too small. 
Despite this, we know that the cost of any circuit with less than 5\% of magic gates is less than the cost of a circuit with 5\% of magic gates, and more than the cost of a circuit with no magic gates. As these two costs are found to be close, this gives a good estimate of 
the cost of any circuit with less than 5\% of magic gates.

As the arguments we made to get the numerical value on the right-hand side of Eq.~(\ref{eq:p_Lack_mut-3}) are rather simplistic, it is worth considering what happens if one takes a different numerical value on the right-hand side of Eq.~(\ref{eq:p_Lack_mut-3}).
A simple numerical analysis indicates that a change in the choice of this value only causes a logarithmic change in the $\mu$s.
For example, we replaced $10^{-31}$ with $10^{-39}$ in Eq.~(\ref{eq:p_Lack_mut-3}) (e.g., consider a logical circuit with $10^8$ times more gate-operations), and found that it only changes the value of $\mu_{\rm A}$ from 1.30 to 1.36. Thus,  we expect that the failure multiplicities in the last column of Table~\ref{table:failure-multiplicity} are representative of a broad range of realistic situations.

The above analysis allows us to plot the blue curves in Fig.~\ref{fig:total-resources}. They show that sharing state preparation can reduce the total physical qubit cost required for the computation by a couple of orders of magnitude (for $K=5$).  It also makes the relative cost of magic states smaller; the solid and dashed blue curves are almost on top of each other in Fig.~\ref{fig:total-resources}, corresponding to the fact that a circuit to implement an algorithm with a realistic number of magic gates only requires about 50\% more resources at level-$K$ than an algorithm of equivalent size that has no magic gates.

\subsection{Qubit cost for a flag fault-tolerant scheme}
\label{sec:example-flag-qubit_EC}

\begin{figure*}
\centering   
\includegraphics[width=\textwidth]{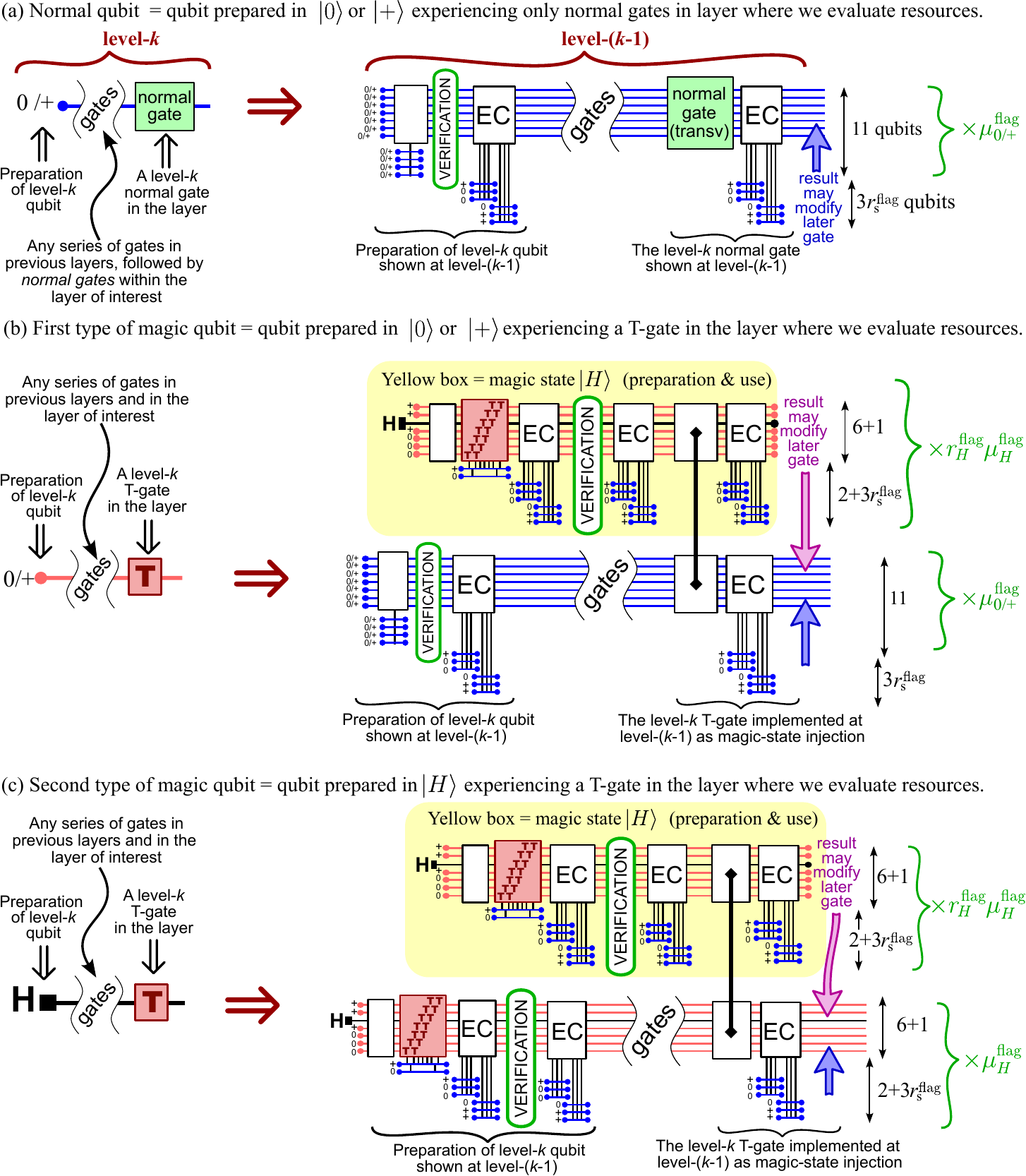}
\caption{Simplified sketch of the error-correction circuitry for 7-qubit code with the flag-qubit approach, showing the aspects necessary to identify the qubit cost. The states at level $k$ (to the left of the red arrows) are shown in terms of states at level $k-1$ (to the right of the red arrows). Rectangles are blocks of gate-operations, whose details are in other figures in this work. 
The colored lines, arrows and symbols like ``0/+'' or ``T'' all have the same meaning as in Fig.~\ref{fig:sketch-Steane}.  However, there is now also a black horizontal line indicating a qubit that starts in $|H\rangle$ and then experiences a $T$-gate. 
} 
\label{fig:sketch-flag-qubit}
\end{figure*}

The main conclusion of the previous subsection is that magic gates are not so costly compared to normal gates.
Thus, it may often be more efficient to minimize the qubit cost for the normal gates, rather than the magic gates. In particular, we observe that the EC gadget employs only normal gates, but it is used repeatedly throughout the entire computation. Having an EC gadget that uses fewer qubits might hence make a much larger impact. This naturally suggests repeating our analysis for the case of the flag-qubit approaches to error correction \cite{chao_quantum_2018,reichardt_fault-tolerant_2021}. We follow the flag fault-tolerant quantum computing scheme of Refs.~\cite{chao_quantum_2018,reichardt_fault-tolerant_2021,chamberland_fault-tolerant_2019}, which requires only 6 ancillary qubits in each EC gadget (see Fig.~\ref{fig:sketch-flag-qubit}a), much less costly than the earlier 28 ancillary qubits in each Steane EC gadget (see Fig.~\ref{fig:sketch-Steane}a).

The detailed circuit diagrams for the flag scheme are given in Figs.~\ref{fig:+_prep_flag}-\ref{fig:Hm_flag}, taken from Refs.~\cite{chamberland_fault-tolerant_2019} and \cite{reichardt_fault-tolerant_2021}. The key differences from the Steane EC scheme of Sec.~\ref{sec:example-Steane_EC} are in the state preparation circuits, which now make use of flag qubits, rather than verification tests, and in the EC gadget that again uses flag qubits. 
Flag-qubit error-correction is argued to have a very similar threshold to Steane error-correction, see Appendix~\ref{supp:sec:toy_model} so for simplicity we take its threshold to be the same as that of Steane error-correction; $p_{\text{thres}}=2 \times 10^{-5}$.
Fig.~\ref{fig:sketch-flag-qubit} is a simplified circuit diagram for the flag scheme; it shows the qubit costs at  level-$(k-1)$ for normal and magic qubits at level $k$, while hiding details not needed when evaluating qubit costs. Note that this scheme implements the magic gate $T_Y \equiv \exp(-i \pi/8 Y)$ \cite{chamberland_fault-tolerant_2019}, rather than the usual $T$ gate $\exp(-i \pi/8 Z)$. They are computationally equivalent up to normal gates (applying appropriate Clifford gates around $T_Y$ can convert $T_Y$ to a $T$-gate), so this makes no difference in terms of resource costs. We will hence continue to refer to the magic gate in the flag scheme as a $T$ gate.

We observe that the circuits are most easily written in terms of three types of qubit;
one type of normal qubit and two types of magic qubit.
The normal qubit (Fig.~\ref{fig:sketch-flag-qubit}a) is one that starts in the state $|0\rangle$ or $|+\rangle$ and only experiences normal gates (Clifford gates).
The first type of magic qubit is one that starts in the state $|0\rangle$ or  $|+\rangle$, and experiences at least one magic gate ($T$-gate) at some point in the calculation. The second type of magic qubit is one that starts in the state $|H\rangle$ and experiences a magic gate ($T$-gate) at some point in the calculation. Every qubit starting in $|H\rangle$ will experience a magic gate, as explained after Eq.~\eqref{eq:beta_K1_K2} below. Magic qubits starting in $|H\rangle$ require different resources compared to magic qubits starting in $|0\rangle$ or $|+\rangle$: this is why two distinct types of magic qubits must be introduced in the recursion for flag-qubit schemes.
Let us define $Q_\opnormal$ as the number of normal qubits at level-$k$ of concatenation,
$Q_{\opmagic}$ as the number of magic qubits of the first type, and 
$Q_{\opmagic;H}$ as the number of magic qubits of the second type.
Then, if there are $K$ levels of concatenation, we have
\begin{eqnarray}
\left( \begin{array}{c} \Qphys_\opnormal \\ \Qphys_\opmagic  \\ \Qphys_{\opmagic;H}\end{array}\right) 
&=& \mathbf{M}_{\rm flag}^K 
\left( \begin{array}{c} \Qalgo_\opnormal \\ \Qalgo_\opmagic  \\ 0\end{array}\right).
\label{eq:recursion_flag}
\end{eqnarray}
The $\Qalgo_{\opmagic;H}=0$ entry in the vector on the right-hand side arises because the abstract circuit for the algorithm of interest is written in terms of gates acting on $|0\rangle$, as indicated in our description of $\cC_{\mathrm{algo}}$ in Sec.~\ref{sec:C_algo}, and so contains no $|H\rangle$ states. 
The matrix 
$\mathbf{M}_{\rm flag}$ can be derived in a similar manner to that in earlier examples. We find,
\begin{eqnarray}
\mathbf{M}_{\rm flag} = 
\left( \begin{array}{ccc} 
\lambda^{\rm flag}_\opnormal & \lambda^{\rm flag}_\opnormal +\beta K_1 & \beta K_2  \\ 
0 & 6K_1 & 6K_2 \\
0 & K_1 & K_2
\end{array}\right)\,.
\label{eq:recursion_matrix_flag}
\end{eqnarray}
The two zeros in the first column mean that $\mathbf{M}_{\rm flag}$'s top-left matrix element is one of its eigenvalues; it is the one associated with normal qubits, 
\begin{align}
\lambda^{\rm flag}_\opnormal=11\muClFlag + 3r^{\rm flag}_s=39.
\label{eq:lambda_easy-flag}
\end{align}
For compactness, we have defined 
\begin{eqnarray}
\beta &\equiv& 2+3r^{\rm flag}_s  \ = \ 8 \nonumber\\
K_1 &\equiv& r^{\rm flag}_H \mu^{\rm flag}_H  \ = \ 8\\
K_2 &\equiv& (1+r^{\rm flag}_H)\mu^{\rm flag}_H
 \ = \ 12  \nonumber
 \label{eq:beta_K1_K2}
\end{eqnarray}
where we take $\mu^{\rm flag}_H$ for the case without shared state preparation (see Table~\ref{table:failure-multiplicity}). The assumptions allowing us to derive the time and failure multiplicities in Tables \ref{table:failure-multiplicity} and \ref{table:time-multiplicity} are explained at the end of this section.  
Note the matrix element  $K_1$ in the bottom row of the matrix $\mathbf{M}_{\rm flag}$; it is responsible for there being $|H\rangle$ states at lower concatenation levels even though there are none at the logical level (level $K$). This is a consequence of any $T$-gate at the logical level (level $K$) requiring a magic state that contains an $|H\rangle$ at level $(K-1)$; see Fig.~\ref{fig:sketch-flag-qubit}(b), which in turn requires an $|H\rangle$ at level $(K-2)$, etc.

In addition to the eigenvalue in Eq.~(\ref{eq:lambda_easy-flag}), 
$\mathbf{M}_{\rm flag}$ has two eigenvalues given by the two-by-two block in the lower-right corner of $\mathbf{M}_{\rm flag}$. These are the eigenvalues for magic qubits. One of these two eigenvalues is zero and the other is 
\begin{align}
\lambda^{\rm flag}_\opmagic=6K_1+K_2 \ =\  (7r^{\rm flag}_H+1)\mu^{\rm flag}_H=60. 
\end{align}
The zero eigenvalue plays no role in the scaling as we increase the number of levels of concatenation $K$; it corresponds to a $K$-independent term.
Hence, the $K$-scaling for normal qubits goes like $(\lambda^{\rm flag}_\opnormal)^K$, 
while the $K$-scaling for magic qubits goes like $(\lambda^{\rm flag}_\opmagic)^K$.

Now we see that, unlike the earlier scheme with Steane EC gadgets (Sec.~\ref{sec:example-Steane_EC}), this flag scheme has the magic-qubit eigenvalue larger than the normal one: $\lambda^{\rm flag}_\opmagic > \lambda^{\rm flag}_\opnormal$. 
This places it in the scaling regime where magic qubits increasingly dominate over normal qubits as $K$ is increased, until the proportion of resources for magic gates goes to infinity as $K\to \infty$. However, the fact that $\lambda^{\rm flag}_\opmagic$ is only about $50 \%$ larger than $\lambda^{\rm flag}_\opnormal$ means that the two will remain comparable for practically relevant $K$s, such as $K\leq 5$ \cite{2_footnote-K=5}.

To quantify this, the method in App.~\ref{sec:scaling-general-details} allows us to find the total physical qubit cost for these parameters. For $K$ levels of concatenation,
the total physical qubit cost is
\begin{eqnarray}
\Qphys_{\rm total:flag} 
&=& 39^K  \Qalgo_\opnormal 
\nonumber \\
& & + \frac{82\times60^K - 65
\times 39^K}{15}\,\Qalgo_\opmagic
\label{eq:Qphys-total-flag}
\end{eqnarray}
where the terms that go like $39^K$ and $60^K$ come from $\lambda_\opnormal$ and $\lambda_\opmagic$ respectively.
Hence, the ratio of resources for an abstract circuit of interest to resources for the same abstract circuit in which all magic gates have been replaced by normal gates at the algorithmic level (level $K$), as defined in Eq.~(\ref{Eq:def-R}) is
\begin{eqnarray}
R_{\rm flag}(K) &=&  \frac{\Qalgo_\opnormal}{\Qalgo_{\rm total}} +\frac{82\times60^K - 65\times39^K}{15\times39^K}\, \frac{\Qalgo_\opmagic}{\Qalgo_{\rm total}} . \qquad
\label{eq:Rflag}
\end{eqnarray}
This grows exponentially with $K$, but for $K=5$ this corresponds to a magic gate ($T$-gate) being about 43 times more costly than a normal gate (Clifford gate). So a worst-case algorithm that has a timestep in which all logical (level-$K$) are magic gates (i.e., with 
$\Qalgo_\opmagic=\Qalgo_{\rm total}$) will require about 43 times the physical qubit cost of an equivalent-sized algorithm with no magic gates.
However, many typical algorithms  have only 5\% of gates being magic gates (see App.~\ref{sec:T-gate-proportion}) for which we can take 
$\Qalgo_\opmagic/\Qalgo_{\rm total} \simeq 0.1$ (as explained below Eq.~\ref{eq:Steane-C}), the magic gates (with their associated magic states) only consume about 5.2 times as many physical qubit cost as the normal gates.

Such a factor of five is clearly not negligible; it is seen as the difference between the solid and dashed green curves in Fig.~\ref{fig:total-resources}. However, it is tiny compared to the factor of $10^4$ reduction in resources achieved (at $K=5$) by switching from Steane EC gadgets to flag-qubit gadgets
(compare the black and green curves in Fig.~\ref{fig:total-resources}).

\subsection{Shared state preparation for the flag-qubit approach} 
\label{sec:flag+shared}
Now we turn to the case of flag scheme with shared state preparation, for which we will be more approximate than elsewhere in this work.

The logic of treating shared state preparation for flag-qubits is the same as for the Steane EC gadgets. However, such an evaluation would require knowing a certain factor
 $J^\text{flag}_\text{EC-repeat}$ which contributes to the factor $D$ in Eq.~\eqref{eq:Ngate-level-1}.
 We recall that $D$ is the number of timesteps at level $(k-1$) required to implement a normal gate at level-$k$, and this depends on the number of repetitions of the flag-qubit EC gadget that are required,  a number which we call $J^\text{flag}_\text{EC-repeat}$. A look at the circuit from Fig.~\ref{fig:EC_gadget_flag} shows that $D=1+12\times J^\text{flag}_\text{EC-repeat}$ (1 time step to implement the transversal gate, $12\times J^\text{flag}_\text{EC-repeat}$ to perform error-correction).

The exact value of $J^\text{flag}_\text{EC-repeat}$ varies depending on details of the error-correction gadget.
Typically, its value is fairly close to 1, with $J^\text{flag}_\text{EC-repeat} =$ 3, 4 or 5 appearing in the literature \cite{tansuwannont2023adaptive,pato2024optimization}.
However, it turns out that we do not need the precise value of $J^\text{flag}_\text{EC-repeat}$, because it only enters our failure multiplicities, and they depend very weakly on $J^\text{flag}_\text{EC-repeat}$.
Let us take the same example of implementing Shor's cracking of RSA-2048 as in Sec.~\ref{sec:steane-shared} (so $K=5$, $Q_{\text{total}}^{\text{algo}} \sim 10^4$, $D^{\text{algo}} \sim 10^{10}, \kappa=0.01$).
 If we take an under-estimate with $J^\text{flag}_\text{EC-repeat}=1$, we get $\mu^{\rm flag}_{0/+} = 1.14$
 and $\mu^{\rm flag}_H = 1.24$.
 If we then take an {\it extreme} overestimate with $J^\text{flag}_\text{EC-repeat}=1000$, we get 
 $\mu^{\rm flag}_{0/+} = 1.19$ and $\mu^{\rm flag}_H = 1.31$.
 Thus, varying $J^\text{flag}_\text{EC-repeat}$ between 1 and 1000 only causes a few percent change in the failure multiplicities, 
 $\mu^{\rm flag}_{0/+}$ and $\mu^{\rm flag}_H$. 
 For the sake of caution, we put the larger of these values in Table ~\ref{table:failure-multiplicity}, and then use them for the orange curves in Fig.~\ref{fig:total-resources}, knowing that this can only be an overestimate.
 
This shows that sharing the state preparation reduces the physical qubit cost by nearly two orders of magnitude.  Then reducing the cost of the magic states can never reduce physical qubit cost more than about a factor of two for $K\leq 5$ \cite{2_footnote-K=5}.
Furthermore, achieving that factor of two would require finding an optimization that makes magic gates no more costly than normal gates, which currently seems unrealistic. This again shows that, while magic-states should be optimized, other optimizations are more important. Hence, magic states should be considered as the last resource to optimize.

\section{Strong effect of EC gadget size on the physical qubit cost}
\label{sec:toy_model}

\begin{figure}
\centering   
\includegraphics[width=0.85\columnwidth]{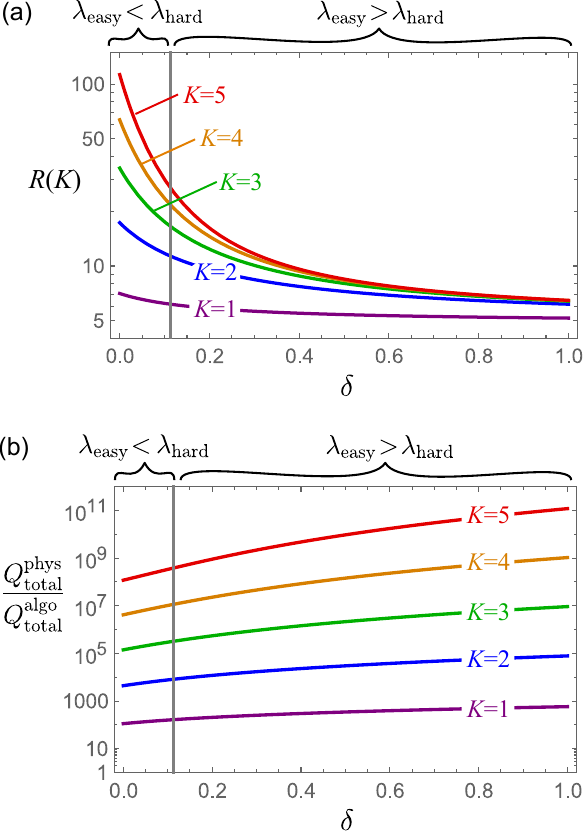}
\caption{Plots of the toy-model in Sec.~\ref{sec:toy_model}, where we vary the size of the error-correction gadget by varying $\delta$. Here $\delta=1$ corresponds to the Steane EC gadget in Sec.~\ref{sec:example-Steane_EC}, and reducing $\delta$ reduces the size of the EC gadget to the point where $\delta=0$ corresponds to a fictional case where the EC gadget requires no ancillary qubits at all. All curves are for the (unrealistic) worst-case scenario where the circuit is made only of magic gates at the level of the abstract circuit (level-$K$), so $\Qalgo_\opmagic=\Qalgo_{\rm total}$ and $\Qalgo_\opnormal=0$.  (a) The relative cost of magic gates to normal gates, as quantified by $R(K)$ in Eq.~(\ref{eq:R-toymodel}), which is the ratio of physical qubits in a circuit that contains only magic logical gates to physical qubits in an equivalent-sized circuit that contains only normal logical gates.
(b) The number of physical qubits required to encode a single logical qubit in the algorithm, $\Qphys_{\rm total}\big/\Qalgo_{\rm total}$, as in Eq.~(\ref{eq:total-resources-toymodel}). 
In both plots, the gray vertical line indicates the value of $\delta$ for which $\lambda_\opmagic=\lambda_\opnormal$. 
Comparing the vertical scales of the two plots (which are both logarithmic), we see from (a) that when we reduce $\delta$ there is an increase in the relative cost of magic gates, 
despite this we see from (b) that the total physical qubit cost are reduced by multiple orders of magnitude.
} \label{fig:plot}
\end{figure}

In the previous section, which gave worked examples of applying our matrix approach,
we saw that changing from the Steane scheme (with larger EC gadgets) to the flag-qubit scheme (with smaller EC gadgets) vastly reduces the resources. These reductions are a few orders of magnitude at $K=5$, see Fig.~\ref{fig:total-resources}. 
However, the switch from Steane to flag-qubit schemes does not only reduce the number of qubits in the EC gadget, it also reduces the number of qubits in the magic-state preparation.
So to confirm that the dominant effect is the reduction of the number of qubits in the EC gadget,
we finish this article with a toy model to directly explore how the size of the EC gadget affects the resource costs of normal and magic gates.

For this, we take a toy model based on the matrix approach for the simple case of the  2-by-2 matrix $\mathbf{M}$ in Sec.~\ref{sec:simplest-case}, and allow ourselves to vary its parameters in a continuous manner, without worrying about whether there are fault-tolerant circuits that achieve any given value of the parameters.
Here, we vary $\lambda_\opnormal$ and $b$ for fixed $\lambda_\opmagic$. From this, we find that reducing the size of the EC gadget
causes an increase in the relative weight of resources of the magic gates to normal gates.  Despite this, reducing the size of the EC gadget causes the physical qubit costs (for both normal and magic operations) to be significantly reduced. It is thus always worthwhile finding ways to reduce the size of the EC gadget.

Specifically here, our toy model begins with the parameters for 7-qubit code scheme using Steane EC gadgets with shared state preparation [see Eqs.~\eqref{eq:lambdaEasy_Steane}-\eqref{eq:b_Steane} and surrounding text], but replaces $r_{\rm s}=3$ by $3 \delta$, where we recall that the number of ancillary qubits in the EC gadget is proportional to $r_{\rm s}$. Then $\delta$ is the toy-model parameter that we can vary to change the resources required by the EC gadget, 
In this case,
\begin{subequations}
\begin{eqnarray}
\lambda_\opnormal &=& 97\,\delta +16
\\ 
\lambda_\opmagic &=& 27
\\
b &=& 477\,\delta +87
\end{eqnarray}
\end{subequations}
where we round coefficients to the nearest integer.
When $\delta=1$ these equations coincide with our earlier example of Steane error correction with shared state preparation in Eqs.~(\ref{eq:lambdaEasy_Steane}-\ref{eq:b_Steane}). In contrast, $\delta=0$ corresponds to a likely fictitious situation in which the EC gadget requires no ancillary qubits at all.
Then, Eq.~(\ref{Eq:Q-physical-total}) gives the number of physical qubits necessary for a given algorithm.
In the rare worst-case scenario where all gates are magic gates
so $\Qalgo_\opmagic=\Qalgo_{\rm total}$ and $\Qalgo_\opnormal=0$, this gives the number of physical qubits required to encode a single logical qubit in the algorithm,  
\begin{eqnarray}
\frac{\Qphys_{\rm total}}{\Qalgo_{\rm total}}
&=& 27^K \!+\! (477\,\delta +87)\frac{(97\,\delta +16)^K\!-27^K}{97\,\delta -11}. \ \ 
\label{eq:total-resources-toymodel}
\end{eqnarray}
Similarly, Eq.~(\ref{Eq:R-for-2-by-2}) with $\Qalgo_\opmagic=\Qalgo_{\rm total}$ and $\Qalgo_\opnormal=0$ gives the ratio of physical qubits necessary for a circuit that contains only magic gates at the level of the abstract circuit (level-$K$) compared to the same circuit with all level-$K$ magic gates replaced by level-$K$ normal gates,
\begin{eqnarray}
R(K) = \frac{477\,\delta +87}{97\,\delta -11} -
\frac{380\,\delta + 98}{97\,\delta -11}
\left(\!\frac{27}{97\,\delta +16}\!\right)^K. \ \ 
\label{eq:R-toymodel}
\end{eqnarray}
Fig.~\ref{fig:plot} plots Eqs.~\eqref{eq:total-resources-toymodel} and \eqref{eq:R-toymodel} for various $K$ values, as $\delta$ varies between 0 and 1, thereby varying the size of the EC gadget.

When analyzing Fig.~\ref{fig:plot}, we assume that the EC gadget's capacity to correct errors is independent of $\delta$. In other words, we assume that while the ``cost'' varies with $\delta$ in the manner shown in Fig.~\ref{fig:plot}, the ``benefit'' (the amount of error correction) does not vary with $\delta$.
If both cost and benefit varied with $\delta$, then Fig.~\ref{fig:plot} would not be a fair comparison of different $\delta$s. 
However, we observe that the benefits (capacity to correct errors) do not vary with $\delta$ in our earlier examples of the Steane EC gadget versus the flag-qubit EC gadget; their capacity to correct errors is quantified by the fault-tolerance threshold, $p_{\text{thres}}$, and this is similar for both schemes (see Appendix~\ref{supp:sec:toy_model}). In short, the reason is that a flag-qubit EC gadget contains fewer ancillary qubits, but takes more timesteps, so its number of fault locations in the largest exRec --- which determine $p_{\text{thres}}$ in the manner mentioned below Eq.~(\ref{eq:textbook-scaling_k-basics}) --- is not so different from that for a Steane-EC gadget. Thus, it seems natural to reflect this in our toy-model by assuming the capacity for error-correction to be $\delta$-independent.

With this in mind, we conclude from the plots in Fig.~\ref{fig:plot}(a) that reducing $\delta$ (reducing the number of ancillary qubits required by each EC gadget) increases the relative cost of magic gates to normal gates (as quantified by $R(K)$). Nonetheless, we see from Fig.~\ref{fig:plot}(b) that reducing $\delta$ can cause the total physical qubit cost to be reduced by multiple orders of magnitude.
Minimizing the EC gadget size is thus an excellent way to reduce the physical qubit cost.

\section{Conclusions}
\label{sec:conc}
We have devised a simple, yet powerful, matrix approach to evaluate the physical qubit cost of concatenated fault-tolerant quantum computing. Our approach yields closed-form expressions that remain simple for arbitrary algorithm size and an arbitrary number of levels of concatenation, especially for recursively concatenated schemes, as outlined in Sec.~\ref{sec:summary}. This makes it a useful tool for comparing physical qubit costs between different schemes and their variants, as we have done in the examples of Fig.~\ref{fig:total-resources}. While we focused on specific examples in the main text, our approach is applicable for more general schemes, as explained further in Appendix~\ref{sec:formalism_general_case}.

We have seen from our analysis that the physical qubit cost is determined only by a small number of parameters (only three in the simplest case) which can be extracted from the circuit designs of the scheme in question. 
However, to get reliable values of these parameters, for accurate estimates of qubit costs, there are two particularly critical factors: the time and failure multiplicity factors. These account for, respectively, the costs associated with the time taken to prepare ancillary qubit states and the additional costs needed to compensate for failed verification of quantum states. For the latter, we had to analyze the impact of these failures on the fault-tolerance threshold. The multiplicity factors can have a large impact on the physical qubit cost: For the examples in Fig.~\ref{fig:total-resources}, reducing the failure multiplicity by sharing state preparation resources reduces the physical qubit cost by orders of magnitude. 

Understandably, the optimizations most effective at reducing the physical qubit cost are those that affect all operations, rather than optimizations restricted to magic states. While one can naturally expect that optimizations affecting all operations are more efficient than optimizations only affecting magic-states, the surprising conclusion of our work is that the former can reduce the costs by several orders of magnitude while the latter only by a marginal amount. This is particularly surprising given that numerous works in the literature emphasize specific optimizations that only affect magic-states. This is also surprising given the current trend in the field to design algorithms that minimize the use of magic operations (these being those operations that require implementation using magic states).
We have shown how the switch from the Steane EC gadget to a flag-qubit scheme significantly lowers the physical qubit cost. 
This is almost entirely due to the significantly lower cost of the EC gadget in the flag-qubit scheme compared to the Steane EC gadget.
In contrast, the optimization of magic states can bring only marginal improvements. The best (i.e., least costly) case in our Fig.~\ref{fig:total-resources}  is flag-qubits with shared state preparation. In that case, even if one could invent a method to \textit{completely} eliminate the additional cost of magic states (so that magic and normal gates have equal costs), it would only reduce the physical qubit cost by a factor of two; a small gain compare to the other improvements shown in Fig.~\ref{fig:total-resources}.

While no optimization is too small to be ignored,
we thus conclude that magic-state optimizations are less important than one might think. Instead, optimizing ingredients that go into every operation (such as EC gadgets) can reap the
most benefits. We have shown how resource sharing can reduce failure multiplicities, leading to sizable cost reductions.  One can also expect that circuit designs that reduce time multiplicities to also have significant impact. Our matrix approach elucidates how each ingredient affects the physical qubit cost, and provides a means of identifying the most worthwhile optimizations to minimize the physical qubit costs of concatenated fault-tolerant quantum computing schemes.

\section{Acknowledgements}

We warmly thank A. Auff\`eves, J.H. Chai, C. Vuillot for many useful discussions and inputs.

This work was supported by 
the European Union (EU) Horizon 2020 research and innovation program via 
``Quantum Large Scale Integration in Silicon'' (QLSI) (Grant No. 951852), 
the Singapore National Research Foundation (NRF) and French National Research Agency (ANR) joint project ``QuRes'' (Grant No. ANR-21-CE47-0019; NRF2021-NRF-ANR005), 
the Merlion Project (Grant No. 7.06.17), 
the ANR program “Investissements d’avenir” (Grant No. ANR-15-IDEX-0002), 
the Laboratories of Excellence (Labex) Laboratoire d’Alliances Nanosciences-Energies du Futur (LANEF),
and the ``Quantum Optical Technologies'' project within the International
Research Agendas program of the Foundation for Polish Science cofinanced by the EU European Regional Development Fund. 
H.K.N. acknowledges the support of the National Research Foundation, Singapore and A*STAR under its CQT Bridging Grant. M.F.A also acknowledges the funding from the European Union's Horizon Europe Research and Innovation programme under the Marie Sklodowska-Curie Actions \& Support to Experts programme (MSCA Postdoctoral Fellowships) - grant agreement No. 101108284.
R.W. also acknowledges the support of the ``OECQ project'' (Contract DOS0226235/00) financed by the French state (via France 2030) and Next Generation EU (via France Relance), and the  ``Q-Loop'' project of the French national quantum strategy within France 2030.

\appendix

\section{Proportion of magic gates ($T$-gates) in typical algorithms}
\label{sec:T-gate-proportion}

\begin{table*}[t]
\bgroup
\def\arraystretch{1.7}
\begin{tabular}{|l|l|l|}
\hline
\ {\bf Computing Task}  &\ {\bf Reference \& assumptions} & \ 
\parbox[t][8mm]{4.5cm}{{\bf Proportion of $T$-gates}\\ $P^{\rm algo}_\opmagic$}
\\
\hline \hline
\ \ \parbox[t][15mm]{6cm}{\raggedright Shor's algorithm for cracking an RSA $n$-bit encryption key.} \ 
&\ \parbox[t]{6cm}{\raggedright Eqs.~(39-41) of Ref.~\cite{Zalka1998Jun}. If Toffolis are implemented as Cliffords+$T$-gates in the usual manner: 
 $\Qalgo_{\rm total}=5n$, $N_T=364 n^3$, $D=5400 n^2$.}\ 
& \ $P^{\rm algo}_\opmagic = 0.0135$
\\
\hline
\ \ \parbox[t][13mm]{6cm}{\raggedright Elliptic curve discrete logarithm problem
(a crucial part of Shor's algorithm) for 256, 384 \& 521 bits elliptic curves.} \ 
& \ \parbox[t]{6cm}{\raggedright Table 1 of Ref.~\cite{Haner2020Apr} has results optimized for low $T$-count, low $\Qalgo_{\rm total}$ or low depth. See also Ref.~\cite{Banegas2021}.}\ 
& \ \parbox[t]{4.5cm}{\raggedright
$P^{\rm algo}_\opmagic < 0.001$  \ (low $T$-count)\\
$P^{\rm algo}_\opmagic < 0.001$ \ (low $\Qalgo_{\rm total}$)\\
$P^{\rm algo}_\opmagic < 0.0125$ \ (low depth) }
\\
\hline
\ \ \parbox[t][12mm]{6cm}{\raggedright Quantum linear-system algorithm for electromagnetic scattering cross-section of a 2D target.}\ 
& 
\ \parbox[t]{6cm}{\raggedright Ref.~\cite{Scherer2017Jan}, including oracles in its Table 2} \ 
& \ $P^{\rm algo}_\opmagic < 10^{-8} $\\
\hline
\ \ \parbox[t][9mm]{6cm}{\raggedright Finance: Estimates for a quantum advantage in derivative pricing.}
& \ \parbox[t]{6cm}{\raggedright Table 1 of Ref.~\cite{Chakrabarti2021Jun} indicates the $T$-count and $T$-depth for various examples.} & \ $P^{\rm algo}_\opmagic < 0.05$ \ (all examples)\\
\hline
\ \parbox[t][9mm]{6cm}{\raggedright Quantum chemistry: Obtaining energies of FeMoco.} & 
\ \parbox[t]{6cm}{\raggedright Table 1 of Ref.~\cite{Reiher2017Jul} with results for serial and parallelized implementations.} \ 
& \ \parbox[t]{4cm}{\raggedright
$P^{\rm algo}_\opmagic < 0.01$ \ (serial)\\
$P^{\rm algo}_\opmagic \sim 0.5$ \ (parallel)} \ 
\\
\hline
\end{tabular}
\egroup
\caption{The proportion of the algorithm's gates that are $T$-gates for various computing tasks in the literature. These $T$-gates correspond to magic gates in our 7-qubit code examples (since they are the gates requiring magic-state injection). This proportion of magic gates is take as $P^{\rm algo}_\opmagic =N_T/(\Qalgo_{\rm total} D)$, where $N_T$ is the number of $T$-gates, $\Qalgo_{\rm total}$ is number of logical qubits, and $D$ is the logical depth.}
\label{supp:table:algo}
\end{table*}

At various points in this work we need a reasonable estimate of the number of magic gates that occur in the layer of the algorithm (i.e. level-$K$ timestep) with the most magic gates.  This is because it is typically the layer of the algorithm
that requires the largest number of physical qubits, what we call $\Lmax$, and so determines the number of physical qubits that the quantum computer must have.
In the body of the text we quantify this as the proportion of magic gates in that step of the algorithm, $\Qalgo_\opmagic \big/ \Qalgo_{\rm total}$, where $\Qalgo_{\rm total}$ is the total number of qubits in the algorithm, and $\Qalgo_\opmagic$ is the number of magic qubits at that timestep with the most magic qubits
(with magic qubits being those involved in magic gates at that timestep).

In principle, we could plunge in the circuit detail for each algorithm in the literature to find the largest value of 
$\Qalgo_\opmagic \big/ \Qalgo_{\rm total}$ in that algorithm.
However, here we prefer to get simple estimates from easily available information summarized in Table~\ref{supp:table:algo}.
That table gives the proportion of gates in the algorithm that are magic gates (where magic gates are $T$-gates in our examples) for various computing task, and calls the proportion $P^{\rm algo}_\opmagic$.  
Table~\ref{supp:table:algo} shows that most algorithm's implementations have a very low proportion of $T$-gates (with the exception of the last line of the table); this proportion is a few percent or less. We will now use $P^{\rm algo}_\opmagic$ to estimate $\Qalgo_\opmagic \big/ \Qalgo_{\rm total}$ for such computing tasks.

To get a lower limit on $\Qalgo_\opmagic \big/ \Qalgo_{\rm total}$, we can assume each timestep has the same number of magic gates, then we have 
\begin{eqnarray}
\frac{\Qalgo_\opmagic}{\Qalgo_{\rm total}} \sim P^{\rm algo}_\opmagic. 
\end{eqnarray}
Thus, the lower bound on $\Qalgo_\opmagic \big/ \Qalgo_{\rm total}$
is a few percent or less in most algorithms.

To get an upper limit on $\Qalgo_\opmagic \big/ \Qalgo_{\rm total}$, we can assume there are a few timestep where all qubits experience magic gates, and many timesteps where there are no magic gates.  Then the proportion of timesteps with magic gates
would then be $P^{\rm algo}_\opmagic$. 
We do not believe that any of the examples in Table~\ref{supp:table:algo} have the property, but we cannot rule it out as a worst case scenario.
In this unlikely case $\Qalgo_\opmagic \big/ \Qalgo_{\rm total}=1$, however it is easy to reduce it \cite{Gidney2020Dec}. For example,
to reduce $\Qalgo_\opmagic \big/ \Qalgo_{\rm total}$,
we imagine taking that one timestep of the algorithm that involve all magic gates, and spreading it out over twenty timesteps, where each of those twenty timesteps only involves 5\% of the qubits having magic gates (while the rest of the qubits do nothing).  At the same time, any timesteps that involve less than 5\% of magic gates will be left unchanged. As the proportion of the algorithm's timesteps that are spread in this manner is only $P^{\rm algo}_\opmagic$, this will only increase the algorithm's depth by a factor of 
\begin{eqnarray}
20 P^{\rm algo}_\opmagic + (1-P^{\rm algo}_\opmagic)
\label{eq:depth-factor}
\end{eqnarray}
As $P^{\rm algo}_\opmagic$ is small, this is a modest increase in depth.
For example, in those algorithms where $P^{\rm algo}_\opmagic=0.01$, we can ensure that
$\Qalgo_\opmagic \big/ \Qalgo_{\rm total}\leq 0.05$ at the cost of 
increasing the algorithm's depth by $\leq19\%$.
In those algorithms where $P^{\rm algo}_\opmagic=0.05$, we can ensure that
$\Qalgo_\opmagic \big/ \Qalgo_{\rm total}\leq 0.05$ at the cost of doubling the algorithm's depth.

However, we emphasize that these are worst case scenarios, and the increase in depth is likely to be much less than that given above. This is because it is rare that all magic gates occur in the same timestep. If this were to occur, it would be rare that one cannot reorganize the operations so that the magic gates are spread over multiple timesteps,  but qubits that are not doing magic gates are doing other gates-operations useful for the algorithm (rather than just waiting doing nothing). 
This would ensure the depth is less than if all such qubits just waited, as assumed in Eq.~(\ref{eq:depth-factor}).  Indeed, developing methods to do this would be an excellent way of reducing the qubits resource costs of an algorithm in which many magic gates (gates requiring magic states) must be done in the same timestep. One such method is discussed in Ref.~\cite{Gidney2020Dec}. In such cases,  re-designing the circuit to spread magic gates more evenly in time could be more important for resource efficiency than reducing the physical qubit cost of each magic state. 

Whatever the scenario, we estimate that such relatively modest increases in depth
as given in Eq.~(\ref{eq:depth-factor}) will only very rarely impact the number of qubits required by error-correction. To understand why, we recall that the goal of 
fault-tolerance is to make the error per  level-$K$ logical operation, $p_K$, small enough that the total algorithm only fails a small proportion of the time; let us assume we can accept that this proportion to 
be $\epsilon_\text{algo}$. Then, if the algorithm contains $N_\text{algo}$ level-$K$ operations, we require that   $p_K N_\text{algo} \leq \epsilon_\text{algo}$. Now recalling that $p_K$ is given by Eq.~(\ref{eq:textbook-scaling_k}), and considering fault-tolerant codes with $t=1$ (like those in the examples considered here), we get 
\begin{eqnarray}
    K = \left\lceil \ \frac{1}{\ln(2)}\ln \left( \frac{\ln\left(p_\text{thres}N_\text{algo}\big/\epsilon_\text{algo}\right)}{\ln(p_\text{thres}/p_0)} \right)  \ \right\rceil \ ,
    \label{Eq:necessary-K}
\end{eqnarray}
where $\lceil \cdots \rceil$ is a ceiling function \cite{ceiling}. 
The argument of this ceiling function varies extremely slowly with $N_\text{algo}$ --- it scales like $\ln\left(\ln\left(p_\text{thres}N_\text{algo}\big/\epsilon_\text{algo}\right)\right)$ 
--- and 
so a change of $N_\text{algo}$ by the factor in Eq.~(\ref{eq:depth-factor}) (which is typically $\lesssim 2$)  is very unlikely to change the argument of the ceiling function enough that $K$ increases by one.
Hence, 
in the vast majority of cases, $K$ will be completely unchanged when the depth is increased by the factor in Eq.~(\ref{eq:depth-factor}). Then this depth increase has no effect on the physical qubit cost.

In those rare cases, where increasing the depth by the factor in Eq.~(\ref{eq:depth-factor}) causes $K$ to increase by one, the physical qubit cost will increase significantly (as seen in Fig.~\ref{fig:total-resources}). This would be a case where $K$ was already extremely close to increasing by one anyway; for example a slightly larger $p_0$ would also make $K$ increase by one.  In such cases, one strategy may be able to keep $K$ unchanged, and compensate by accepting a larger value of $\epsilon_{\rm algo}$. This requires no increase in the physical qubit cost, but one must run the circuit a few more times before being sure that it is giving the correct result.  To be more precise, we remark that if one runs the circuit multiple times, all runs of the circuit that are error-free give the same correct result, but each run with a random error will give a different wrong result.
Hence, even if a fairly high proportion of the runs fail due to a random error (say $\epsilon_{\rm algo}\sim 0.2$), one can identify the correct result by running the circuit a few $1/\epsilon_{\rm algo}$ times (for example, 10-15 times if $\epsilon_{\rm algo}\sim 0.2$). Only one of the calculation's results will be repeated multiple times in these runs, and that will be the error-free result.

The conclusion of this section is that 
for most algorithms in  
Table~\ref{supp:table:algo} with $P^{\rm algo}_\opmagic$ of a few percent or less, it is reasonable to assume that the abstract circuits that implements algorithms of interest will only have about 5\% of magic gates in the layer with the most 
magic gates, and this is what we do in the body of this article.

\section{matrix approach in the general case}
\label{sec:scaling-general-details}

\subsection{Definition and estimation of the number of qubit types}
\label{sec:def_qubits_type}
When we observe that a fault-tolerant scheme requires multiple types of normal qubits or multiple types of magic qubits, the scaling is not described by Eq.~(\ref{eq:recursion_T_Steane-1}), which only has one type of each. That is when we need the general case, based on a matrix $\mathbf{M}$ of larger dimensions.  

However, while we have no simple way to say {\it a priori} how many types of qubit a given fault-tolerant scheme requires (it is necessary to look at the details of the circuits), we can explain the procedure we followed in our examples.
We first identify every combination of initial states, gate-operations, and measurements that is required by the scheme in question. We then group the qubits into ``qubit's types''. A type of qubit is defined as a specific combination of state-preparation, gates, and measurements that a qubit can go through. Then, the (i,j) element of the matrix $\mathbf{M}$ represents the number of qubits of type "i" at level-(k-1) that are required inside one qubit of type "j" at level-k. This approach suggests that there could have a huge number of qubit's type (because many combinations of state-preparation gates and measurements are possible). In practice, however, one can often group the qubits into a small number of qubit's type ($2$ or $3$ in the examples we considered in the main text). This is because many qubit's types are equivalent. We will say that two qubit types "$j_0$" and "$j_1$" are equivalent if, for any $K$, it is sufficient to know \textit{the sum} of the number of level-$K$ $j_0$ and $j_1$ qubits there are (and the number of qubits of type $j_{i}$ for $i \notin \{0,1\}$) to deduce the number of level-0 qubits they would require. In this case, it is not useful to separate the qubits "$j_0$" and "$j_1$" into two different types: we can merge them into a single one, meaning that the matrix $\mathbf{M}$ can have its dimension reduced by one. In different terms, two qubit's type are equivalent if, for any $K$, they require the same number of level-0 qubits. For instance, in our flag and Steane EC-gadgets examples, for any $K$, every level-$K$ normal qubit initialized in either $\ket{0}$ or $\ket{+}$ and measured in the Pauli $X$ or $Z$ bases requires the same number of level-0 qubits: it is not necessary to separate them into different qubit's type. A natural example of where two types are not equivalent is the difference between normal and magic qubits: because level-$K$ magic qubits require state injection, they will usually require more level-0 qubits than normal qubits, hence, they must be classified into different types. It can also be that two magic qubits (or two normal qubits) do not belong to the same type. We have such examples for the flag scheme, with the introduction of two types of magic qubits: see Eq.~\eqref{eq:recursion_matrix_flag} where the shape of $\mathbf{M}$ indicates that the second type of level-$k$ magic qubits requires a different number of level-$(k-1)$ qubits than the first type of magic qubits. Finding a minimum number of qubit's type thus requires a careful look at the circuit in order to identify the few types of qubits that require a different number of level-0 qubits, with the help of sketches like Figs. \ref{fig:sketch-Steane} or \ref{fig:sketch-flag-qubit}. Often, the resulting matrix will be of low dimension. 

Outside of this section, when we specify a number of types of qubits, we assume that the reduction described here has already been performed in the sense that all equivalent types of qubits have already been merged into a single type.

\subsection{Formalism for the general case}
\label{sec:formalism_general_case}
Here we give details of how to generalize the matrix approach introduced in Sec.~\ref{sec:simplest-case} to cases where there are multiple types of normal and magic qubits (for instance because some normal gates require more qubits than other normal gates, or because there are different type of magic gates, each requiring a different number of qubits to prepare their magic state). We first generalize our approach in Appendix~\ref{sec:generalizing_matrix_structure}, before giving relatively general results about the cost of a circuit composed of magic gates in the limit of large $K$, compared to a circuit only composed of normal gates, in concatenated schemes, in Appendix~\ref{sec:asymtotic_general_case}. These results illustrate the flexibility of our matrix approach, showing how it can be used as a tool to analyze concatenation schemes that are more complicated than the examples treated in detail in this article.

\subsubsection{Generalizing the matrix structures}
\label{sec:generalizing_matrix_structure}
Here, we use the main assumption that our work is built on, namely, assumptions 1 $\&$ 2 in Sec.~\ref{sec:main-assumptions-definitions}. 
Then in general, there may be $j_\opnormal$ types of normal qubits at level $k$, each requiring different resources at level $(k-1)$, and $j_\opmagic$ types of magic qubits, each requiring different resources at level $(k-1)$.

To explain how to deal with this case, we make a few assumptions to keep the explanation compact, these include the assumptions about the circuit $\cC_\mathrm{algo}$ in 
Sec.~\ref{sec:C_algo}. We also assume that the magic gates are single-qubit magic gates (like the $T$-gate in our examples in Sec.~\ref{sec:FTQC7qb}). If this is not the case the following argument will not fundamentally change, but it will be a bit uglier (particularly if there are magic two-qubit gates as well as magic single-qubit gates).

In this case the matrix recursion relation looks like that for the  2-by-2 matrix $\mathbf{M}$ in Eq.~(\ref{eq:recursion_T_Steane-1}), but now it contains a larger matrix. It is written as 
\begin{align}
    \begin{pmatrix}
        Q_\opnormal(k-1) \\
        Q_\opmagic(k-1)
    \end{pmatrix} =\mathbf{M} \begin{pmatrix}
        Q_\opnormal(k) \\
        Q_\opmagic(k)
    \end{pmatrix}, 
\end{align}
where
now 
$Q_\opnormal(k)$ is a vector with $j_\opnormal$ elements, whose $i$th element is the $i$th type of normal qubit (at level $k$). 
Similarly, $Q_\opmagic(k)$ is a vector with $j_\opmagic$ elements, 
whose $i$th element is the $i$th type of magic qubit  (at level $k$). 
This means that 
$\mathbf{M}$ is the $(j_\opnormal+j_\opmagic)$-by-$(j_\opnormal+j_\opmagic)$ matrix that takes the form
\begin{eqnarray}
\mathbf{M} = 
\left( \begin{array}{cc} \mathbf{M}_\opnormal & \mathbf{B}  \\ 0 & \mathbf{M}_\opmagic \end{array}\right). 
\label{Eq:M-general-case}
\end{eqnarray}
whose $ij$th element is the number of level-$(k-1)$ qubits of type $i$ necessary for a level-$k$ qubit of type $j$.
Here $\mathbf{M}_\opnormal$ is a $j_\opnormal$-by-$j_\opnormal$ matrix, 
$\mathbf{M}_\opmagic$ is a $j_\opmagic$-by-$j_\opmagic$ matrix, 
and $\mathbf{B}$ is a rectangular $j_\opnormal$-by-$j_\opmagic$ matrix.
The zero in $\mathbf{M}$ is a $j_\opmagic$-by-$j_\opnormal$ null matrix. It is there because normal qubits at level $k$ are only made from normal qubits at level $(k-1)$. 
The matrix $\mathbf{M}$  thus has $\left[(j_\opnormal+j_\opmagic)^2-j_\opnormal j_\opmagic\right]$ positive elements, so the resources will depend on $\left[(j_\opnormal+j_\opnormal)^2-j_\opnormal j_\opmagic\right]$ parameters of the error-correction circuitry.
These can be extracted from a given fault-tolerant scheme in the manner outlined at the end of Sec.~\ref{sec:resources_for_basic_components}.

To find the resources with $K$ levels of concatenation, we have (in analogy with Eq.(\ref{eq:recursion_T_Steane})),
\begin{eqnarray}
\left( \begin{array}{c} \Qphys_\opnormal \\ \Qphys_\opmagic \end{array}\right) 
&=& \mathbf{M}^K
\left( \begin{array}{c} \Qalgo_\opnormal \\ \Qalgo_\opmagic \end{array}\right).
\label{eq:recursion_T_Steane_app}
\end{eqnarray}
Here
$\Qalgo_\opnormal$ and $\Qalgo_\opmagic$ are vectors whose elements are the number of logical (level-$K$) qubits of that subtype in the bottleneck layer of the algorithm, $\Lmax$.
Then
$\Qphys_\opnormal$ and $\Qphys_\opmagic$ are the same vectors for the number of physical (level-0) qubits in the layer $\Lmax$.

In this case, it is unlikely that there is an easy way to find the bottleneck layer, $\Lmax$, which we recall corresponds to the timestep of the logical (level-$K$) circuit that requires the most physical (level-0) qubits. Thus, the method used in Sec.~\ref{sec:layer} is likely to be too simplistic. It is likely that we will have to use our matrix approach to calculate the physical qubit cost for each plausible candidate for the bottleneck layer, and only the result of these calculations will reveal to us which one is the true bottleneck layer, $\Lmax$.

At this point, we assume $\mathbf{M}_\opnormal$, $\mathbf{M}_\opmagic$ and $\mathbf{M}$ are all diagonalizable and we diagonalize $\mathbf{M}$ in the textbook manner to get
$\mathbf{M} = \mathbf{U} \mathbf{M}_{\rm diag} \mathbf{U}^{-1}$. Here, $\mathbf{M}_{\rm diag}$ is the diagonal matrix of eigenvalues.
The zero in $\mathbf{M}$'s lower-left corner means that the eigenvalues of $\mathbf{M}$ are those of $\mathbf{M}_\opnormal$ and $\mathbf{M}_\opmagic$. We chose $\mathbf{M}_{\rm diag}$ so that its first $j_\opnormal$ diagonal elements are eigenvalues of $\mathbf{M}_\opnormal$, and its remaining $j_\opmagic$ diagonal elements are eigenvalues of $\mathbf{M}_\opmagic$.
Then $\mathbf{U}$'s $j$th column is the right eigenvector for $\mathbf{M}_{\rm diag}$'s $j$th eigenvalue.
The number of physical qubits for $K$ level of concatenation is then given by
\begin{eqnarray}
Q_{\rm total}^{\rm physical} 
&=& \sum_{i,j,j'=1}^{j_\opnormal+j_\opmagic} \left( \mathbf{U}_{ij'} \left[\mathbf{U}^{-1}\right]_{j'j}Q_j^{\rm algo} \right) \, \lambda_{j'}^K\, .\qquad \
\label{eq:Qtot-arbitrary-sized-matrix}
\end{eqnarray}
Here, $\lambda_{j'}$ for $1 \leq j' \leq j_\opnormal$ correspond to the eigenvalues of $\mathbf{M}_\opnormal$ and $\lambda_{j'}$ for $j_\opnormal < j' \leq j_\opnormal+j_\opmagic$ to the eigenvalues of $\mathbf{M}_\opmagic$.
This is the general analytic result for the total physical qubit cost that is required when there are multiple types of normal and magic gates.

Now we turn to evaluating $R$, we recall that it was previously defined as a ratio; it is the physical qubit cost required by a given algorithm, divided by the resources required by a fictitious algorithm of the same size in which all level-$K$ magic operations are replaced by level-$K$ normal operations. Equivalently, this means we replace the magic qubits by normal qubits at level-$K$. However,  this replacement is ambiguous whenever there are multiple types of normal qubits at level-$K$ with different physical qubit costs, unless we specify which type of level-$K$ normal qubits replaces the level-$K$ magic qubits.  
So we define $R$ as the physical qubit cost required by a given algorithm, divided by the resources required a fictitious algorithm of the same size in which all magic qubits are replaced by normal qubits of type $m$. To evaluate the resources for this fictitious algorithm we first note that the number of magic gates in the bottleneck layer $\Lmax$ of the real algorithm is $Q_\opmagic^{\rm algo}= \sum_{i=j_{\opnormal+1}}^{(j_\opnormal+j_\opmagic)} Q_i^{\rm algo}$. Thus, in this layer, the fictional algorithm (where all magic gates are replaced by normal gates of type 
$m$) has $Q_j^{\rm algo}$ replaced by 
$Q_j^\text{fict}$
where 
\begin{eqnarray}
Q_j^\text{fict} &=& 
\left\{ \begin{array}{cll} 
Q_j^\text{algo} +\delta_{jm} Q_\opmagic^{\rm algo} & \mbox{for} &  j\leq j_\opnormal
\\
0  & \mbox{for} &  j >  j_\opnormal
\end{array}\right.
\label{eq:Q_j^fict}
\end{eqnarray}
where $\delta_{jm}$ is a Kronecker delta-function.   
In this case the numerator of $R$ is given by Eq.~(\ref{eq:Qtot-arbitrary-sized-matrix}), while the denominator of $R$ is given by Eq.~(\ref{eq:Qtot-arbitrary-sized-matrix}) with $Q_j^{\rm algo}$ replaced by 
$Q_j^\text{fict}$. Here, we write $R$ in an equivalent manner as
\begin{eqnarray}
R &=& \frac{
\sum_{i,j,j'=1}^{j_\opnormal+j_\opmagic} \left( \mathbf{U}_{ij'} \left[\mathbf{U}^{-1}\right]_{j'j}Q_j^{\rm algo} \right) \, \lambda_{j'}^K
}
{
\sum_{i,j,j'=1}^{j_\opnormal} \left( \widetilde{\mathbf{U}}_{ij'} \left[\widetilde{\mathbf{U}}^{-1}\right]_{j'j}Q_j^\text{fict} \right) \, \lambda_{j'}^K
}\, , \qquad
\label{Eq:R-general-matrix-case}
\end{eqnarray}
where the denominator contains the matrix $\widetilde{\mathbf{U}}$ which diagonalizes $\mathbf{M}_\opnormal$. This is because the denominator assumes a circuit with only normal qubits, hence its physical qubit's count can be found through $\mathbf{M}_\opnormal$ alone.

\subsubsection{Behavior of $R$ in the large-$K$ limit}
\label{sec:asymtotic_general_case}

Now we look at the behavior of $R$ in the large $K$ limit, to see if a circuit containing magic operations requires more physical qubits than a circuit exclusively composed of normal operations, when there is a large number of concatenations.
For this simple, but useful, conclusions can be derived from our matrix approach by looking at the eigenvalues of $\mathbf{M}$ in the numerator and denominator of $R$ in \eqref{Eq:R-general-matrix-case}. 
We note that the numerator of Eq.~(\ref{Eq:R-general-matrix-case}) contains eigenvalues for both magic and normal qubits, while the denominator contains only eigenvalues for normal qubits.
If the largest eigenvalue appearing in the numerator of $R$ is larger than the largest one appearing on the denominator, $R$ will diverge as $K$ increases. If not, $R$ will go to a finite constant value at $K\to\infty$.

Note that the choice of $m$, i.e., the choice of the type of normal qubits that replace magic qubits to make the fictional circuit, changes the denominator in Eq.~(\ref{Eq:R-general-matrix-case}),
hence it changes the value of $R$. However, 
so long as we choose $m$ among the types of normal qubits that are already in the original abstract circuit implementing the algorithm, then the choice of $m$ simply changes the value of the prefactors in
the denominator, without changing any eigenvalues. This does not change how the denominator scales as $K\to \infty$, so the choice of $m$ does not change whether $R$ diverges or remains constant as $K\to \infty$.

This shows the interest of our matrix approach in more complicated situations with multiple types of normal and magic qubits. 
It tells us the physical qubit cost in the large $K$ limit can very easily be identified from inspecting the eigenvalues of the matrix $\mathbf{M}$. While we do not explore the general case for finite $K$ here, we do a detailed analysis of such an example in  Sec.~\ref{sec:example-flag-qubit_EC}, where $\mathbf{M}$ is a 3-by-3 matrix because that is a 7-qubit scheme with one type of normal qubit, but two types of magic qubit.
\begin{figure*}
\centering   
\includegraphics[width=0.80\textwidth]{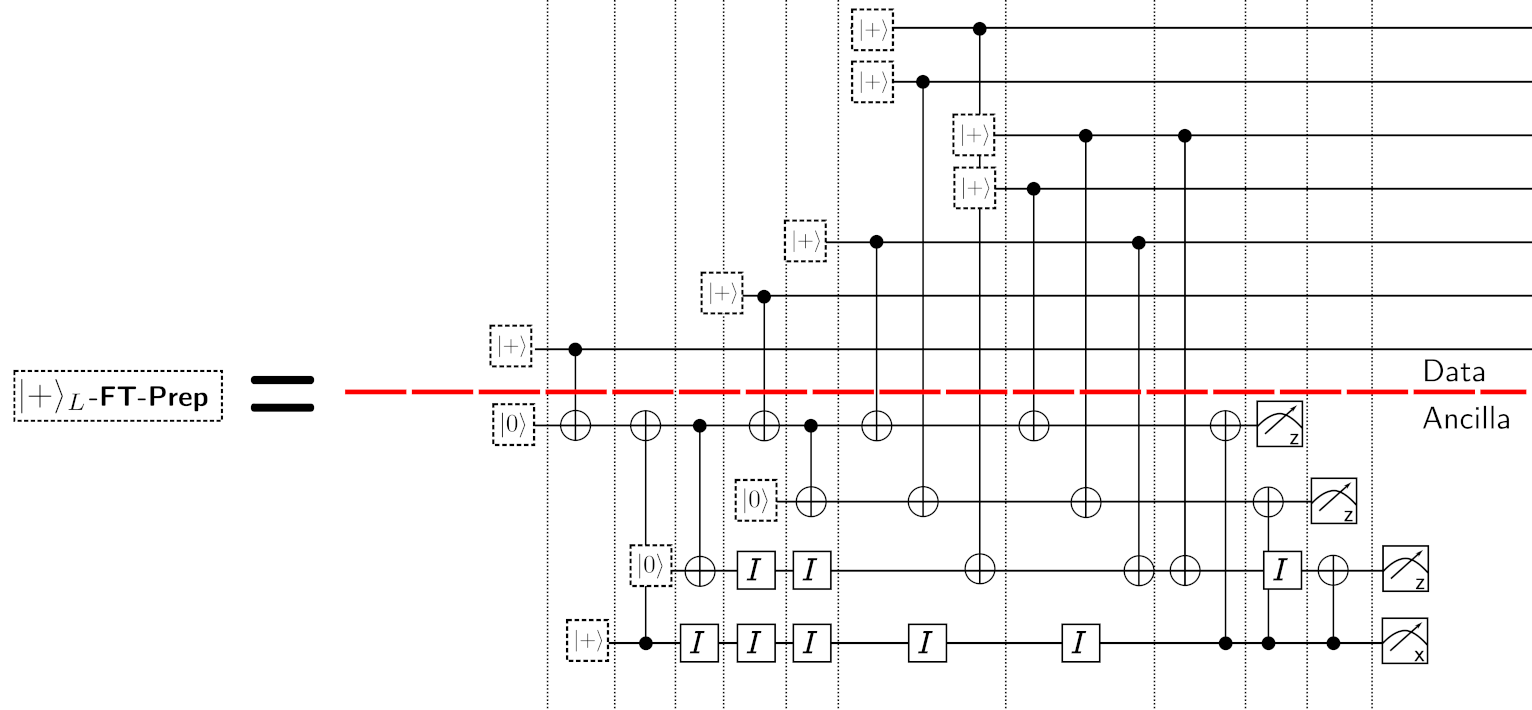}
\caption{Fault-tolerant preparation of the logical $\ket{+}$: $\ket{+}\!\textrm{-FT-Prep}$ encoded in 7-qubit code (top $7$ qubits), with a verification based on flag qubits, from Refs \cite{chamberland_fault-tolerant_2019} and \cite{chao_quantum_2018}. The verification requires $4$ ancillary qubits, at the bottom, see \cite{chao_quantum_2018} and Appendix C.1 of \cite{chamberland_fault-tolerant_2019}. The bottom ancillary qubit is the flag qubit: if it is measured in $\ket{-}$, the state is rejected. There are numerous identity gates in this circuit (timesteps at which the qubit does nothing except stores its state for a later gate), however we mark with ``$I$'' those identity gates whose failure could lead to a state rejection, and thereby contribute to the state's  failure probability. The fault-tolerant preparation of the logical $\ket{0}\!\textrm{-FT-Prep}$ can be found by swapping the roles of Pauli $X$ and $Z$ everywhere in the circuit, i.e. swapping target and control for all $\cnot$s, exchanging the role of $X$ and $Z$ measurements, and $\ket{0}$ and $\ket{+}$ preparation. }\label{fig:+_prep_flag}
\end{figure*}
\begin{figure*}
\centering   
\includegraphics[width=0.85\textwidth]{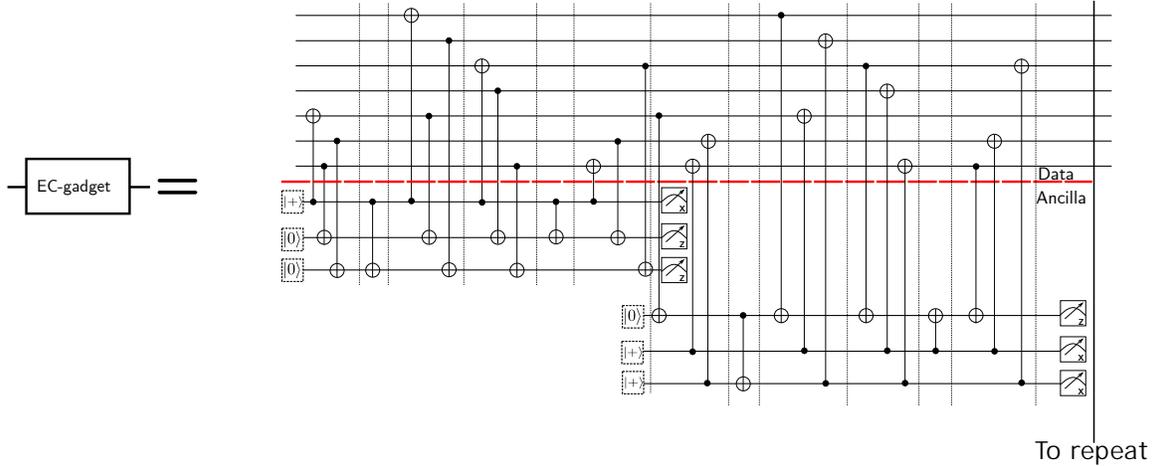}
\caption{Error correction gadget based on flag qubits. The top $7$ lines represent the 7-qubit code encoded qubit. It requires $3$ ancillary qubits to extract the syndrome, but $3 r^{\text{flag}}_{s}=6$ in practice due to the fact the ancillary states take multiple timesteps to be fault-tolerantly prepared, in particular for $k>1$. Hence, the two pairs of $3$ ancillary qubit at the bottom are extracting the $6$ stabilizers of 7-qubit code (see \cite{reichardt_fault-tolerant_2021} for the original proposal). To guarantee fault-tolerance, this circuit is repeated a few times, however the number of repetitions, $J^\text{flag}_\text{EC-repeat}$, does not appear in our evaluation of physical qubit cost. This is because our resource estimates would depend on $J^\text{flag}_\text{EC-repeat}$ in a negligible manner, as explained in Sec.~\ref{sec:flag+shared} and at the end of Sec.~\ref{sec:time_mult_flag}. Hence, we neglect this dependence.
} 
\label{fig:EC_gadget_flag}
\end{figure*}
\begin{figure}
\centering   
\includegraphics[width=0.85\columnwidth]{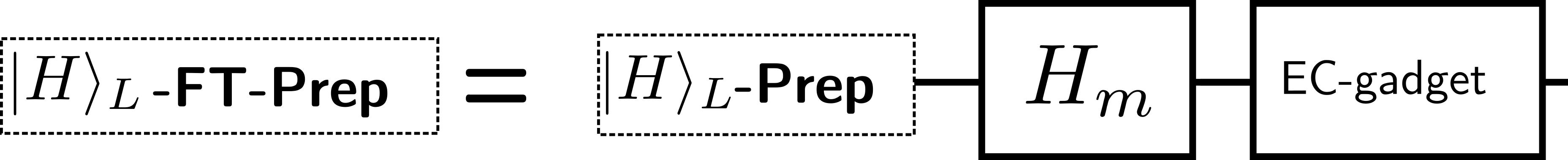}
\caption{Fault-tolerant preparation of the magic state 
$\ket{H}\!\textrm{-FT-Prep}$, taken from \cite{chamberland_fault-tolerant_2019}. 
It takes as an input the state $\ket{H}\!\textrm{-Prep}$ whose preparation is 
shown inn Fig.~\ref{fig:H_prep_flag}, while the circuit for the gadget $H_m$ is in Fig.~\ref{fig:Hm_flag}, and the circuit for the EC gadget is that already shown in Fig.~\ref{fig:EC_gadget_flag}.
Note that the formal definition of the state $\ket{H}=T_Y\ket{0}$. 
\label{fig:H_FT_Prep_flag}}
\end{figure}
\begin{figure}
\centering   
\includegraphics[width=0.9\columnwidth]{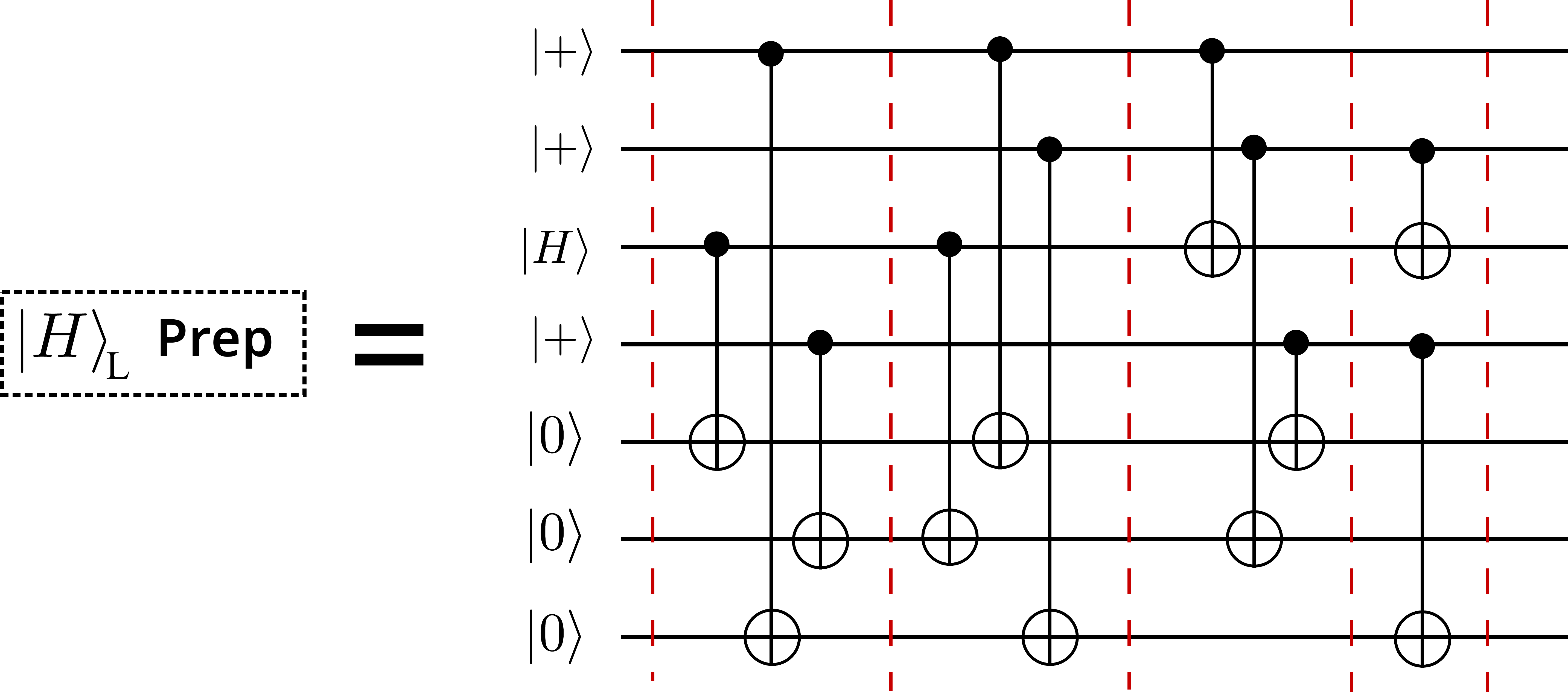}
\caption{Non-fault-tolerant preparation of the state $\ket{H}\equiv T_Y \ket{0}$ in 7-qubit code. This preparation is not fault-tolerant as a single fault for any of the $\cnot$s (or idling locations) can introduce more than one error on the logical qubit. The circuit in Fig.~\ref{fig:H_FT_Prep_flag} allows to make the preparation fault-tolerant, according to the description provided in the text of the appendix. \label{fig:H_prep_flag}}
\end{figure}
\begin{figure}
\centering   
\includegraphics[width=0.95\columnwidth]{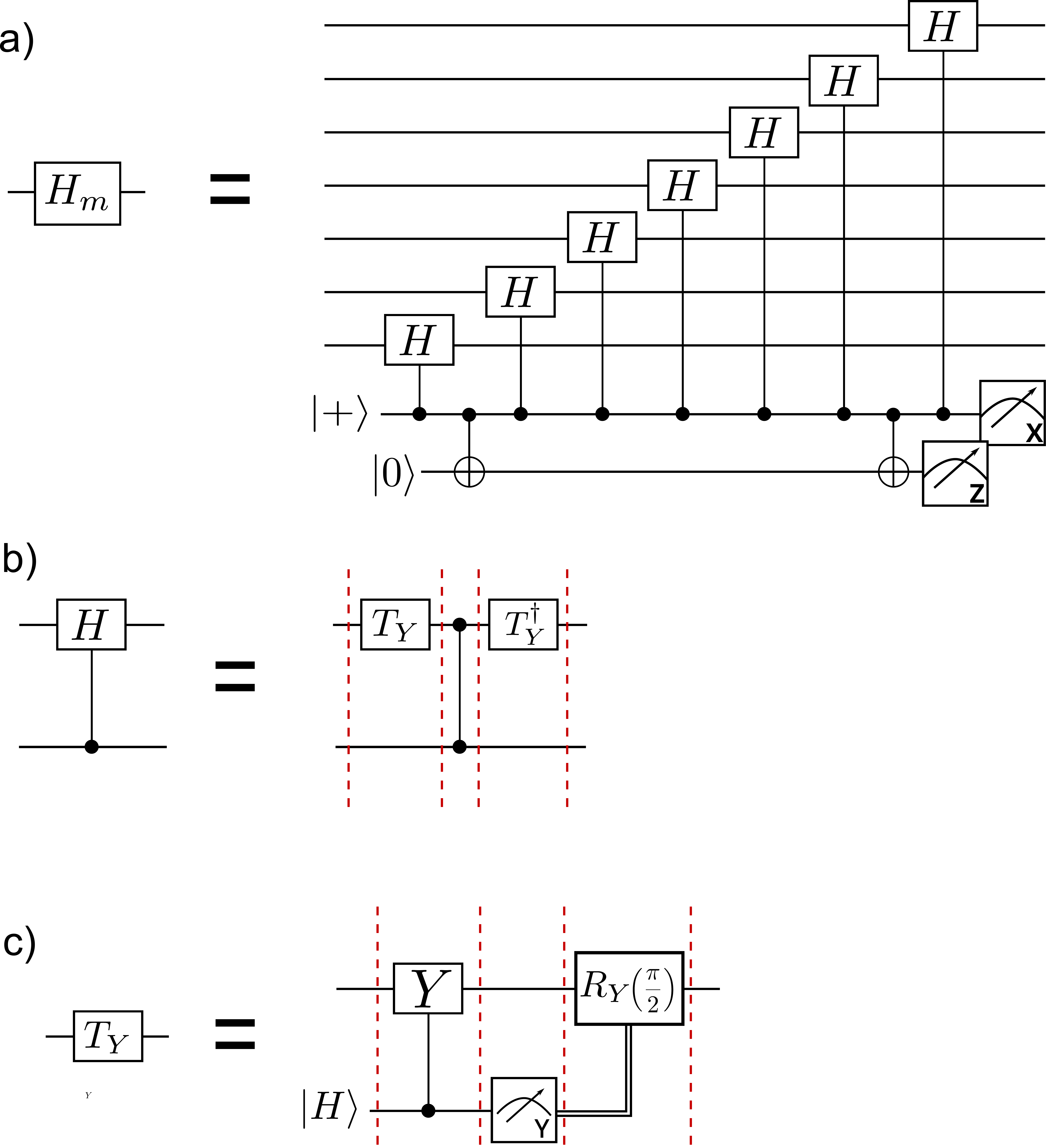}
\caption{a) The top $7$ qubits represent a logical qubit encoded in 7-qubit code. The $2$ ancillary qubits at the bottom are doing a flag fault-tolerant measurement of the logical Hadamard operator of the logical qubit: the top one will provide the outcome of the logical Hadamard measurement while the bottom one is the flag, used to make the overall preparation of the $\ket{H}$ state (shown in Fig.~\ref{fig:H_FT_Prep_flag}) fault-tolerant.
b) The controlled Hadamard is decomposed as indicated here. State injection for the $T_Y$ and $T^{\dagger}_Y$ gates will be done for $k \geq 1$. To keep notations light, the $T_Y$-gate is simply labeled $T$ in the parts referring to flag qubits. Yet, it does represent the $T_Y=\exp(-i \pi/8 Y)$ gate. Note that $T_Y$ is computationally equivalent to the "usual" $T$-gate ($\exp(-i \pi/8 Z)$): they are the same rotation, but performed on two different Pauli axes ($Y$ or $Z$). c) State injection gadget allowing to implement $T_Y$ on the data qubit (top qubit). The gate $T_Y^{\dagger}$ can be implemented by applying an extra Clifford $R_Y(\pi/2)$ on the top qubit after $cY$. $T_Y^{\dagger}$ will have the same effective duration than $T_Y$ as the extra $R_Y(\pi/2)$ can be combined with the classically controlled feedback which either implements a logical identity or $R_Y(\pi/2)$. For $k=0$, $T_Y$ (or $T_Y^{\dagger}$) can directly be implemented at the physical level (i.e., no need to perform state injection).  
\label{fig:Hm_flag}}
\end{figure}

\section{Circuit diagrams and parameters for flag qubits}

Here we give some details on how we arrive at the values of parameters used to evaluate the physical qubit cost for flag-qubit error correction in Sec.~\ref{sec:example-flag-qubit_EC}.
For completeness, we give the circuit diagrams for various parts of the flag-qubit fault-tolerant scheme in Figs.~\ref{fig:+_prep_flag}-\ref{fig:Hm_flag}; these circuits are copied from Refs.~\cite{chao_quantum_2018,reichardt_fault-tolerant_2021,chamberland_fault-tolerant_2019}.  Many parameters can be read directly off these circuits, just as for the Steane EC gadgets, and these are summarized in Fig.~\ref{fig:sketch-flag-qubit}. However, the time multiplicity and failure multiplicities require a little more explanation, so we address them here. 

As mentioned in Sec.~\ref{sec:example-flag-qubit_EC} the magic gate in the flag-qubit scheme is $T_Y \equiv \exp[-i \pi/8 Y]$ (while the usual $T$-gate is $T_Z \equiv \exp[-i \pi/8 Z]$). However, we simply call it a T-gate, because $T_Y$ and $T_Z$ are computationally equivalent (they represent the same rotation but around a different Pauli axis).

\subsection{Time multiplicities for flag-qubits}
\label{sec:time_mult_flag}
We need to estimate the time multiplicity coefficients shown in Table \ref{table:time-multiplicity}. There $r_s^{\text{flag}}$ is the time multiplicity for the ancillary states used in the flag EC gadget, while $r_H^{\text{flag}}$ is the time multiplicity for the $\ket{H}$ magic-state. To be more precise, we assume that the flag EC gadget uses a pool of $3$ ancillary qubits that it might re-use later. This pool corresponds to the first $3$ ancillary qubits in Fig.~\ref{fig:EC_gadget_flag} (meaning that if $r_s^{\text{flag}}=1$, we would only have $3$ ancillary qubits in Fig.~\ref{fig:EC_gadget_flag}, not $6$). The time multiplicities are estimated through a formula $r^{\text{flag}}_i=\lceil \tau_{\text{i,anc}}/\tau_{i, \Delta \text{anc}} \rceil$, where $\tau_{\text{i,anc}}$ is the time to prepare, use and measure the $i \in \{s,H\}$ type of ancilla, and $\tau_{i, \Delta \text{anc}}$ is the minimum time separating two consecutive usage of this ancilla. We recall that the $T$-gate performed with magic-state injection lasts for $2 \tau_L$ (see the state-injection gadget in Fig.~\ref{fig:hard_gate}), and two $T$-gates are always separated by one Clifford, so we have $\tau_{H, \Delta \text{anc}}=3\tau_L$.  Now, each $r_i$ can in practice be fully determined from the circuits we provide in Figs.~\ref{fig:+_prep_flag}, \ref{fig:EC_gadget_flag}, \ref{fig:H_FT_Prep_flag}, \ref{fig:H_prep_flag}, \ref{fig:Hm_flag}. There here we acknowledge a crucial point; the EC gadget in Fig.~\ref{fig:EC_gadget_flag} need to be repeated $J^\text{flag}_\text{EC-repeat}$ times to guarantee fault-tolerance. However, for any for any $J^\text{flag}_\text{EC-repeat}\geq 1$, a numerical estimate of each $r_i^{\text{flag}}$ shows that the values in Table \ref{table:time-multiplicity} are sufficient to guarantee that enough qubits are in the computer (without needing  the exact value of $J^\text{flag}_\text{EC-repeat})$. We say "sufficient", because, strictly speaking, the values in Table \ref{table:time-multiplicity} are upper bounds on the required time multiplicities. 

\subsection{Failure multiplicities for flag-qubits}
Here we explain the assumptions that we used in order
to find the values of $N^{\text{flag}}_{0/+}$ and $N^{\text{flag}}_{H}$ given in Table~\ref{table:failure-multiplicity} (they correspond to, respectively, the first and second line of the fourth column of the flag-qubit section in Table~\ref{table:failure-multiplicity}). For this, we need to find $N^{\text{flag}}_{0/+}$ and $N^{\text{flag}}_{H}$. They are then used in Eq.~\eqref{eq:rob-m-inequality-simpler} to determine the values of failure multiplicities, $\mu^{\text{flag}}_{0/+}$ and $\mu^{\text{flag}}_{H}$, given in the penultimate column of  Table~\ref{table:failure-multiplicity}.

We start with $N^{\text{flag}}_{0/+}$, which is estimated from the state preparation circuit shown in Fig.~\ref{fig:+_prep_flag} taken from Ref.~\cite{chamberland_fault-tolerant_2019}. This circuit prepares the state $\ket{+}$. The same circuit prepares $\ket{0}$ under permutation of Pauli $X$ and $Z$ everywhere in the circuit (see the caption of Fig.~\ref{fig:+_prep_flag}), so the following explanation for $\ket{+}$ preparation applies equally for $\ket{0}$ preparation. 
To guarantee fault-tolerance, this circuit relies on post-selecting the states for which the flag is measured in $\ket{+}$ (the bottom ancillary qubit in Fig.~\ref{fig:+_prep_flag}). This guarantees that a single fault leads to at most a weight-one $X$ error and a weight one $Z$-error on the output state, see Ref.~\cite{chao_quantum_2018}. Hence, such a post-selection satisfies the axioms of a 1-prep (see for instance Appendix A of Ref.~\cite{chamberland_fault-tolerant_2019} for CSS codes such as the 7 qubit code considered here). 
This means that $N^{\text{flag}}_{0/+}$ is the number of ways a single fault could imply that the result of the verification fails --- with verification failing when the flag (the bottom ancillary qubit in Fig.~\ref{fig:+_prep_flag}) ending up in the state $\ket{-}$ (rather than $\ket{+}$). For this, we count all the gates, state preparation and identity gates for which one fault could make the flag qubit measured in $\ket{-}$. From Fig.~\ref{fig:+_prep_flag}, we have $11$ state preparations, $15$ $\cnot$s, $9$ identity gates, and $1$ measurement that could lead to verification failure: it gives $N^{\text{flag}}_{0/+}=36$.

We move on with $N^{\text{flag}}_{H}$. Here, we assumed $N^{\text{flag}}_{H}=\sqrt{2/p_{\text{thres}}}$. This is motivated by the fact $1/p_{\text{thres}}$ counts the number of malignant pair of fault-locations in the biggest exRec of the computation. A malignant pair of fault-location, is a pair of fault-locations for which the introduction of faults introduces an uncorrectable error on the logical qubit of the computation. While not all pairs of fault-locations are necessarily malignant, the numbers are usually sufficiently close for our purpose, since we recall that $\mu_H^{\text{flag}}$ only depends \textit{logarithmically} on $N^{\text{flag}}_{H}$ (see Eq.~\eqref{eq:rob-m-inequality-simpler}), hence an order of magnitude estimate is sufficient. To illustrate the legitimacy of our assumption, we can mention that the number of fault-locations in the largest exRec for Steane EC gadget is $575$, while $\sqrt{2/p_{\text{thres}}}=316$: these are close estimate given the logarithmic dependence of the failure multiplicities.

\section{Values of the error-correction threshold}
\label{supp:sec:toy_model}

Throughout this work, we assume that the error-correction threshold is similar for different error-correction schemes, even when they require very different physical qubit costs.  This greatly simplifies our analysis, because it means we can compare the physical qubit costs of different error-correction schemes, while assuming the ``benefits'' of those schemes are about equivalent; in other words, that each scheme has about the same probability of successfully carrying our any algorithm for given $K$ and given $p_0$. This is the case if they all have similar fault-tolerance thresholds (see Sec.~\ref{sec:basics-of-concatenation}).

For simplicity, we assume that all error-correction schemes considered in this work (Steane, flag-qubit, and our toy-model) have the same threshold as for Steane error-correction \cite{Aliferis2005Apr}; $p_{\text{thres}}=2 \times 10^{-5}$. 
However, this assumption needs justifying, as it may seem surprising that a flag-qubit EC gadget has a similar capacity to correct errors as a Steane EC gadget, when the flag-qubit gadget is much smaller.
Yet, the literature strongly implies that this is the case.
In short, this is because flag-qubit EC gadgets require less ancillary qubits, but use them for more timesteps, so the number of fault-locations that determine the threshold appears to be similar to  that for Steane EC gadgets.

Having said that, there is not yet a consensus in the literature on the {\it true} threshold for flag-qubits.
Such a {\it true} threshold would be one that applies for any gate operation and a reasonably-general noise model
($p_{\text{thres}}=2 \times 10^{-5}$ for Steane EC gadgets under an adversarial independent stochastic noise model \cite{Aliferis2005Apr} is such a true threshold).
Instead, different works (listed below) calculate what are called {\it pseudothresholds} for flag-qubits; these are thresholds that only apply for specific gate-operations exposed to specific noise, and are in principle only applicable for a single level of concatenation. As expected, each gate-operation and noise model gives different pseudothresholds. The works listed below show that 
some of these pseudothresholds for flag-qubits EC gadgets are better than the equivalent pseudothreshold for Steane EC gadgets, and some are worse.
However, most of these pseudothresholds for flag-qubits EC gadgets are the same order of magnitude as the equivalent one for Steane EC gadgets. We therefore assume, as a reasonable working hypothesis, that the true threshold for flag-qubits EC gadgets is also similar to that for Steane EC gadgets, so for simplicity we take it to be the same.

To justify the above claim about the pseudothresoholds, we now list one by one various works, and the pseudothresholds that they found for particular gate operations and noise models.
Ref.~\cite{Chamberland2018Feb} evaluated the pseudothreshold for a quantum memory (i.e. a logical identity gate) depolarizing noise model and different noise strength for identity gates compared to two-qubit gates. When the identity gates had a comparable noise strength to two-qubit gates, the pseudothreshold for flag-qubit EC gadgets was $3.39 \times 10^{-5}$, while that for the  Steane EC gadget was $6.29 \times 10^{-4}$, which is about eighteen times bigger.
When the noise of the identity gates is 100 times less than that of the two-qubit gate, the pseudothreshold for flag-qubit EC gadgets was $1.41 \times 10^{-5}$, while that for Steane EC gadgets was $3.84 \times 10^{-5}$, which is more than two time bigger.  This suggests that the advantage in one scheme compared to the other heavily depends on the noise model.
Ref.~\cite{Chao2018Aug} did numerical simulations for a quantum memory with depolarizing noise. They defined the pseudothreshold differently, through the leading order term in the expression of the logical error-rate. They then found this the pseudothreshold for flag-qubit EC gadgets was about $2\times 10^{-3}$, while that for Steane EC gadgets was $1.4\times 10^{-3}$. So a pseudothreshold defined in this way is very similar for the two types of EC gadgets, while, this time, the flag-qubit EC-gadget has a slightly bigger pseudothreshold than Steane.

Ref.~\cite{Liou2023Feb} went beyond a quantum memory, and considered the pseudothreshold for a logical $\cnot$ gate. 
Again they took depolarizing noise, and found a pseudothreshold for flag was $3.02 \times 10^{-5}$. This is smaller than the equivalent pseudothreshold for Steane EC gadgets in Ref.~\cite{Aliferis2005Apr}, which was $9.3 \times 10^{-5}$.  

All these results strongly suggest that the threshold for the two types of error-correction are similar, despite the fact they have very different physical qubit costs.
As this work is concentrating on the analysis of physical qubit costs, it is a natural simplification to assume the true threshold is the same in all cases. 
As the true threshold is closer to that for logical $\cnot$ gates than to that for quantum memory, it can be expected that it is a bit below  the pseudothreshold in Ref.~\cite{Liou2023Feb}. Hence, it seems reasonable to assume the flag-qubit error-correction threshold is about $p_{\rm thres} = 2\times 10^{-5}$, the same as the Steane error-correction threshold for an adversarial independent stochastic noise model \cite{Aliferis2005Apr}.

\bibliography{biblio}
\end{document}